\documentclass[3p,12pt]{elsarticle}
\usepackage[fleqn]{amsmath}
\usepackage{cases}
\usepackage{booktabs}
\usepackage{multirow}
\usepackage{xcolor}
\usepackage[normalem]{ulem}
\usepackage{tabularx}
\usepackage{siunitx}

\usepackage{comment}
\usepackage[ruled, linesnumbered]{algorithm2e}
\usepackage{algpseudocode}
\usepackage{grffile}
\usepackage{rotating}
\usepackage{lineno}
\usepackage{hyperref}
\usepackage{graphicx,enumerate}
\usepackage{amssymb}
\allowdisplaybreaks[4]
\usepackage{bm}
\usepackage[caption=false]{subfig}
\usepackage{caption,color}
\usepackage{subeqnarray}
\usepackage{multirow}
\usepackage{lscape}
\usepackage{rotating}
\usepackage{graphics}
\usepackage{epstopdf}
\usepackage{threeparttable,booktabs,tabularx}
\journal{Journal of Computational Physics}
\usepackage{floatrow}
\usepackage{floatrow}\newfloatcommand{capbtabbox}{table}[][\FBwidth]
\floatname{algorithm}{Algorithm}
\biboptions{numbers,sort&compress} 

\newcommand{\bn}{\bm n}

\newcommand{\bq}{\bm q}

\newcommand{\bu}{\bm u}

\newcommand{\bx}{\bm x}

\newcommand{\bxi}{\bm \xi}

\newcommand{\myd}{\;\mathrm{d} }

\begin{document}
\begin{frontmatter}

\title{GSIS-ALE for moving boundary problems in rarefied gas flows} 

\author{Jianan Zeng}
\author{Yanbing Zhang}
\author{Lei Wu\corref{mycorrespondingauthor}}
\ead{wul@sustech.edu.cn}

\address{Department of Mechanics and Aerospace Engineering, Southern University of Science and Technology, Shenzhen 518055, China}


\begin{abstract}
Multiscale rarefied gas flows with moving boundaries pose significant challenges to the numerical simulation, where the primary difficulties involve robustly managing the mesh movement and ensuring computational efficiency across all flow regimes.
Build upon recent advancements of the general synthetic iterative scheme (GSIS), this paper presents an efficient solver to simulate the large displacement of rigid-body in rarefied gas flows. The newly developed solver utilizes a dual time step method to solve the mesoscopic kinetic and macroscopic synthetic equations alternately, in an arbitrary Lagrangian-Eulerian framework. Additionally, the overset mesh is used and the six degree-of-freedom rigid body dynamics equation is integrated to track the motion of solids. 
Four moving boundary problems encompassing a wide range of flow velocities and gas rarefaction are simulated, including the periodic pitching of airfoil, particle motion in lid-driven cavity flow, two-body separation in supersonic flow, and three-dimensional lunar landing, demonstrating the accuracy and efficiency of the GSIS in handling multi-scale moving boundary problems within an overset framework.
\end{abstract}


\end{frontmatter}
\section{Introduction}\label{sec:1}

In the realm of aerospace engineering, unsteady flows with moving objects are often encountered. For instance, during the lift-off and touchdown of the Chang'e lunar exploration missions~\cite{wei2023illumination}, the spacecraft frequently modulates its jet thrust to adjust its orientation. 
The plume-induced hysteresis effects, coupled with the nonlinear unsteady aerodynamic influence of the jet propulsion, are critical factors that must be meticulously investigated.
Due to the high cost of experiments, numerical simulation has become indispensable for understanding these flows. 
While the traditional computational fluid dynamics based on the Navier-Stokes (NS) equations is well developed to handle moving object problems, the rarefaction effects 
associated with the low pressure environment and spatially-fast-varying macroscopic quantities make the numerical simulation a huge challenge, as the complicated Boltzmann equation or simplified kinetic model equations, which are defined in a six-dimensional phase-space, should be solved. 

The kinetic equations can be solved by the stochastic~\cite{bird1994molecular} and deterministic~\cite{broadwell1964study,chu1965kinetic,Aristov2001} methods. 
The stochastic direct simulation Monte Carlo (DSMC) method is widely used~\cite{bird1994molecular}, yet deterministic discrete velocity methods~\cite{yang1995rarefied,li2004study} have demonstrated growing potential for solving the simplified kinetic models~\cite{bhatnagar1954model,holway1966new,shakhov1968approximate} and the Boltzmann equation~\cite{Tcheremissine2006,wu2013deterministic,Wu2015JCP}.
However, the splitting of streaming and collision processes in both methods can be computationally intensive for near-continuum flows, as the spatial cell size and time step are constrained by the mean free path and mean collision time, respectively.
Furthermore, the DSMC is an explicit time-marching scheme, which is often impractical for  scenarios with moving boundaries (e.g., two-body separation), as the characteristic flow time may exceed the computational time step by several orders of magnitude.
These challenges have spurred the development of multiscale methods to overcome limitations on cell size and time step, such as the unified gas kinetic scheme (UGKS)~\cite{xu2010unified,zhu2016implicit}, the Fokker-Planck DSMC method~\cite{gorji2011fokker}, the discrete UGKS (DUGKS)~\cite{guo2013discrete,yuan2021multi},  the asymptotic NS preserving Monte Carlo method~\cite{fei2023}, and the general synthetic iterative scheme (GSIS)~\cite{su2020can, Su2020SIAM, zhang2023efficient, zeng2023CICP}.

In recent years, we have seen sporadic developments of multiscale methods for moving boundary problems. For instances, Chen \textit{et al.} evaluated the moving face flux in a rest reference frame and developed an arbitrary Lagrangian-Eulerian (ALE)-UGKS for simulating a rotating ellipse and a Crookes radiometer in rarefied gases~\cite{chen2012unified}. Wang \textit{et al.} incorporated the ALE technology and a mesh deformation method into the DUGKS for small movement problems, such as the fluid-structure interaction in microelectromechanical systems~\cite{wang2019arbitrary,wang2022investigation,wang2023arbitrary}. 
Recently, the DUGKS is coupled with the immersed boundary method to simulate moving boundary problems~\cite{tao2018combined,he2024thermal}.
Shrestha \textit{et al.} used the DSMC method as the fluid solver and the Newton-Euler method to solve the rigid body motion equations, enabling the simulation of rigid body motion in rarefied gases, such as a thermophoretic motion of a particle of complex shape~\cite{shrestha2015numerical}. Most studies are limited to small displacement motions using mesh transformation, which have not demonstrated the potential for handling multi-body separation problems using body-fitted meshes. Moreover, these works adopt explicit time-marching methods, where the uniform global time step is determined by the smallest cell size and the global Courant-Friedrichs-Lewy (CFL) number. Consequently, the global time step may become extremely small due to the refinement of cell sizes in near-wall regions, such as the leading edge or the tip of a wing. Although the multiscale property alleviates the restriction of spatial cell size, the computational efficiency in unsteady flow simulations is still constrained by the minimum time step, which may be on the order of the local collision time.

To overcome these limitations, this research extends the GSIS to tackle moving boundary problems in rarefied gas flows from the continuum to free-molecular flow regimes. The innovation lies in its capacity to integrate the ALE with a moving mesh within an unstructured overset framework, enabling the resolution of intricate relative motion across the full spectrum of gas rarefaction and a broad spectrum of velocities. 
Additionally, by inheriting the fast convergence and asymptotic preserving characteristics~\cite{Su2020SIAM}, the implicit solver notably reduces the constraints on cell size and time step.
Furthermore, the proposed GSIS-ALE methodology offers the advantage of simplicity, meaning that anyone well-versed in the ALE method within traditional computational fluid dynamics can easily write the GSIS-ALE code~\cite{zhang2023efficient}.

The rest of the paper is organized as follows. In Section~\ref{sec:2}, the Shakhov kinetic model is introduced, and the associated macroscopic synthetic equation in GSIS is presented. In Section~\ref{sec:3}, the general framework of the GSIS-ALE and the overset mesh method are described. Numerical tests to assess the accuracy and efficiency of GSIS-ALE are performed in Section~\ref{sec:4}. In Section~\ref{application}, the GSIS-ALE is applied to study the two-body separation in supersonic flows and three-dimensional lunar landing. Finally, conclusions and outlooks are provided in Section~\ref{sec:5}.
\section{Mesoscopic kinetic equation and GSIS}\label{sec:2}

In this section, the Shakhov kinetic model~\cite{shakhov1968approximate} is employed to describe the dynamics of monatomic gases, and the synthetic equation in GSIS-ALE is introduced.

\subsection{The Shakhov kinetic equation}

We denote $L_0, T_0$, and $\rho_0$ as the reference flow length, temperature, and mass density, respectively. The reference velocity is defined as $v_0 = \sqrt{k_BT_0/m}$, where $k_B$ is the Boltzmann constant, and $m$ is the molecular mass. The Knudsen number $\text{Kn}$ and the rarefaction parameter $\delta_{rp}$ are defined as:
\begin{equation}\label{eq:defined_kn}
    \text{Kn} = \frac{\mu(T_0)}{\rho_0 L_0}\sqrt{\frac{\pi m}{2k_B T_0}}, \quad \delta_{rp} = \sqrt{\frac{\pi}{2}}\frac{1}{\text{Kn}},
\end{equation}
where $\mu$ is the shear viscosity obtained using the inverse power-law potential, following the relation $\mu(T)=\mu(T_0)\left(T/T_0\right)^{\omega}$, with $\omega$ being the viscosity index. For argon, $\omega=0.81$.

The velocity distribution function $f(\bx, \bxi, t)$ is used to describe the states of monatomic gas, where $t$ is the time, $\bx$ is the spatial coordinate, and $\bxi$ is the molecular velocity. Macroscopic quantities, such as the mass density $\rho$, flow velocity $\bm{u}$, temperature $T$, pressure tensor $\bm{P}$ (normalized by $\rho v_0^2$) and heat flux $\bm{q}$ (normalized by $\rho v_0^3$), are obtained by taking the moments of velocity distribution function $f$:
\begin{equation}\label{eq:getmoment}
\begin{aligned}
    &\rho = \langle 1,f \rangle,\quad
    \rho\bm{u} = \langle {\bxi},f \rangle, \quad
    \frac{3}{2}\rho T=\left\langle \frac{1}{2}|\bxi-\bm{u}|^2, f \right\rangle,\\
    &\bm{P} =\langle (\bxi-\bm{u}) \otimes (\bxi-\bm{u}), f \rangle,\quad
    \bm{q} =\left\langle \frac{1}{2}|\bxi-\bm{u}|^2 (\bxi- \bm{u}), f \right\rangle,
\end{aligned}
\end{equation}
where the operator $\langle h,\psi\rangle=\int h\psi \myd \bxi$ is defined as the integral of $h\psi$ over the whole velocity space $\bxi$. 

The Shakhov kinetic equation is used to describe the evolution of velocity distribution function:
\begin{equation}\label{equ:generalModel}
\frac{\partial f}{\partial t}+\bxi\cdot \frac{\partial f}{\partial \bx} = \frac{g-f}{\tau},
\end{equation}
where the local dimensionless relaxation time is
\begin{equation}
    \tau(\bx)= \frac{1}{\delta_{rp}} \times \frac{T(\bx)^{\omega -1}}{\rho(\bx) },
\end{equation}
and the local reference distribution function is
\begin{equation}\label{reference VDF}
\begin{aligned}
g(\bx)= g^{eq}(\bx)\left[1+\frac{2\bq(\bx) \cdot (\bxi - \bu(\bx))}{15 \rho(\bx)T^2(\bx)}\left(\frac{ |\bxi - \bu(\bx)|^2}{2T(\bx) }-\frac{5}{2}\right)\right],
\end{aligned}
\end{equation}
with $g^{eq}(\bx)$ being the local  equilibrium (Maxwellian) VDF:
\begin{equation}\label{equi_VDF}
g^{eq}(\bx)= \rho(\bx) \left(\frac{1}{2\pi T(\bx)}\right)^{3/2}
\exp\left(-\frac{|\bxi - \bu(\bx)|^2}{2T(\bx)}\right).
\end{equation}

\subsection{ALE-type kinetic equation and boundary conditions}


The ALE framework is utilized to simulate moving boundary problems. During unsteady motion, the geometric information of the physical mesh, such as the coordinate of nodes, the normal vector of surfaces, and the volume of cells, may change temporally. A surface motion velocity $\bm{v}$ that modifies the convective flux is introduced. By integrating Eq.~\eqref{equ:generalModel} in a time dependent control volume $V$ of the finite volume mesh and applying the Gauss theorem, we have
\begin{equation}\label{equ:kinetic_ale}
   \frac{\partial }{\partial t}\int_{V} f\myd V+\oint_{\partial V}(\bxi-\bm{v})\cdot \bm{n} f \myd S  = \int_{V}\frac{g-f}{\tau}\myd V, 
\end{equation}
where $\partial V$ represents the boundaries set of the control volume $V$, $S$ represents the area of the surface, and $\bm{n} = \bm{S}/|\bm{S}|$ is the normal direction of the surface. Note that the surface motion velocity only affects the convective term on the left-hand side of the mesoscopic (and consequently, macroscopic) equations. The Eulerian form of the governing equations can be regarded as a special case of the ALE form equations, when the mesh motion velocity is zero. 

The diffusive boundary condition is applied at the wall, where gas molecules are reflected diffusely from the moving wall in thermodynamic equilibrium. The velocity distribution of gas molecules at the moving wall is given by:
\begin{equation}
    f_{\text{wall}} = \left\{
    \begin{array}{cc}
        f_{\text{in}}, &  \bm{n}\cdot(\bxi-\bm{v}_{\text{wall}}) \geq 0, \\
        g_{\text{wall}}^{\text{eq}}, & \bm{n}\cdot(\bxi-\bm{v}_{\text{wall}}) < 0. 
    \end{array}
    \right.
\end{equation}
Here, $\bm{v}_{\text{wall}}$ represents the wall velocity, $ f_{\text{in}} $ is the velocity distribution function of gas molecules incident on the wall, computed using a second-order interpolation method; $g_{\text{wall}}^{\text{eq}}$ is the fully thermally equilibrium of gas molecules reflected from the wall under specified wall temperature and velocity, with the density determined by the non-penetration condition:
\begin{equation}
    \int_{\bm{n}\cdot(\bxi-\bm{v}) \geq 0} (\bxi-\bm{v}_{\text{wall}}) f_{\text{in}} \mathrm{d}\bm{\xi} = -\int_{\bm{n}\cdot(\bxi-\bm{v}_{\text{wall}})  < 0} (\bxi-\bm{v}) g_{\text{wall}}^{\text{eq}} \mathrm{d}\bm{\xi}.
\end{equation}
For the far-field boundary used in this paper, the distribution function is set to be locally Maxwellian corresponding to the gas with density $\rho_\infty$, velocity $\bm{u}_{\infty}$ and temperature $T_{\infty}$.


\subsection{Macroscopic synthetic equations}


Under the ALE framework, by taking the velocity moments of the kinetic model~\eqref{equ:generalModel}, the governing equations for conservative variables $\bm{W}=[\rho, \rho\bm{u}, E]^{\intercal}$ are obtained as follows:
\begin{equation}\label{eq:macroscopic_equation_2}
    \frac{\partial \bm{W}}{\partial t} + \frac{\partial \bm{F}}{\partial \bx} = 0, 
    \quad\text{with} \quad 
    \bm{F} = \bm{F}_c-\bm{v}\bm{W} + \bm{F}_v,
\end{equation}
where $E=\frac{3}{2}\rho T+\frac{1}{2}\rho u^2$ denotes the kinetic energy, and  $\bm{F}_c$ and $\bm{F}_v$ represent the convective and viscosity fluxes, respectively:
\begin{equation}\label{eq:macro_flux_define}
\bm{F}_c= [\rho \bu, \rho \bu\bu+p, \bu E]^{\intercal}, \quad \bm{F}_v = [0, \bm{\sigma}, \bm{\sigma}\cdot\bm{u}+\bm{q}]^{\intercal},  
\end{equation}
with $\bm{\sigma} = \bm{P} - \rho T \bm{\mathrm{I}}$ being the deviatoric stress tensor, and $\bm{\mathrm{I}}$  the identity matrix.

Equation~\eqref{eq:macroscopic_equation_2} is not closed, since expressions for the stress $\bm{\sigma}$ and heat flux $\bm{q}$ are still unknown. The exact stress tensor and heat flux without any truncation should be calculated directly from velocity distribution function, rather than a simple first-order linear constitutive relations obtained by the Chapman-Enskog expansion~\cite{chapman1990mathematical}. In GSIS, the stress and heat flux are separated into the NS constitutive relations and higher-order terms (HoTs) in the follow form~\cite{zeng2023CICP}:
\begin{equation}\label{eq:full_constitutive}
\begin{aligned}
    &\bm{\sigma} = \underbrace{-\mu \left(\nabla\bm{u}+\nabla\bm{u}^{\mathrm{T}}-\frac{2}{3}\nabla\cdot\bm{u}\mathrm{I}\right)}_{ \bm{\sigma}^{\text{NS}}} + \text{HoT}_{\bm{\sigma}},\\
    &\bm{q} = \underbrace{-\kappa \nabla T}_{\bm{q}^{\text{NS}}} + \text{HoT}_{\bm{q}}.
\end{aligned}
\end{equation}
where $\bm{\sigma}^{\text{NS}}$ and $\bm{q}^{\text{NS}}$ are the conventional NS constitutive relations describing the linear dependence of shear stress and heat flux on the gradient of velocity and temperature, respectively; and the corresponding nominal shear viscosity $\mu$ and heat conductivity $\kappa$ are given by,
\begin{equation}
    \mu=\rho T \tau, \quad \kappa = \frac{c_p}{ \Pr}\mu.
\end{equation}
with $c_p = 5 / 2$ being the heat capacity at constant pressure. The HoTs describe the rarefaction effects, which is calculated by subtracting the parts of NS constitutive relations from the moments of the velocity distribution functions:
\begin{equation}\label{eq:getHoTs}
\begin{aligned}
    \text{HoT}_{\bm{\sigma}} &= \left \langle(\bxi-\bu)\otimes (\bxi-\bu)-\frac{\mathrm{I}}{3}|\bxi-\bu|^2, f\right \rangle  -\bm{\sigma}^{\text{NS}},\\
    \text{HoT}_{\bm{q}} &= \left \langle \frac{1}{2}(\bxi-\bu)|\bxi-\bu|^2, f\right \rangle -\bm{q}^{\text{NS}}.
\end{aligned}
\end{equation}

Note that the NS constitutive relations in Eqs.~\eqref{eq:full_constitutive} and \eqref{eq:getHoTs} are calculated at different iteration steps. Consequently, these terms cannot cancel each other out until the final steady state is reached. As demonstrated in numerical simulations, this kind of approach facilitates fast convergence in the entire range of gas rarefaction~\cite{Su2020SIAM, Zhu2021JCP, zeng2023general}.
Moreover, it enpowers the asymptotic preserving property, releasing the constraint on the spatial cell size and time step~\cite{zeng2023CICP} in the (near) continuum flow regime. 

\section{ALE-type GSIS with overset mesh}\label{sec:3}


Detailed parallel implementation of the GSIS in the Eulerian framework for steady state solutions can be found in Ref.~\cite{zhang2023efficient}.
To address the time-dependent moving boundary problems, this work employs the remapping-free ALE technique, where the mesh motion velocity $\bm{v}$ modifies the convection terms in meso- and macro-scopic governing equations. The overset mesh assembly and the numerical methods for the mesoscopic kinetic and macroscopic synthetic equations are elaborated below. 

\subsection{Temporal discretization of ALE-type GSIS}

For the unsteady flow evolution with a given time step $\Delta t=t^{n+1}-t^n$, the governing equations are discretized using a second-order time-accurate scheme. Spatial discretization employs the unstructured cell-centered finite volume method. By applying the Gauss theorem to convert the volume integration into a sum of fluxes through surfaces, the spatiotemporal discrete forms of the kinetic and synthetic equations are as follows:
\begin{align}
   & \label{eq:dis_micro_single_time_step}
  \frac{3f_i^{n+1}-4f_i^n+f_i^{n-1}}{2\Delta t} + \frac{1}{V_i}\sum_{j\in N(i)} \xi_{v,n} f_{ij}^{n+1}S_{ij}=\frac{g^{n}_{i}-f^{n+1}_{i}}{\tau^{n}_i},\\ 
  \label{eq:dis_macro_single_time_step}
  &\frac{3\bm{W}_i^{n+1}-4\bm{W}_i^n+\bm{W}_i^{n-1}}{2\Delta t} + \frac{1}{V_i}\sum_{j\in N(i)}\bm{F}_{ij}^{n+1}S_{ij}=0,
\end{align}
where $f_i$ and $\bm{W}_i$ are the averaged distribution function and conservative variables for the discrete cell $i$, respectively. $N(i)$ denotes the set of neighboring cells of $i$, and cell $j$ is one of the neighbors, the interface connecting the cell $i$ and cell $j$ is denoted as subscript $ij$. Hence, $V_i$ is the volume of cell $i$, and $S_{ij}$ represents the area of the cell interface $ij$. $\xi_{v,n} = (\bm{\xi}-\bm{v})\cdot \bm{n}_{ij}$ is the normal component of the relative molecular velocity $\bm{\xi}-\bm{v}$ along normal direction $\bm{n} = \bm{S}/|\bm{S}|$ in  coordinate, and  $\xi_{n}$ will uniformly be used to represent $ \xi_{v,n} $ subsequently, for the sake of simplicity. $\xi_n f_{ij}, \bm{F}_{ij}$ is the interface flux of the velocity distribution function and macroscopic governing equation, respectively.

An additional local numerical time step $\Delta t_{p,i} = t^{m+1} - t^{m}$ is introduced in a dual-time stepping method to solve Eqs.~\eqref{eq:dis_micro_single_time_step} and~\eqref{eq:dis_macro_single_time_step}.
Let $ f^m $ and $ W^m $ denote the intermediate approximate solutions at the inner iteration step $ m $, the above two governing equations can be written as:
\begin{align}
    \label{eq:dis_micro}
&\frac{f^{m+1}-f^{m}}{\Delta t_{p,i}}+\frac{3f_i^{m+1}-4f_i^n+f_i^{n-1}}{2\Delta t} + \frac{1}{V_i}\sum_{j\in N(i)} \xi_{v,n} f_{ij}^{m+1}S_{ij}=\frac{g^{m}_{i}-f^{m+1}_{i}}{\tau^{m}_i},\\ 
  \label{eq:dis_macro}
&\frac{\bm{W}^{m+1}-\bm{W}^{m}}{\Delta t_{p,i}}+  \frac{3\bm{W}_i^{m+1}-4\bm{W}_i^n+\bm{W}_i^{n-1}}{2\Delta t} + \frac{1}{V_i}\sum_{j\in N(i)}\bm{F}_{ij}^{m+1}S_{ij}=0.
\end{align}
Once the inner $ m $-th loop converges, we set $ f^{n+1} = f^{m+1}$. 
To apply a simple matrix-free implicit solving of the above discretized equations, incremental variables $ \Delta f_i^m = f_i^{m+1} - f_i^m $ and $ \Delta\bm{W}_i^m = \bm{W}_i^{m+1} - \bm{W}_i^m $ are introduced for a local pseudo time step $\Delta t_{p,i}$. The delta-form governing equations of Eq.~\eqref{eq:dis_micro} and Eq.~\eqref{eq:dis_macro} for the implicit iterative algorithm can be written as follows:
\begin{align}
\label{eq:delta_form_micro}
&\left( \frac{3}{2\Delta t} + \frac{1}{\Delta t_{p,i}} + \frac{1}{\tau_i^m}\right)\Delta f_i^{m} + \frac{1}{V_i}\sum_{j\in N(i)} \xi_{n}\Delta f_{ij}^{m}S_{ij}=r_i^m,\\
\label{eq:delta_form_macro}
&\left(\frac{3}{2\Delta t}+ \frac{1}{\Delta t_{p,i}}\right)\Delta \bm{W}_i^{m} + \frac{1}{V_i} \sum_{j\in N(i)} \Delta\bm{F}_{ij}^{m}S_{ij}=\bm{R}_i^m.
\end{align}
with
\begin{align}
    r_i^m=&\frac{g^{m}_{i}-f^{m}_{i}}{\tau^{m}_i}-\frac{1}{V_i}\sum_{j\in N(i)} \xi_{n}f_{ij}^{m}S_{ij}-\frac{3f_i^{m}-4f_i^n+f_i^{n-1}}{2\Delta t},\\ 
    \bm{R}_i^m =&-\frac{1}{V_i}\sum_{j\in N(i)} \bm{F}_{ij}^{m}S_{ij} - \frac{3\bm{W}_i^{m}-4\bm{W}_i^n+\bm{W}_i^{n-1}}{2\Delta t}.
\end{align}

Note that $r_i^m$ and $R_i^m$ represent the mesoscopic and macroscopic residuals in the $m$-th step, respectively. Since the implicit fluxes in delta-forms will not effect the steady convergent solution, the mesoscopic and macroscopic fluxes on the cell interface $ij$ on the left hand side of the Eq.~\eqref{eq:delta_form_micro} can be constructed by the first order upwind scheme. The accuracy is completely determined by the calculation of the residual terms. To achieve a second-order spatial accuracy, the interface flux values on the right sides are reconstructed as follow:
\begin{equation}\label{eq:interfaceflux}
\begin{aligned}
    &\xi_n f_{ij}=\frac{1}{2}\xi_n^+f_L + \frac{1}{2}\xi_n^-f_R,\quad \xi_n \Delta f_{ij}= \frac{1}{2}\xi_n^+\Delta f_i + \frac{1}{2}\xi_n^-\Delta f_j,\\
    &F_{ij} = \mathcal{F}(W_L, W_R, S_{ij}), \qquad \Delta \bm{F}_{ij} = \frac{1}{2}[\Delta \bm{F}_i + \Delta \bm{F}_j+\Gamma_{ij}(\Delta\bm{W}_i - \Delta\bm{W}_j)],
\end{aligned}
\end{equation}
where $\xi_n^{\pm}=[1\pm \text{sign}(\xi_n)]$ represents the interface sign directions relative to the cell center value. Specifically, we define $f_{L/R} = f_{i/j} + \phi \nabla f_{i/j} \cdot \bm{x}$, where $\phi$ is calculated using the Venkatakrishnan limiter. The macroscopic flux $\bm{F}_{ij}$ is decomposed into convective and viscous components as $\bm{F}_{ij} = \bm{F}_c + \bm{F}_v$ (see Eq.~\eqref{eq:macro_flux_define}). The visco0us flux $\bm{F}_v$ is computed using central differencing, while the AUSM+UP scheme is employed in the convective flux. The reconstructed macroscopic variables on the left and right sides of the interface are given by $\bm{W}_{L/R} = \bm{W}_{i/j} + \phi \nabla \bm{W}_{i/j} \cdot \bm{x}$, with the limiter $\phi$ calculated using the Venkatakrishnan method. To simplify the numerical fluxes on the left-hand side of Eq.~\eqref{eq:delta_form_macro}, a first-order upwind scheme is employed. Additionally, for the macroscopic terms in the delta-form flux $\Delta \bm{F}_{ij}$, fluxes based on the Euler equations are used to simplify the implicit increments of macroscopic fluxes. Here, $\Gamma_{ij} = |U_n| + c_s + \frac{2\mu}{\rho|\bm{n}_{ij} \cdot (\bm{x}_j - \bm{x}_i)|}$. Since the control volume satisfies $\sum{j\in N(i)}\bm{n}_{ij}A_{ij}=0$, it follows that $\sum_{j\in N(i)}\bm{F}_iA_{ij}=0$, allowing the flux to be directly represented by the convective flux. The flux with subscript $j$ can be written in a matrix-free form as $\Delta \bm{F}_j^m=\bm{F}_c(\bm{W}_j^m + \Delta \bm{W}_j^m) - \bm{F}_c(\bm{W}_j^m)$. Substituting Eq.~\eqref{eq:interfaceflux} into Eq.~\eqref{eq:delta_form_macro}, the implicit governing equations for mesoscopic and macroscopic equation become:
\begin{equation}
\begin{aligned}
    &d_i\Delta f_i^{m} + \frac{1}{2V_i}\sum_{j\in N(i)}\bm{\xi}_n^-\Delta f_j^{m}S_{ij}=r_i^m,\\
    &D_i\Delta \bm{W}_i^{m} + \frac{1}{2V_i}\sum_{j\in N(i)} \left(\Delta\bm{F}_{j}^{m} - \Gamma_{ij}\Delta \bm{W}_j^m\right)S_{ij}=\bm{R}_i^m,
\end{aligned}
\end{equation}
where the matrix elements are
\begin{equation}
    \begin{aligned}
d_i&=\frac{3}{2\Delta t}+\frac{1}{\Delta t_{p,i}} + \frac{1}{\tau_i^m} + \frac{1}{2V_i}\sum_{j\in N(i)}\bm{\xi}_n^+S_{ij},\\
D_i &= \frac{3}{2\Delta t}+\frac{1}{\Delta t_{p,i}}+\frac{1}{2V_i}\sum_{j\in N(i)}\Gamma_{ij}S_{ij}.
    \end{aligned}
\end{equation}

The above equations can be solved by the standard point relaxation method. Additionally, the global physical time step $\Delta t$ is determined by the flow characteristic time, which depends on the specific problem. The local pseudo time step $\Delta t_{p,i}$ in the incremental equations is determined by the CFL number, the local characteristic length $\Delta x_i$, and the maximum molecular velocity $\bxi_{\text{max}}$,
\begin{equation}
    \Delta t_{p,i} = \text{CFL}\frac{\Delta x_i}{|\bm{u}_i + \bxi_{\text{max}}|},
\end{equation}
with $\text{CFL}_{DVM} =10^5$ and $\text{CFL}_{NS}=10^3$ for the kinetic and macroscopic solvers, respectively.


\subsection{Algorithm of GSIS-ALE}

To achieve fast convergence, GSIS alternately solves the mesoscopic kinetic equation and macroscopic synthetic equations, where the solutions of the former provide HoTs and the solutions of the latter facilities the expedited evolution of velocity distribution function in inner iteration. To be specific, the time evolution of GSIS for unsteady flow from $t^n$ to $t^{n+1}$ is given below: 
\begin{enumerate}[\bfseries Step 1]
    \item For a given macroscopic properties $\bm{W}^{n}$ and the velocity distribution function $f^{n}$ in a previous iteration step $n$, set $\bm{W}^{(m=0)}=\bm{W}^n$ and $f^{(m=0)}=f^n$.
    \item When both the macroscopic properties $\bm{W}^{m}$ and the velocity distribution function $f^{m}$ are obtained, $f^{m+1/2}$ can be solved according to the kinetic equation \eqref{eq:dis_micro}, by replacing $m+1$ with $m+1/2$. Note that the superscript $m+1/2$ means that the velocity distribution function solved at this stage have not been corrected by the solutions of synthetic equations. \label{loop1}
    
    \item For given velocity distribution function $f^{m+1/2}$, the corresponding macroscopic variables $\bm{W}^{n+1/2}$ and $\text{HoTs}^{m+1/2}$ are calculated based on Eqs.~\eqref{eq:getmoment} and \eqref{eq:getHoTs}, respectively. Then, the synthetic equations \eqref{eq:macroscopic_equation_2} with the constitutive relations~\eqref{eq:full_constitutive} are solved to obtain $\bm{W}^{m+1}$, where the associated macroscopic boundary conditions are evaluated from $f^{m+1/2}$ as well.
    
    \item The converged solution of macroscopic variables of the synthetic equations $\bm{W}^{m+1}$ is used in the next step of kinetic solver to calculate the equilibrium statement $g^{\text{eq}}$. The velocity distribution function is modified to incorporate the change of macroscopic properties by changing the equilibrium part from $g^{\text{eq}}(\bm{W}^{m+1/2})$ to $g^{\text{eq}}(\bm{W}^{m+1})$:
    \begin{equation}\label{eq:updatef}
            f^{m+1} = f^{m+1/2} + \left[g^{\text{eq}}\left(\bm{W}^{m+1}\right) - g^{\text{eq}}\left(\bm{W}^{m+1/2}\right)\right].
    \end{equation}
    \item Convergence is checked through the macroscopic properties $\bm{W}^{m}$. If the convergence criterion is not met, the process iterates back to Step \ref{loop1} and repeats above loops until convergence is achieved. Upon successful convergence, the solution vectors $\bm{W}^{n+1}$ and $f^{n+1}$ are updated using the newly obtained values of $\bm{W}^{m+1}$ and $f^{m+1}$.
    \item Update the mesh displacement using either the rigid body degree-of-freedom equations or a prescribed trajectory, as detailed in Section~\ref{dof}. Then, at the new time $t^{n+1}$, reassemble the mesh and update the donor cells and mesh motion velocity, as detailed in Section~\ref{overset_assembly}.
\end{enumerate}

\begin{figure}[t]
    \centering
    \includegraphics[width=0.8\textwidth,clip = true]{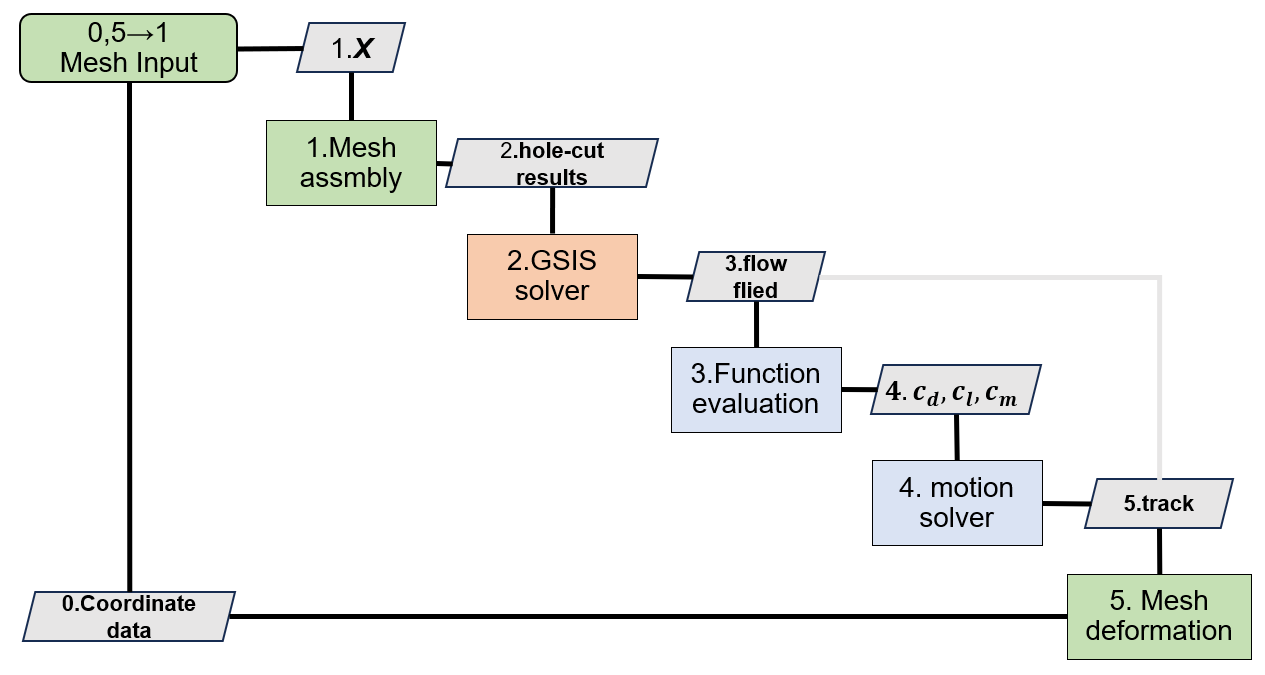}
    \caption{Design structure matrix for GSIS-ALE algorithm.}
    \label{fig:dsm}
\end{figure}
The above steps will be repeated until the evolution process is completed. This process can be addressed by the design structure matrix in Fig.~\ref{fig:dsm}, which shows that the mesh assembly strategy, numerical solver, trajectory generator, and mesh deformation are the primary components of this overset framework. The mesh assembler handles the initial meshes, updating the edge and cell types of the component meshes. The GSIS solver is then used to compute the converged solution at a given physical time step, determining the aerodynamic characteristics of the rigid body. This information is then passed to the rigid body motion module to generate the subsequent trajectory. In addition to solving the trajectory through rigid body motion equations, a predefined trajectory can also be utilized. Finally, based on the new trajectory, the mesh coordinates are updated at each time step.

\subsection{Rigid body motion equation}\label{dof}

The motion of the rigid body follows a given trajectory or be free. When the object is in free motion~\cite{shrestha2015numerical}, the force and torque acting on the rigid body due to the fluid must be determined in advance. Consider a rigid body $B$ with center of mass $\bm{X}$. The force $\bm{\mathcal{F}}$ and torque $\bm{\mathcal{T}}$ exerted on the body from the fluid are given by:
\begin{align}
    \bm{\mathcal{F}} &= -\oint_{\partial S} \sigma \cdot \bn \myd S, \label{eq:force_define}\\
    \bm{\mathcal{T}} &= -\oint_{\partial S} (\bm{x} - \bm{X})\times (\sigma \cdot \bn) \myd S,
\end{align}
where $\partial S$ represents the boundary of the object, $\bn$ is the outward normal vector to the interface, and $\sigma$ is the stress tensor in the fluid. The motion of rigid body $B$ is solved by the Lagrangian method, where the center-of-mass velocity $\bm{v}$  and angular velocity $\bm{\omega}_b$ of rigid object are governed by the equations of degrees of freedom as:
\begin{align}
    M\frac{\myd \bm{v}}{\myd t} &= \bm{\mathcal{F}}, \label{eq:dof_1}\\
    \bm{I}\frac{\myd \bm{\omega}_b}{\myd t} +\omega_b\times(\bm{I} \cdot \bm{\omega}_b) &= \bm{\mathcal{T}},\label{eq:dof_2}
\end{align}
$ M$ represents the mass of the rigid body, and $\bm{I}$ represents the inertia tensor with respect to the body coordinate axes. For a square object with side length $ L$ and density $ \rho$, they are given by $ M = \rho L^2$ and $ I_{zz} = \frac{1}{6} \rho L^4$, respectively. Integrating the above equation over time and using the first-order backward Euler discretization, the displacement $X$ and angular $\Phi$ increment of the rigid body after $n$ time steps are obtained as:
\begin{align}
    \Delta \bm{X} &= \bm{v}\Delta t + \frac{\bm{\mathcal{F}}^n}{2M}\Delta t^2,\label{eq:track_01}\\
    \Delta \bm{\Phi} &= \bm{\omega}_b\Delta t + \frac{\bm{\mathcal{T}}^n - \bm{\omega}_b^n \times (\bm{I} \cdot \bm{\omega}_b^{n})}{2\bm{I}}\Delta t^2.\label{eq:track_02}
\end{align}
Here, Eq.~\eqref{eq:track_01} can be directly computed in the ground coordinate system, while Eq.~\eqref{eq:track_02} is calculated in the body coordinate system. In the above equation, physical quantities with the subscript $b$ represent quantities in the body coordinate system. After obtaining the angular increment in the body coordinate system, the attitude angle of the rigid body can be derived through the transformation matrix between the body coordinate and the ground coordinate. Note that the term $ \bm{\omega}_b \times (\bm{I} \cdot \bm{\omega}_b)$ vanishes in the two-dimensional case, which means that the transformation matrix can be ignored and the angular increment can be directly computed in the ground coordinate system. Based on the change in attitude angle and the displacement of the center of mass, the update of the node positions can be achieved using the quaternion method~\cite{betsch2009rigid}. The geometry information (edge center, edge normal vector, and cell center coordinates) of any $ x \in B$ can be updated, and the edge velocity is given by:
\begin{equation}
    \bm{v}_s = \frac{\bm{x}_s^{n+1} - \bm{x}_s^n}{\Delta t}.
\end{equation}

Through the above steps, given the position, velocity, and force of the center of mass, the six-degree-of-freedom equations are solved to obtain the geometric and velocity information of the grid at the new time step, while simultaneously updating the motion velocity of any interface on the rigid body. 


\subsection{Overset mesh assembly}\label{overset_assembly}


The mesh movement method significantly impacts the simulation accuracy, especially in case of rigid-body large displacement motion. Even in the continuum flows, numerical solutions are already sensitive to the geometric conservation laws~\cite{thomas1979geometric}, non-conservative interpolation~\cite{volkner2017analysis,cui2021high}, and spatiotemporal accuracy compared to small deformation motion.  Mesh movement encompasses a wide array of techniques~\cite{luke2012fast,anderson2005adaptive,yuan2018immersed,tao2018combined}. For examples, spring analogy methods~\cite{degand2002three} move the nodes of mesh cells without altering the mesh topology. However, when mesh deformation leads to negative cell volumes, remeshing method becomes necessary to address the failures caused by large deformations. In contrast to mesh deformation, immersed boundary methods~\cite{tao2022sharp} utilize a fixed background grid and do not move the nodes of the mesh or require remeshing. Instead, this approach reconstructs boundary conditions within the mesh where it intersects the boundary surface to enforce a no-slip boundary condition. The overset method~\cite{horne2019massively} partitions the computational domain into multiple subgrids and facilitates information exchange through grid interpolation. This method possesses several unique features that make it particularly effective for predicting large rigid-body displacements of complex geometric shapes. Firstly, component grids utilize unstructured body-fitted meshes, adept at handling complex geometries and ensuring high-quality local meshes. Secondly, during mesh motion processes, all subgrids do not require regeneration, leading to efficient computations for unsteady problems. Thirdly, compared to stationary meshes, the overset grid method only requires adding an interpolation boundary condition, ensuring compatibility with existing macroscopic and mesoscopic solvers. The overset method has gained considerable attention~\cite{crabill2016high,tang2003overset,hu2021robust} for its application in capturing unsteady flow problems involving relative motion, such as the separation of multistage rockets and the rotational motion of helicopter rotors.

Here the overset mesh technology is utilized based on the cell-centered unstructured finite volume method. The cells are classified into the active (celltype = 1), interpolation ($\text{celltype} = 2/4$), and inactive (celltype = 0) cells, see Fig.~\ref{fig:os_05}. The primitive variables $\bm{W}$ and velocity distribution function $f$ within the first two types of cells are updated through iteration and interpolation, respectively, while those in inactive cells remain unchanged during computation. The interpolation cells form a closed shell encompassing all active cells, which can be regarded as a new boundary condition, see Fig.~\ref{fig:os_06}. The donor cells for interpolation, derived from the active cells of other component meshes, also need to be determined. Take the two-dimensional simulation as an example, the basic steps of the proposed cell-centered overset assembly method based on wall distance are illustrated below:

\begin{figure}
    \centering
    \subfloat[Initial mesh]{\includegraphics[height = 6cm, keepaspectratio]{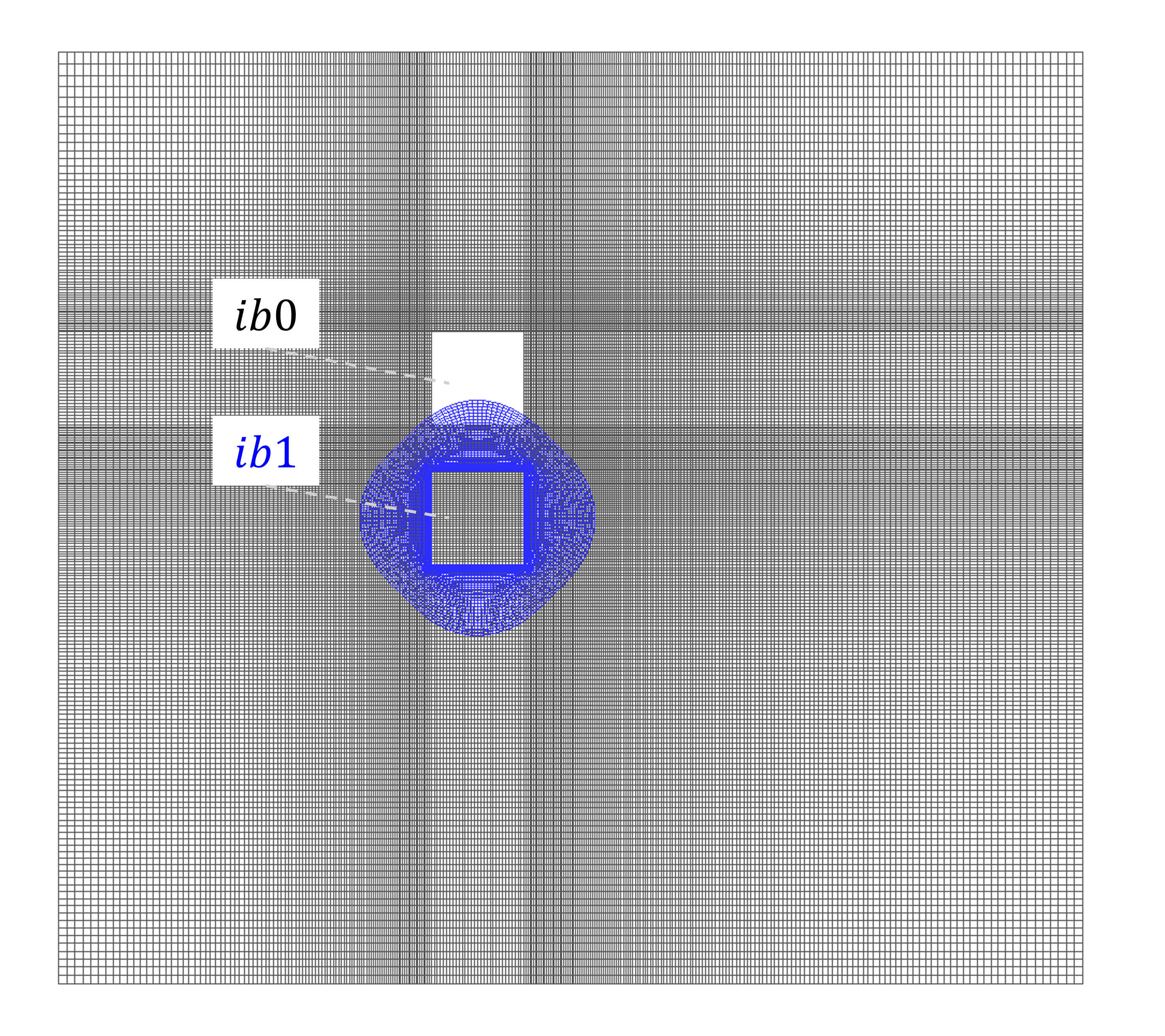}\label{fig:os_01}}
    \subfloat[Preliminary host check]{\includegraphics[height = 6cm, keepaspectratio]{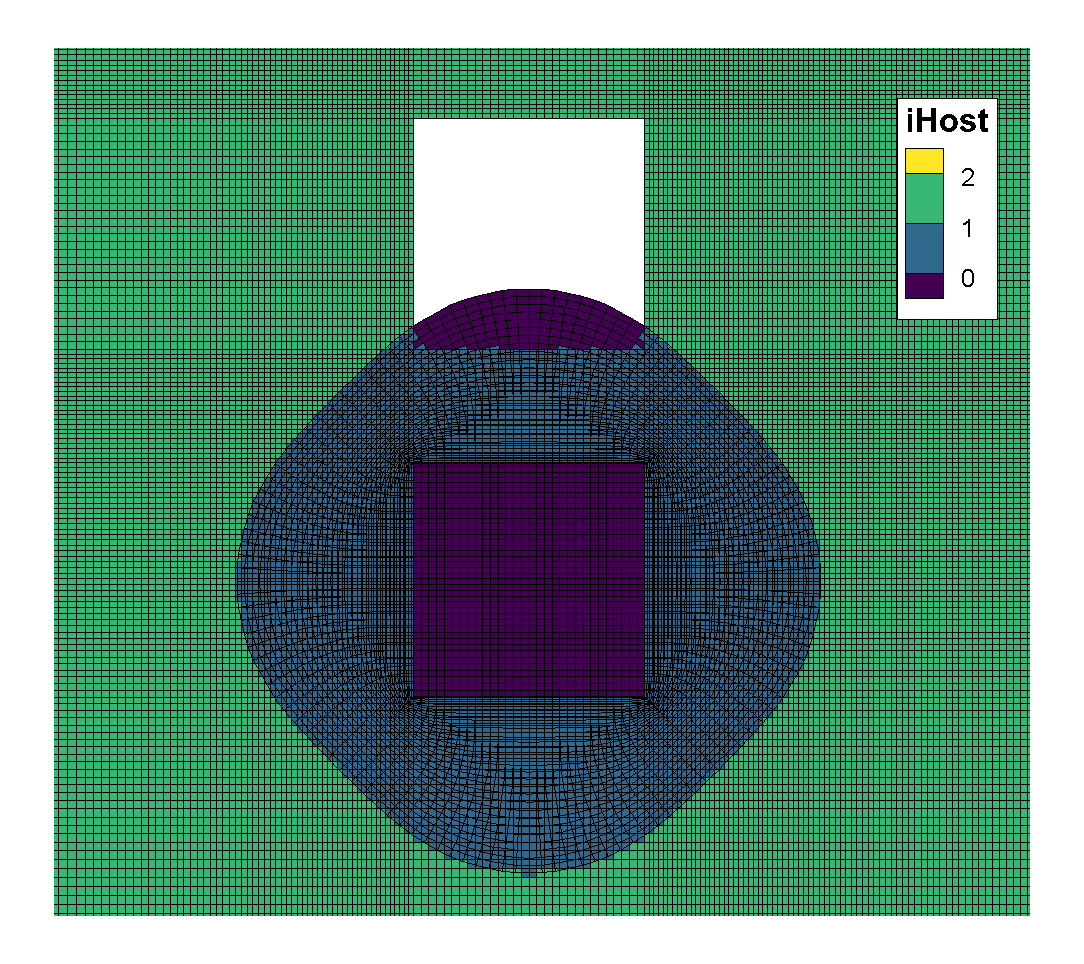}\label{fig:os_02}}\\
    \subfloat[Overset mesh system]{\includegraphics[height = 6cm, keepaspectratio]{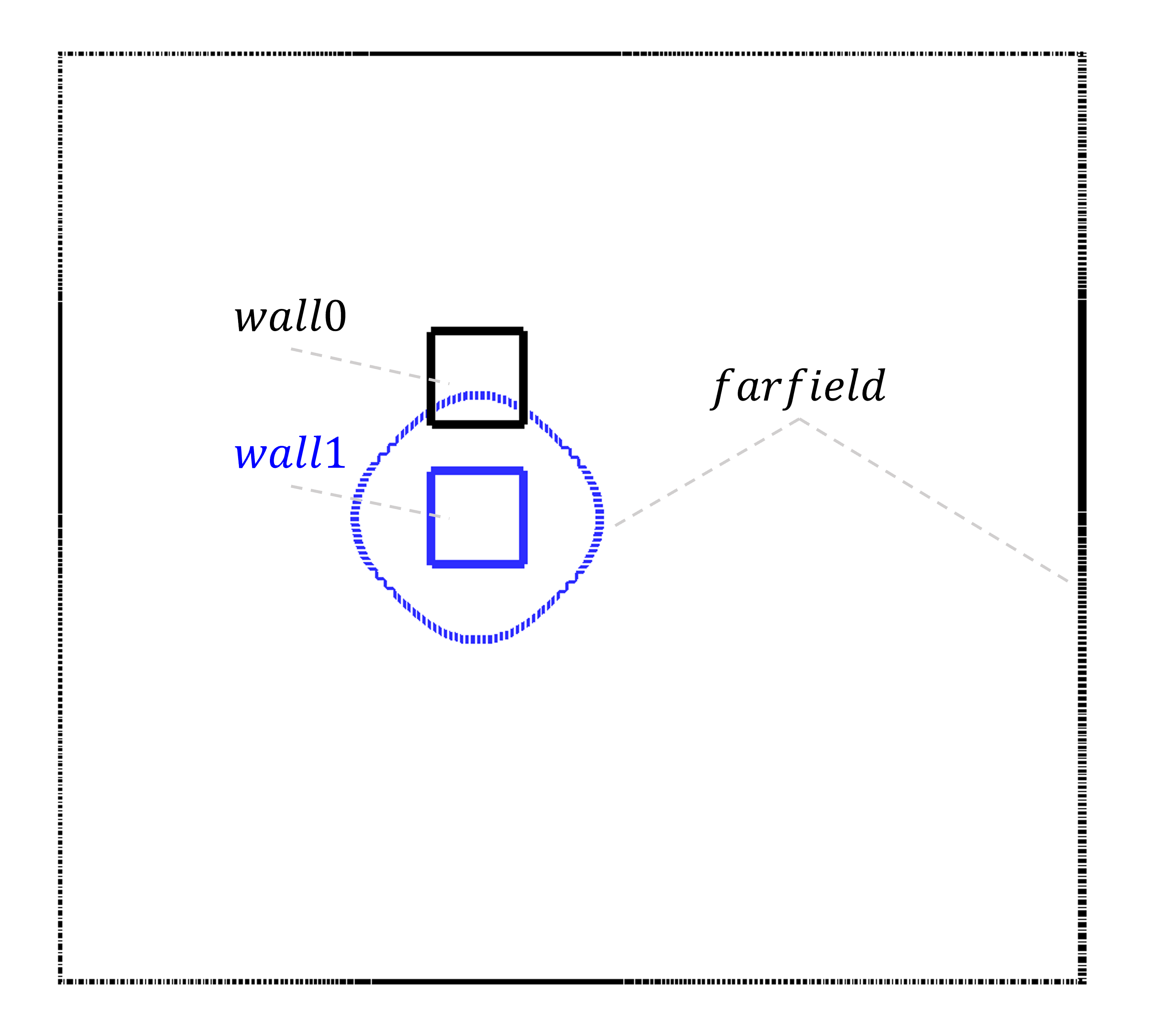}\label{fig:os_03}}
    \subfloat[Wall distance]{\includegraphics[height = 6cm, keepaspectratio]{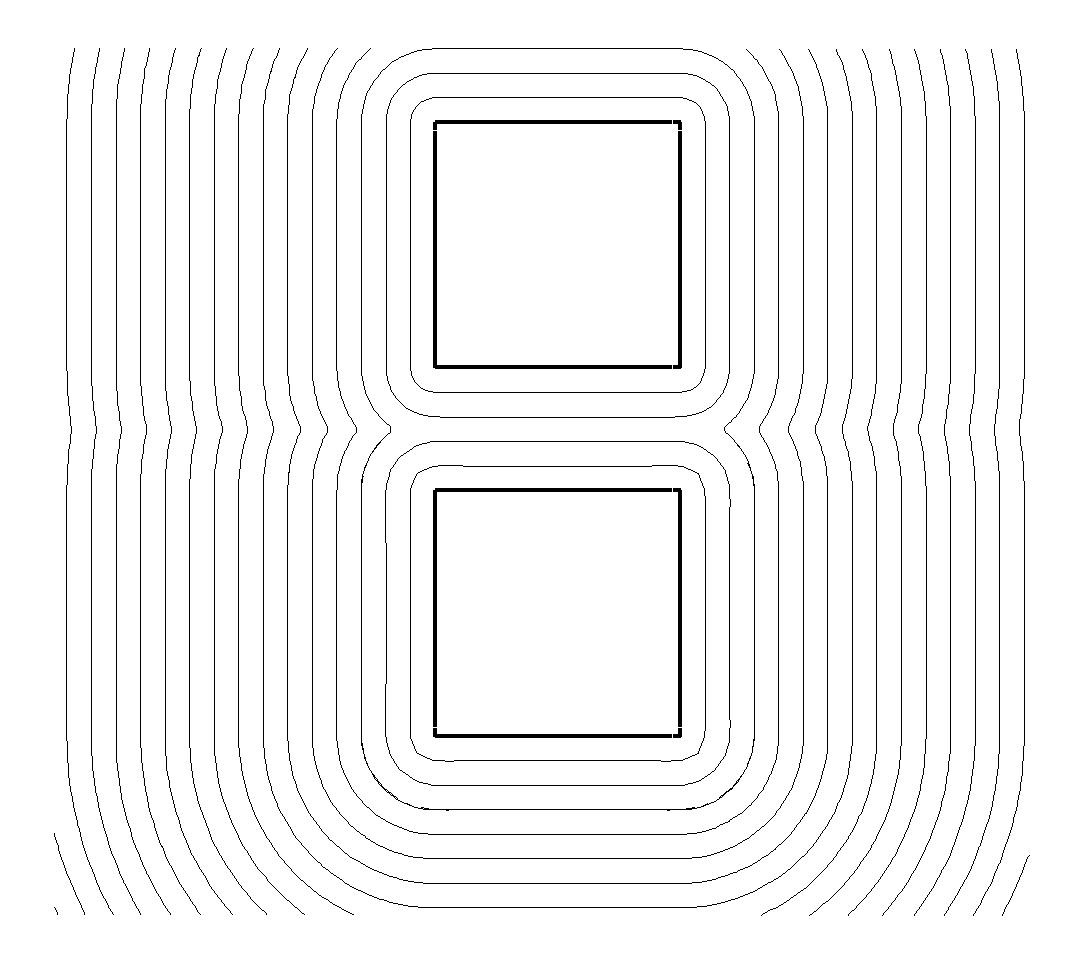}\label{fig:os_04}}
    \\
    \subfloat[Cell classification]{\includegraphics[height = 6cm, keepaspectratio]{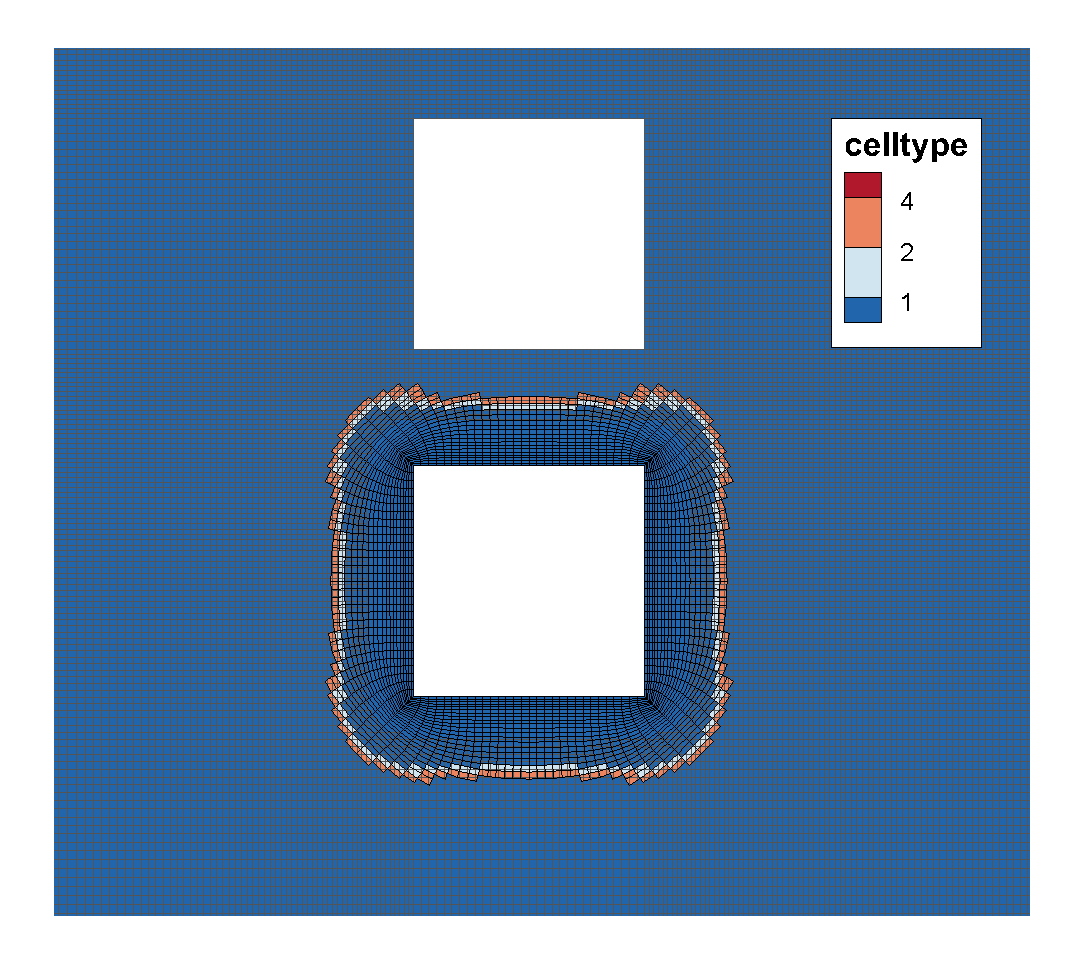}\label{fig:os_05}}
    \subfloat[Edge classification]{\includegraphics[height = 6cm, keepaspectratio]{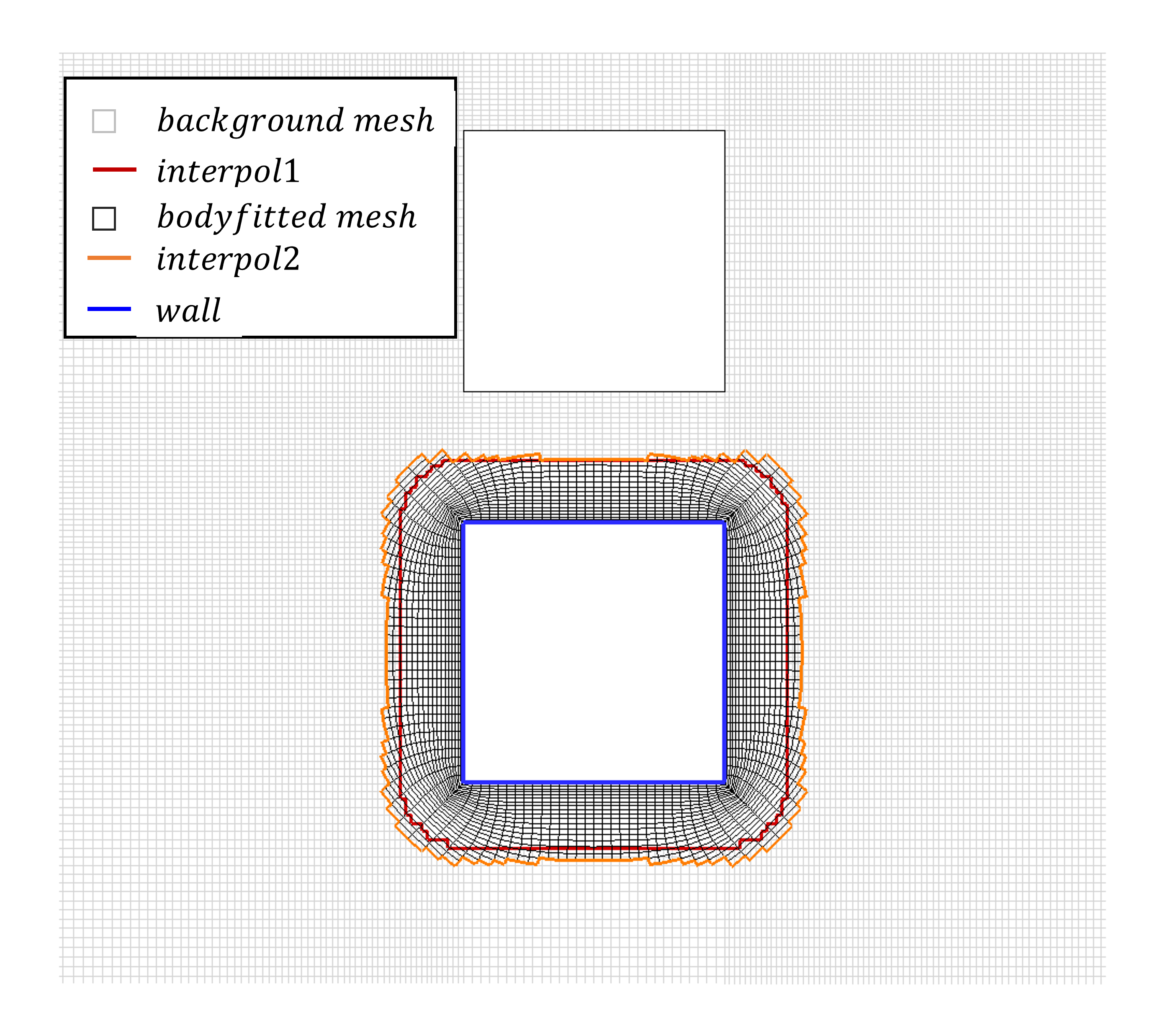}\label{fig:os_06}}
    \caption{Schematic of basic steps of the overset mesh assembly method.}
    \label{fig:os_step}
\end{figure}

\begin{enumerate}[\bfseries Step 1]
    \item Assume there are $N$ component meshes involved in the assembly process, numbered from $ib=0$ to $ib=N-1$, with the background mesh designated as $ib=0$; For instance in Fig.~\ref{fig:os_01}, two cubical meshes overlap with each other, black representing $ib0$ and blue representing $ib1$, where $ib0$ denotes the background grid.
    
    \item Determine whether each cell exists as a donor cell within the component meshes, storing the information in an integer array named $\text{IHost}[N]$. Specifically, for example in Fig.~\ref{fig:os_02}, $\text{iHost}[ib]=0$ indicates that cell $i$ do not have a donor cell in the component mesh numbered $ib$;
    \item Gather wall point information from all component meshes; See Fig.~\ref{fig:os_03}, both component meshes contain wall boundaries, where $wall[ib]$ represents the wall index indices in mesh $ib$.
    
    \item Retrieve the minimum wall distance for each cell relative to all component meshes and store these values in an array named wallDist[N+1]. Columns 0 to $N-1$ store the minimum wall distance for each cell relative to the ib=$N$ component mesh, while column $N+1$ stores the component mesh number corresponding to the global minimum wall distance; Fig.~\ref{fig:os_04} illustrates the minimum global wall distance computed in the two grid systems.
    
    \item Initialize the cell type of all cells to 0. Define cells in components with ib>0 having iHost=0, and cells in the ib=0 background mesh with iHost=2, as inactive cells. Then among cells with iHost=1, cells with wallDist[N]=ib are defined as active cells, while cells with $\text{wallDist}[N]\neq$ib are defined as inactive cells. All remaining cells are defined as active cells;
    
    \item Define the two layers of inactive cells adjacent to active cells as interpolation cells. Mark the interpolation cells with $cellType=2$ and $4$, and construct interpolation stencils between different component meshes for all interpolation cells; Fig.~\ref{fig:os_05} presents the results of the grid classification for the ib0 component. Compared to the initial grid in Fig.~\ref{fig:os_01}, only the interior grid cells with a wall distance less than one-quarter of the cell length are retained. The outermost two layers of the grid are marked as $celltype=2/4$, indicating interpolation cells.
    
    \item Initialize the faceType of all interior interfaces to 1. If either side of an interface has an inactive cell, set $faceType=0$. If one side of the interface has an active cell and the other side has an interpolation cell, set $faceType=-1$. Fig.~\ref{fig:os_06} shows the results after edgeType classification. Comparing these results with the cellType classification in Fig.~\ref{fig:os_05}, it is evident that the interpolation boundaries exist between the interpolation cells and the active cells, forming a closed shell.
\end{enumerate}

The first six steps above are referred to as the hole-cutting method. Typically, wall distance definition is insufficient for the hole-cutting step. Criteria based on mesh size and wall distance may produce isolated cells, i.e., inactive cells surrounded by active cells. To address this issue, an additional reference wall distance can assist in hole-cutting. However, manual inspection is often still necessary to check for any overlooked isolated cells. In this paper, the marked cell information is utilized in constructing the hole-cutting. The basic idea is to define the iHost array to store the cells' host information, indicating whether a host cell exists. Here, iHost=1 signifies that the cell has a host in a component mesh, while iHost=0 and iHost=2 indicate that the cell is outside the component boundary or inside the closed geometry, respectively. The single-ray method can determine whether a cell is within a non-convex closed polygon, but it may fail if the ray passes through a vertex. To improve robustness, multiple orthogonal rays are used (e.g., three rays in three-dimensional space) to jointly determine whether a cell is within a closed geometry for N-dimensional problems.

As the proposed method is based on the finite volume method, the residuals at the cell center are obtained by summing all boundary fluxes. Therefore, the interior edges are further classified into three categories in the final step. The first category includes edges where one side is an interpolation cell and the other side is an active cell, defined as interpolation boundaries. The second category includes edges where both sides are active cells, which remain as internal edges. The remaining edges are defined as inactive edges, meaning no flux calculation is performed. The edge type assignment procedure is conducted after cell type classification. The faceType array stores the status of the interior edges and is initialized to 1. In the edge classification results, interpolation edges and inactive edges will be marked as faceType = -1 and faceType = 0, respectively. Edge classification using cell type information is simple and intuitive. The result of the boundary definition is shown in Fig.~\ref{fig:os_06}. In the non-background component meshes, the defined set of interpolation boundaries forms a closed shell that encompasses all active cells. Compared to a single mesh solver that contains only active cells, introducing a new interpolation boundary achieves information exchange between different component meshes.


\begin{figure}[t]
	\centering
{\includegraphics[height=0.25\textheight,keepaspectratio]{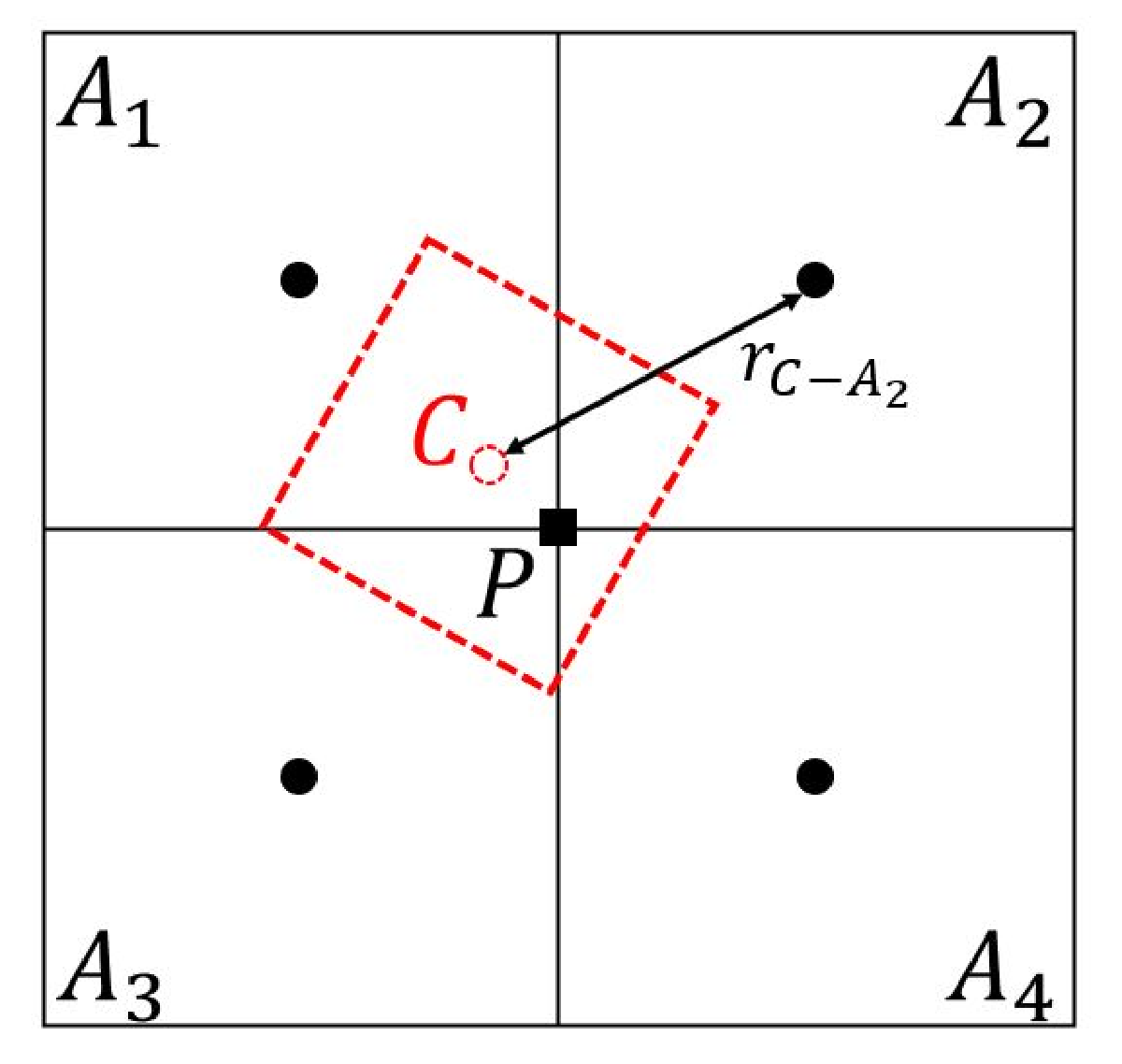}}
	\caption{Determination of interpolation stencils. Interpolation cell C(red dashed line) has its closest neighbor as cell A1 (black solid line) at node P. Using neighboring node information, all neighboring units (A1, A2, A3, and A4) are set as donor cells. }
	\label{fig06:interpolSet}
\end{figure}

Once the mesh assembly is finalized, before iterating on the active cells, the flow field of the interpolation cells should first be updated as boundary conditions. To simplify the algorithm for determining interpolation stencils and save memory usage, only the hosts of interpolation cells are searched, the search regime is limited to active cells in the corresponding component mesh. The interpolation stencils result is depicted in Fig.~\ref{fig06:interpolSet}. Interpolation cell C has its closest neighbor as cell A1 at node P. Using neighboring node information, all neighboring units (A1, A2, A3, and A4) are set as donor cells. Additionally, Information transfer can be broadly categorized into two methods: via variables in cells or via fluxes at interfaces. Each interpolation method can be used, but given its simplicity and popularity, this paper focuses solely on information exchange via cell variables. Fig.~\ref{fig06:interpolSet} shows a 2D overset interpolation setup to depict the exchange of solutions between two overset meshes, where the solution is transferred at the stencils via interpolating equations such as:
\begin{equation}
\phi_C = \frac{\sum_{i=0}^{N_c} (\phi_{A_i} r_{C-A_i}^{-2})} { \sum_{i=0}^{N_c} r_{C-A_i}^{-2}},\quad N_c = 4,
\end{equation}
the variable $\phi_C$ represents the interpolated macroscopic primitive variables $\bm{W}$ or velocity distribution function $f$ at cell $C$, while $r_{C-A_i}$ denotes the distance between cell $C$ and cell $A_i$. The major drawback of this interpolation method is that it does not ensure flux conservation within the interpolation cell. However, it has been shown that this lack of conservation does not degrade the solution quality if the overset occurs in regions without sharp gradients, such as shocks or discontinuities. The test results in Section~\ref{sec:4} confirm the effectiveness of the applied method.

\section{Numerical validations}\label{sec:4}

This section presents two examples to validate the proposed GSIS for the moving boundary problems: pitching NACA 0012 airfoil and particle motion in lid-driven cavity flow. The convergence criterion for the kinetic scheme is that the volume-weighted relative change of the moments(density, velocity, and total temperature) between two iterations must be less than $10^{-6}$,
\begin{equation}
    E^k = \frac{\sqrt{\sum_i(\phi_i^k - \phi_i^{k-1})^2\myd \Omega}}{\sqrt{\sum_i(\phi_i^{k-1})^2\myd \Omega}}\Big|_{max} \leq 10^{-6}, \quad \phi\in(\rho, \bm{u}, T),
\end{equation}
while the convergence criterion for the macroscopic schemes is less than $10^{-7}$, and a maximum iteration step of 100. The CFL number for pseudo time step in the micro and macro solver are $10^8$ and $10^4$, respectively. All those tests are performed in double precision on a workstation with Intel (R) Core(TM) i7-9700K CPU@3.60GHz processors.

The three-dimensional velocity space is reduced to two-dimensional by introducing the reduced velocity distribution functions, see \ref{sec:reduction}. 

\subsection{Pitching NACA 0012 airfoil}

\begin{figure}[t]
    \centering
    \subfloat[full domain]{\includegraphics[scale=0.38,clip = true]{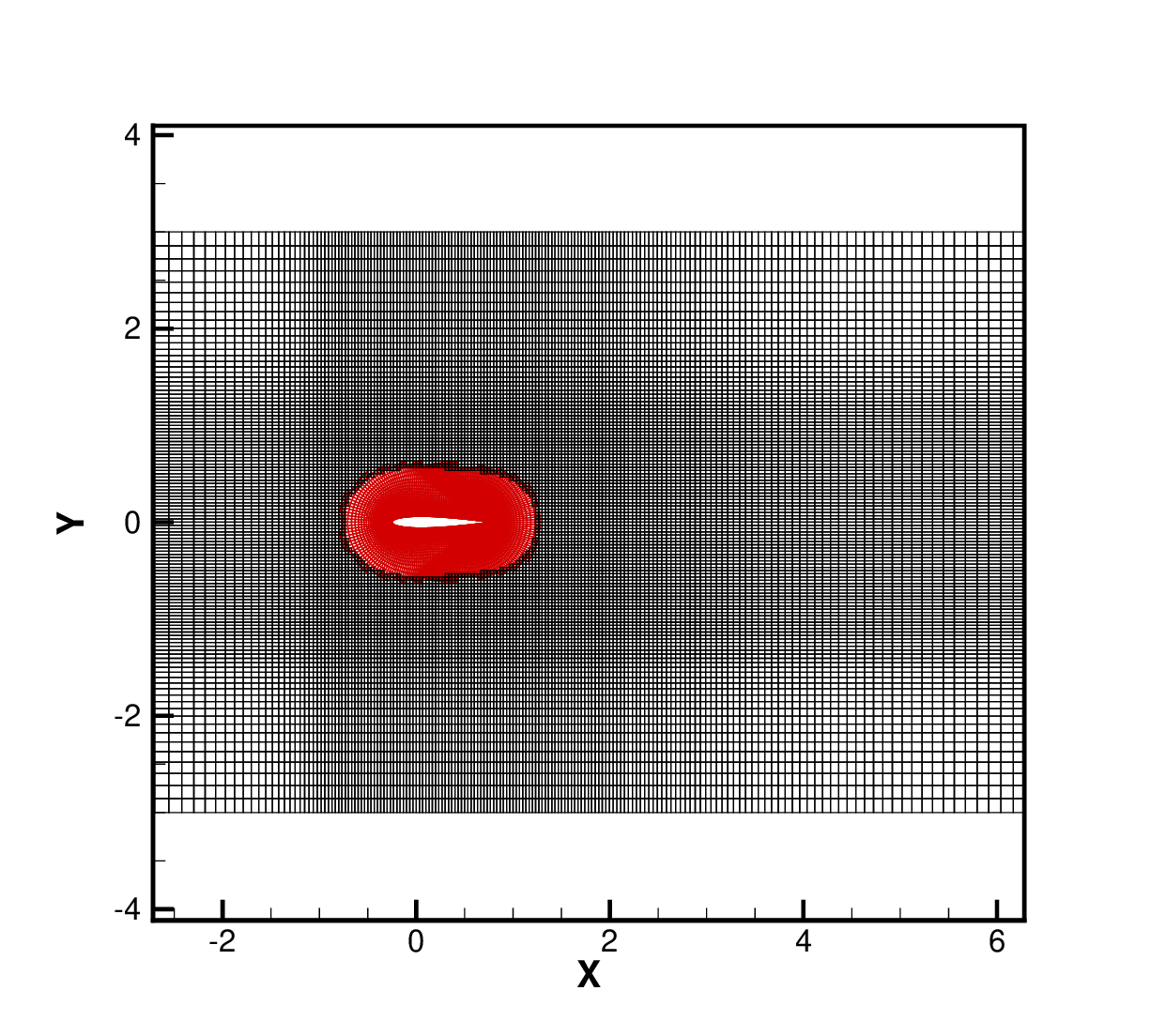}}
    \subfloat[enlarged view near the airfoil]{\includegraphics[scale=0.38,clip = true]{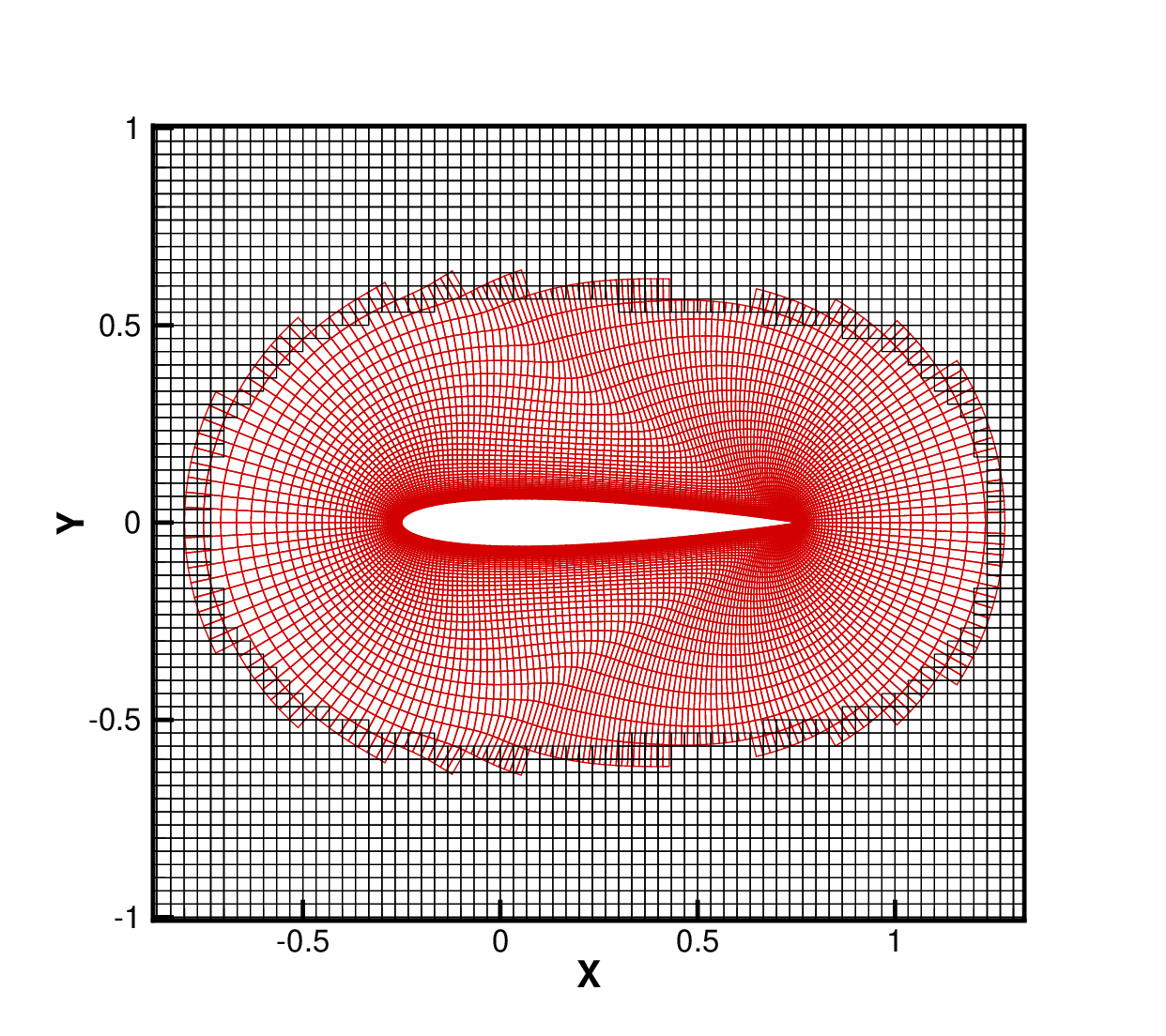}}
    \caption{Assembled overset mesh for the flow simulation around a NACA 0012 airfoil. The initial background mesh includes only the far-field boundaries, while the body-fitted mesh encompasses the wall boundary. The number of grids in the body-fitted mesh (red lines) and the background mesh (black lines) is 20,400 and 13,310, respectively. The airfoil surface is discretized with 300 mesh points, and the height of the first layer mesh is set to 0.001c, where $c$ represents the chord length.}
    \label{fig:naca0012_mesh}
\end{figure}

The flow around a stationary airfoil is first considered to verify the accuracy of the GSIS in subsonic flow. For an inviscid flow (Kn$=1.85\times 10^{-6}$) around the NACA 0012 airfoil, two flow conditions are considered: (a) the Mach number Ma is 0.8 and the angle of attack (AOA) is $\alpha=0^\circ$, and (b) Ma$=0.85, \alpha=1^\circ$. 
Figure \ref{fig:naca0012_mesh} illustrates the assembled physical mesh. The surface of airfoil is characterized by the diffuse boundary condition with a constant temperature $T_w = 300$~K. The inlet and outlet boundaries are positioned at the left and right background region. In each velocity direction, $32$ uniform discrete velocity points are used in in the range $[-10,10]$. In the steady-state simulation, the time step is set to $10^5 t_{ref}$, while in the unsteady simulation, it is set to $1/100$ of the pitching frequency.

\begin{figure}[t]
    \centering
    {\includegraphics[scale=0.3,clip = true]{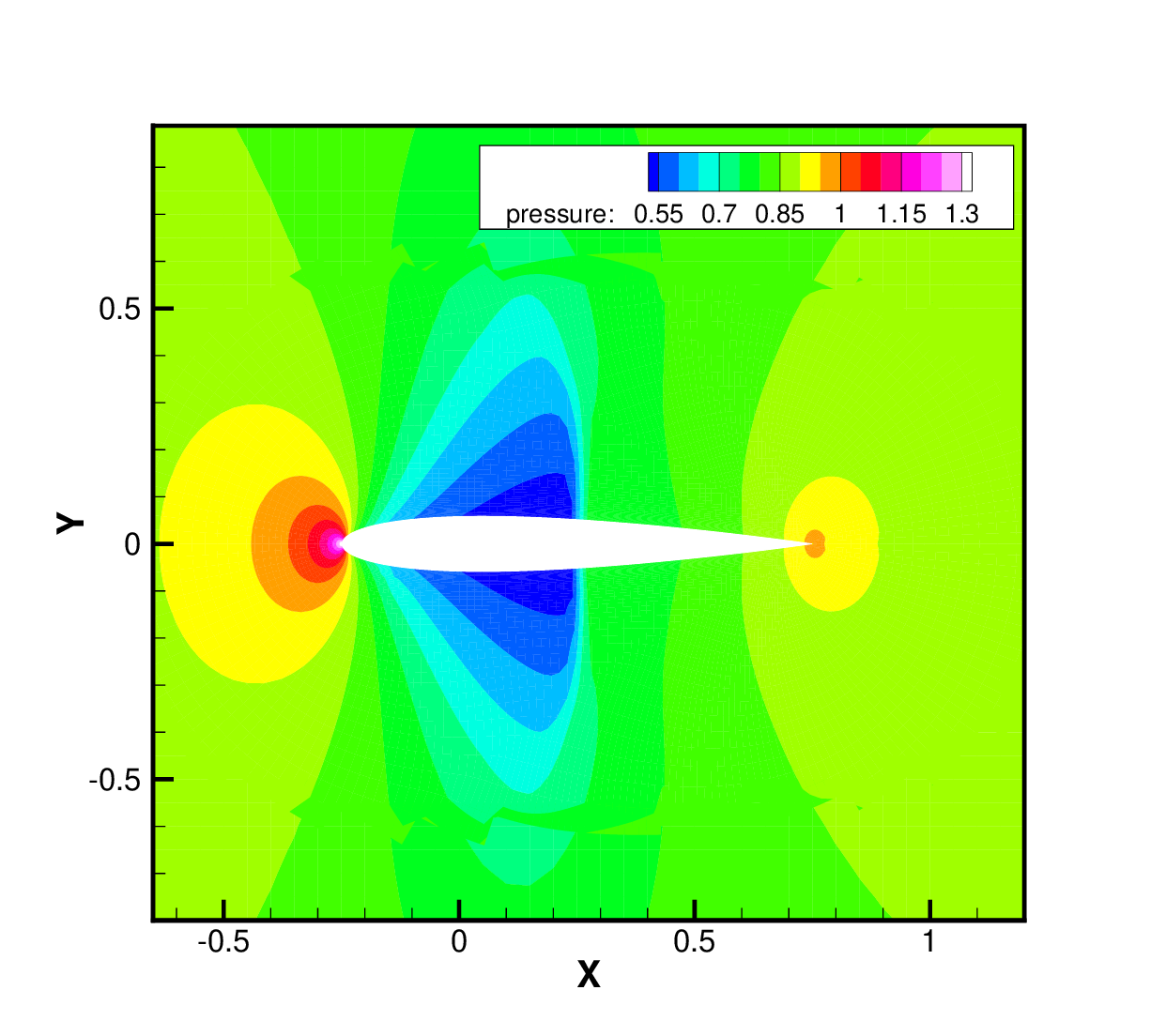}}
    {\includegraphics[scale=0.3,clip = true]{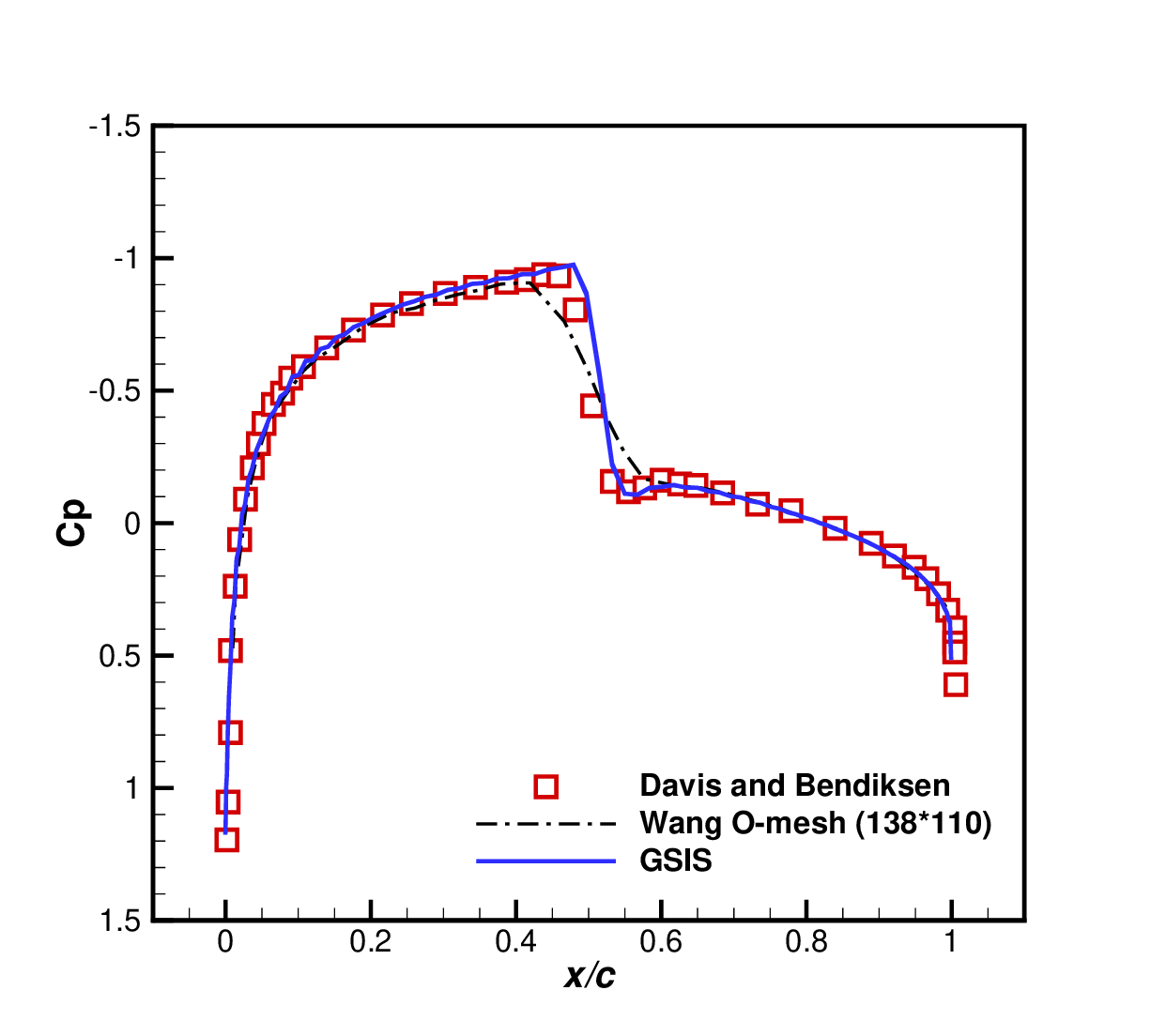}}\\
    {\includegraphics[scale=0.3,clip = true]{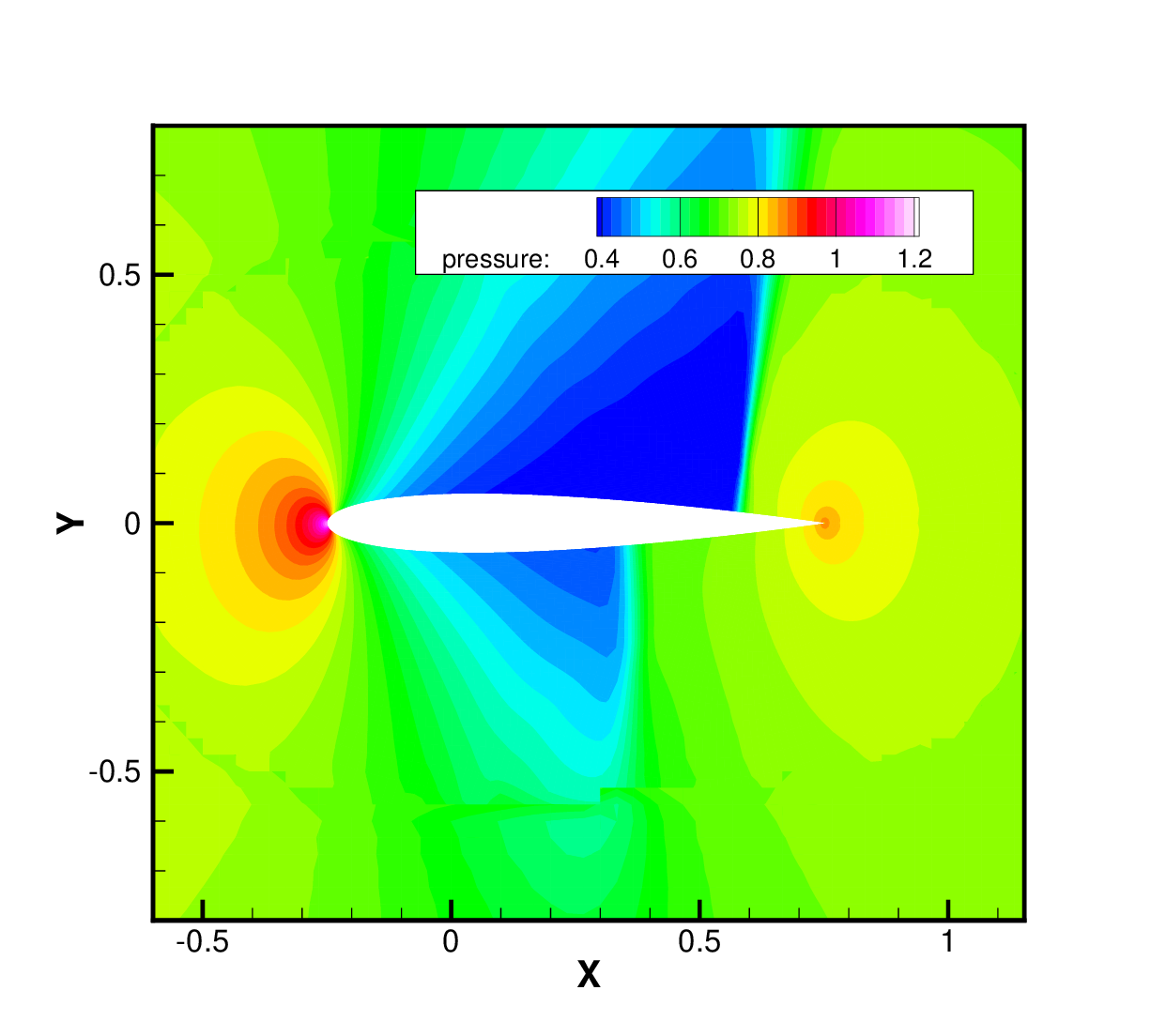}}
    {\includegraphics[scale=0.3,clip = true]{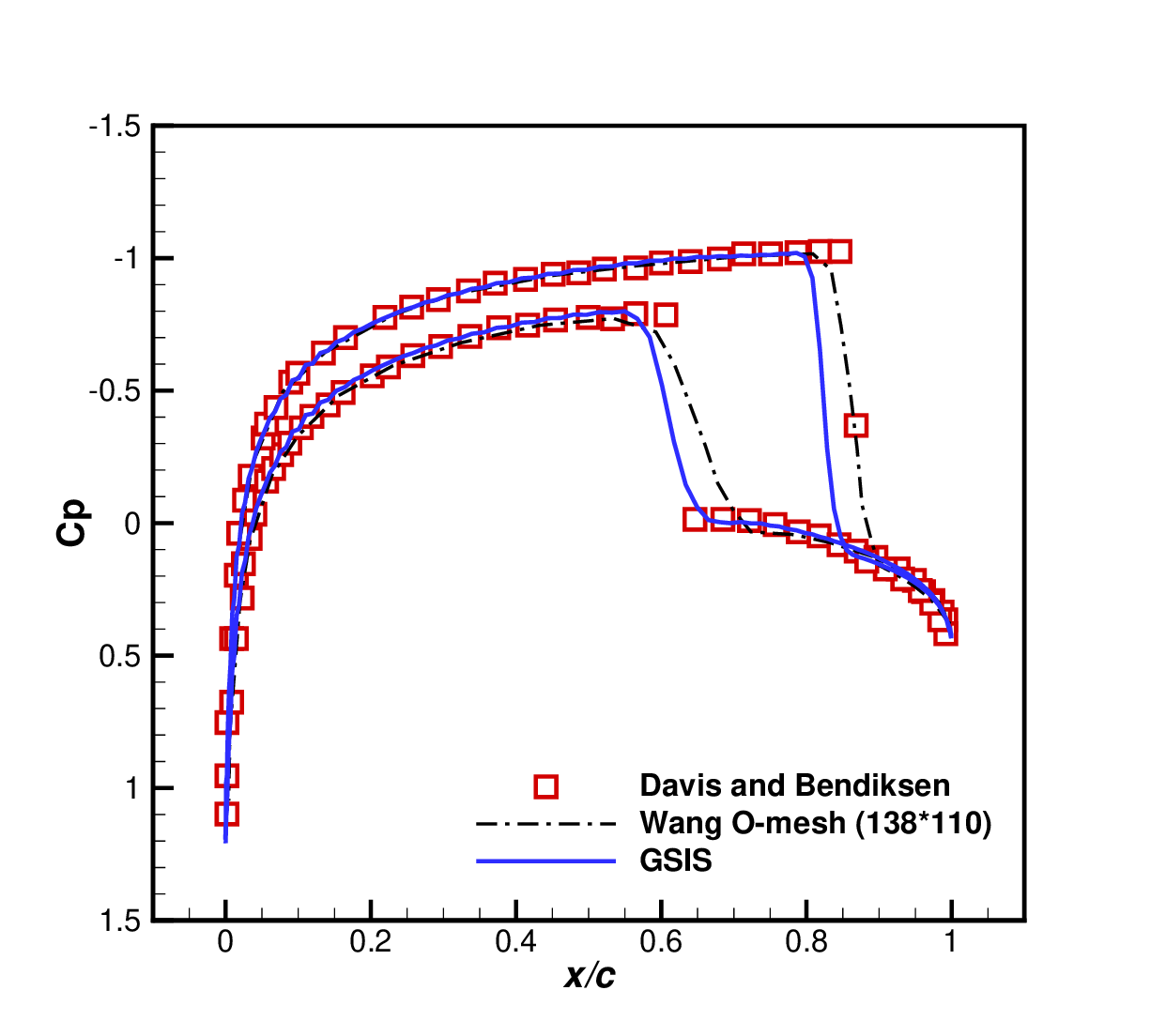}}
    \caption{Pressure contours and surface pressure coefficients for the inviscid flow around the NACA 0012 airfoil, when (top) Ma=$0.8, \alpha = 0^\circ$ and (bottom) Ma=$0.85, \alpha = 1^\circ$. 
    }
    \label{fig:naca0012}
\end{figure}

Figure~\ref{fig:naca0012} shows the pressure contour and airfoil surface pressure coefficient, which is defined as
\begin{equation}
    C_p=\frac{p-p_0}{0.5 \rho_0 U_0^2}.
\end{equation}
In the subsonic flow at different Mach numbers, shock waves are observed on the upper and lower surfaces of the airfoil. And as the AOA increases, the location of the excitation wave on the upper surface is more rearward. For the pressure coefficients on the airfoil surfaces with zero angle of attack, the shock wave locations given by GSIS are in general agreement with the ALE-DUGKS~\cite{wang2019arbitrary}. The thickness of the shock wave obtained by GSIS is sharper, which may be due to the greater numerical dissipation caused by the use of a coarser computational mesh in ALE-DUGKS. When $\text{Ma} = 0.85$, the positions of shock wave at the upper and lower surface obtained from the GSIS  are slightly skewed towards the leading edge, which may be related to the choice of macroscopic flux scheme. Overall, GSIS-ALE results show good agreement with the experiment~\cite{davis1993unsteady} and numerical~\cite{wang2019arbitrary} results.


\begin{figure}[t]
    \centering
    \includegraphics[scale=0.4, clip = true]{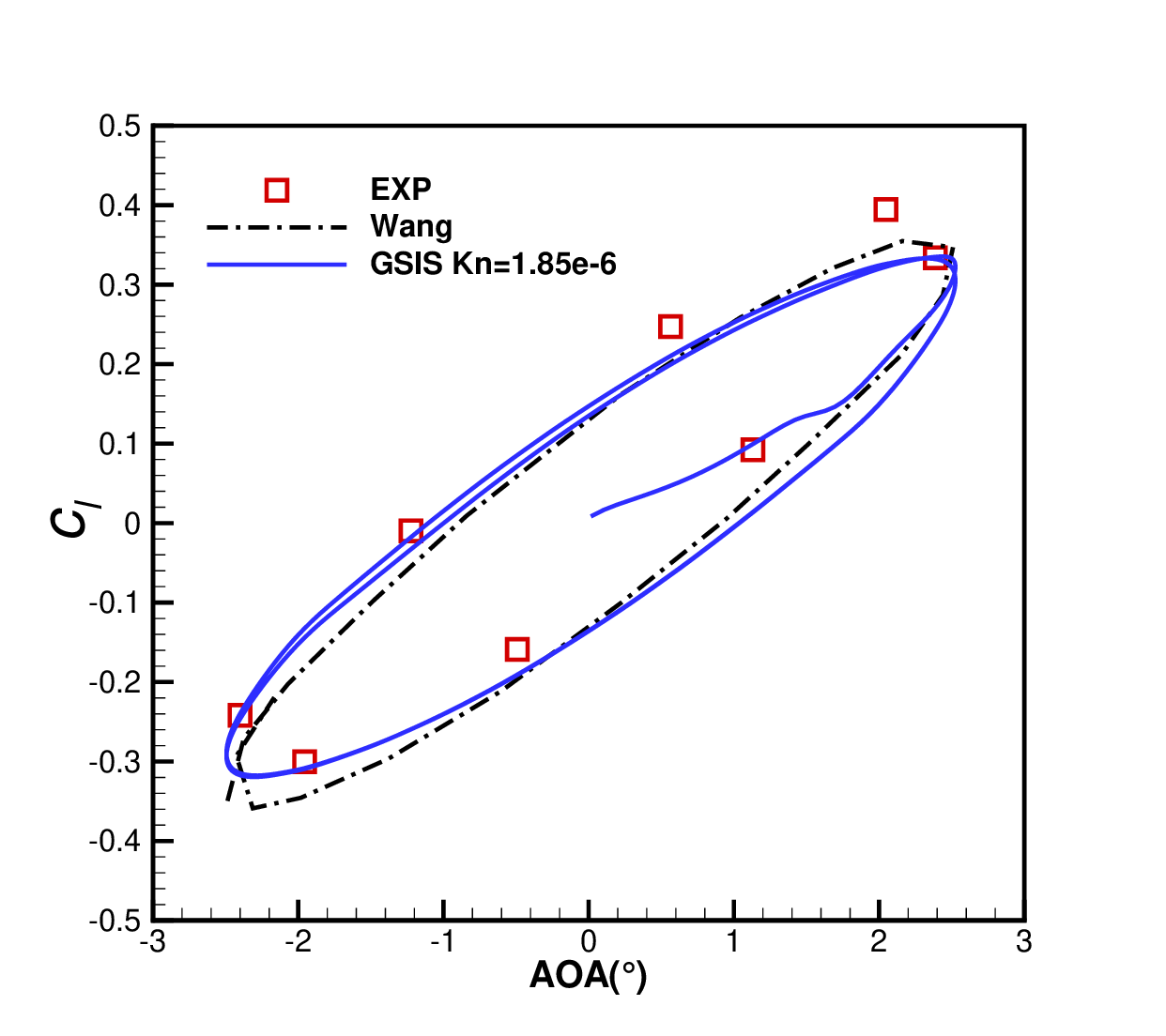}
    \caption{The evolution of lift coefficient $C_l$ under different AOAs for inviscid flow around a pitching NACA 0012 airfoil, when $\text{Ma}=0.755$.
    }
    \label{fig:lift_curve}
\end{figure}

Next, the airfoil pitching problem with inviscid inflow is simulated, which is used to verify the simulation capability of GSIS-ALE for the moving boundary problem. Referring to the setup by Davis \textit{et al}~\cite{davis1993unsteady}, the airfoil pitching point is at $1/4$ chord length, the Mach number of the free-stream is $\text{Ma} = 0.755$, and the change of AOA with time can be expressed as:
\begin{equation}
\begin{aligned}
     \alpha(t) = &0.016^\circ + 2.51^\circ \sin\phi, \quad
       \phi= &\frac{2U_0}{c}k t,
    \end{aligned}
\end{equation}
where $k = 0.0814$ is defined as the deceleration frequency of the pitching motion. The physical mesh in Fig.~\ref{fig:naca0012_mesh} was still used and reassembled at each time step. The converged flow field of the stationary airfoil with $\alpha=0.016^\circ$ was used as the initial computational flow field before the start of the airfoil motion. 

The lift coefficients defined as
\begin{equation}
    C_l = \frac{F_x}{0.5 \rho_0 U_0^2},
\end{equation}
where  $F_x$ is the component of the force defined by Eq.~\eqref{eq:force_define} in the $x$-direction, is shown in Fig.~\ref{fig:lift_curve} as a function of the AOA. A total of five cycles were calculated, and the converged dynamic lift curve is obtained after the second cycle and shows a very clear unsteady effect, i.e., $C_l$ is smaller in the head-up process ($\phi = -90^\circ \rightarrow 90^\circ$) than in the head-down process ($\phi = 90^\circ \rightarrow 270^\circ$). The lift coefficients given by GSIS are closer to the experimental results than the ALE-DUGKS~\cite{wang2019arbitrary} in the downward half of the cycle ($\phi = 180^\circ \rightarrow 360^\circ$), and slightly lower than the DUGKS results in the upward half of the cycle ($\phi = 0^\circ \rightarrow 180^\circ$). 

\begin{figure}
    \centering
    \includegraphics[trim={0 20 50 40},clip,scale=0.3]{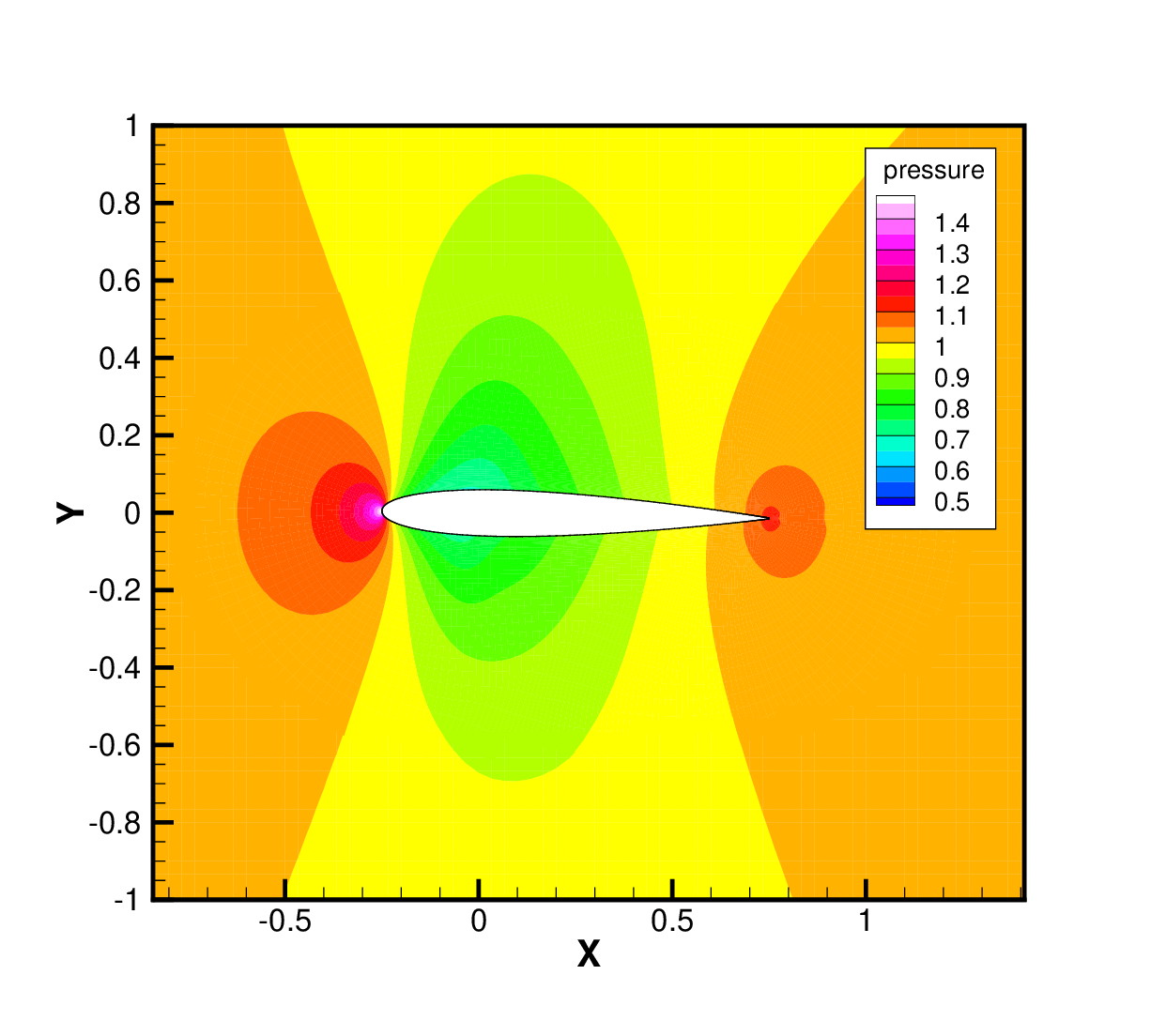}
    \includegraphics[trim={0 20 50 40},clip,scale=0.3]{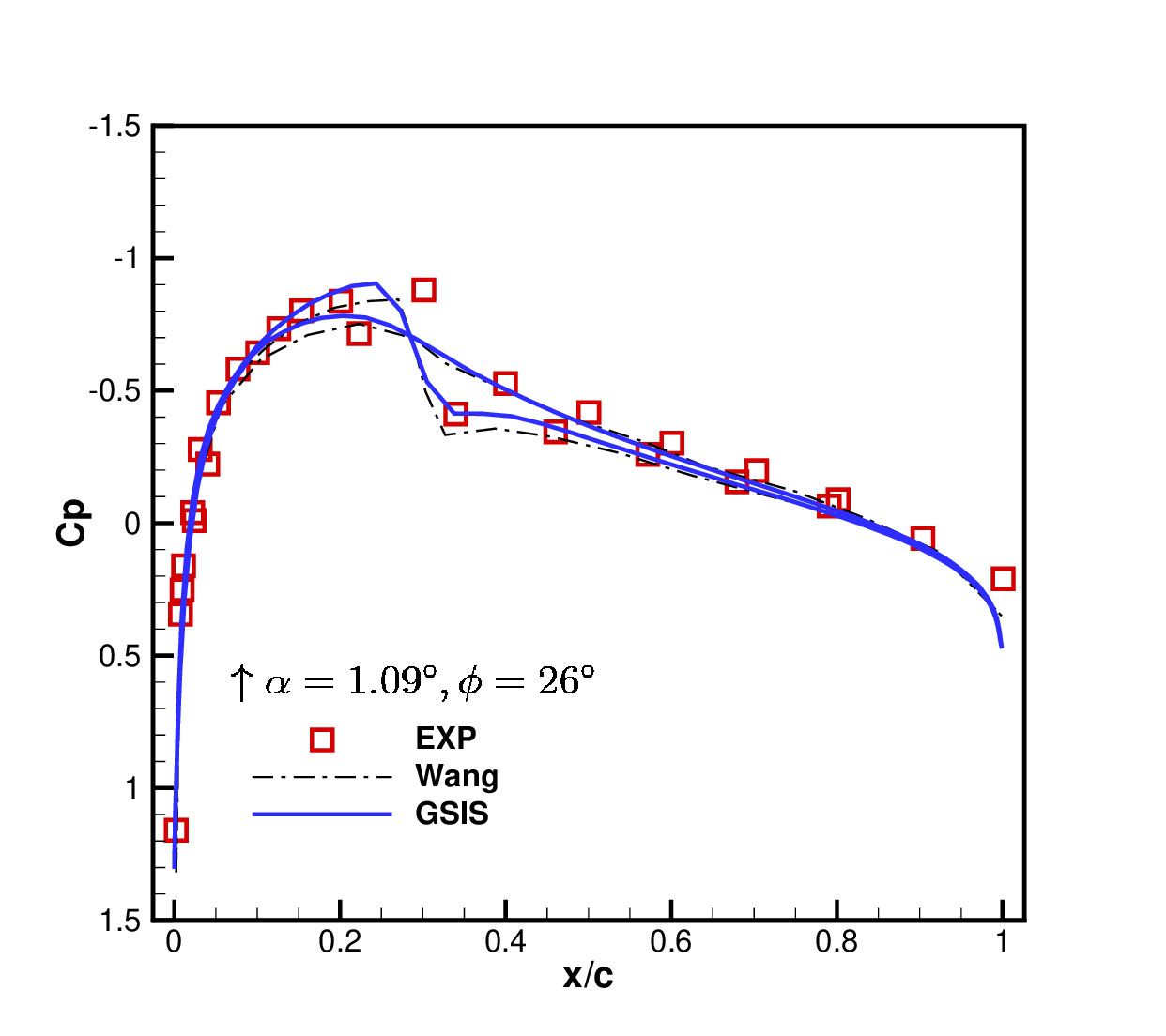}\\
    \includegraphics[trim={0 20 50 40},clip,scale=0.3]{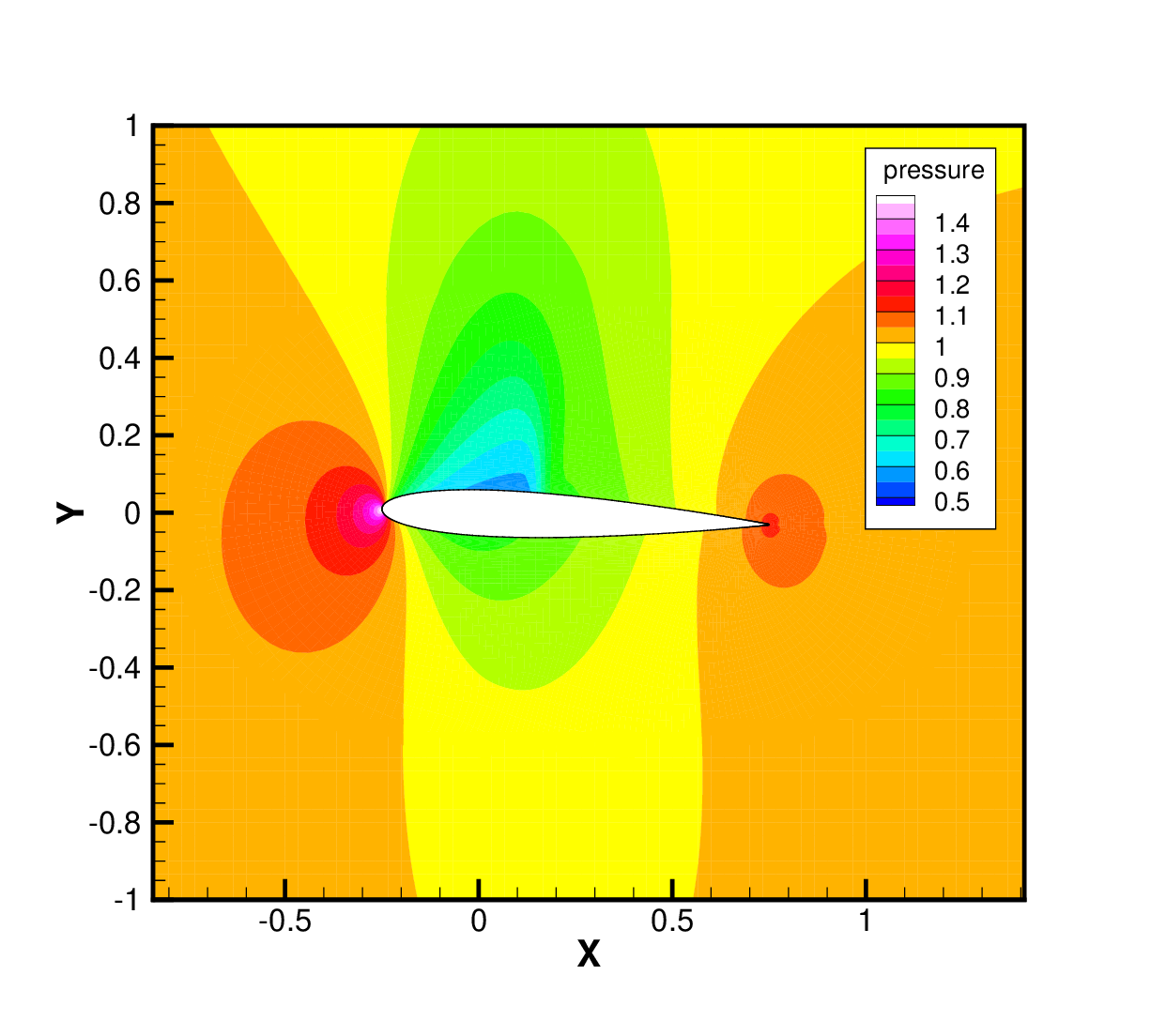}
    \includegraphics[trim={0 20 50 40},clip,scale=0.3]{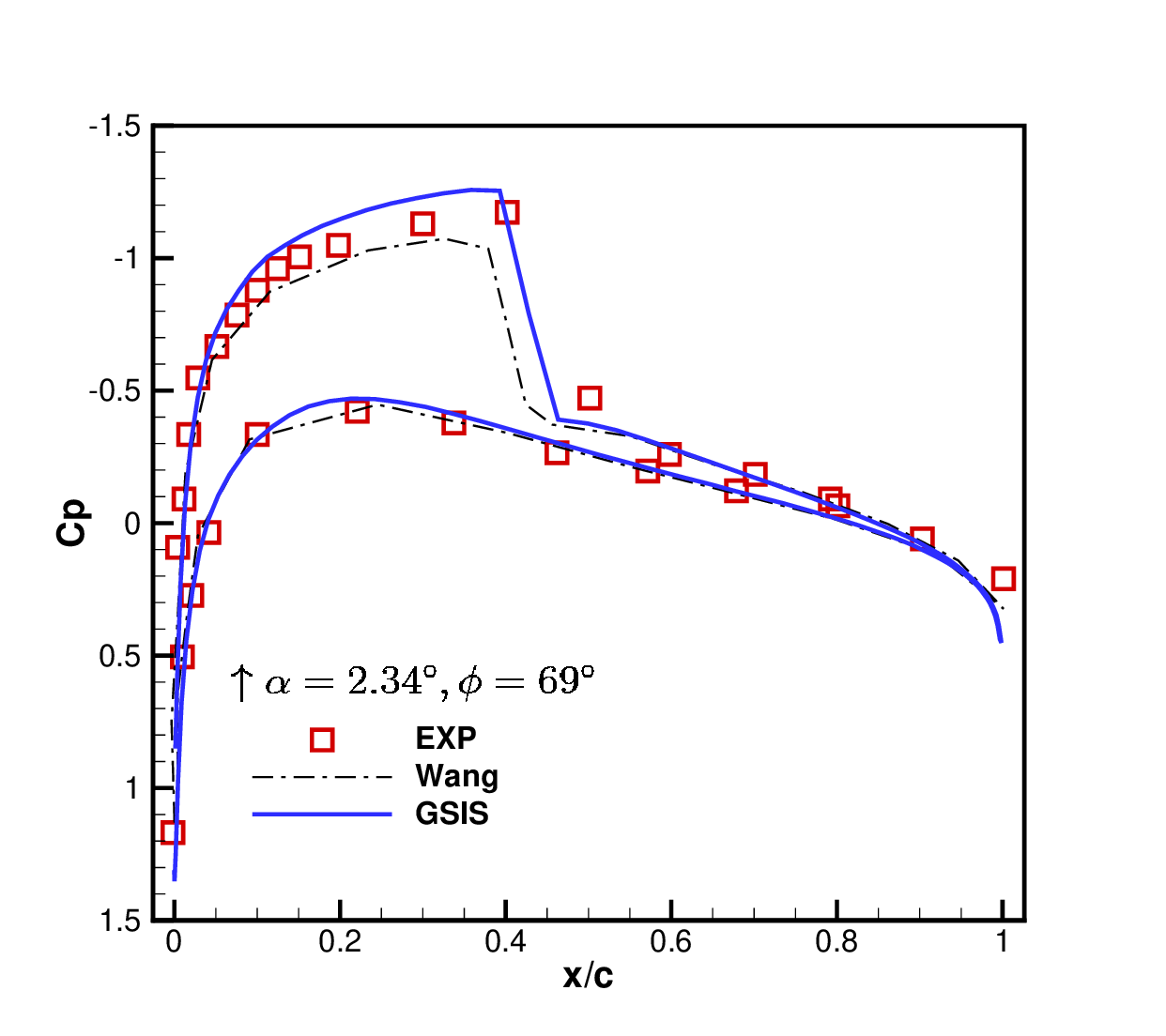}\\
    \includegraphics[trim={0 20 50 40},clip,scale=0.3]{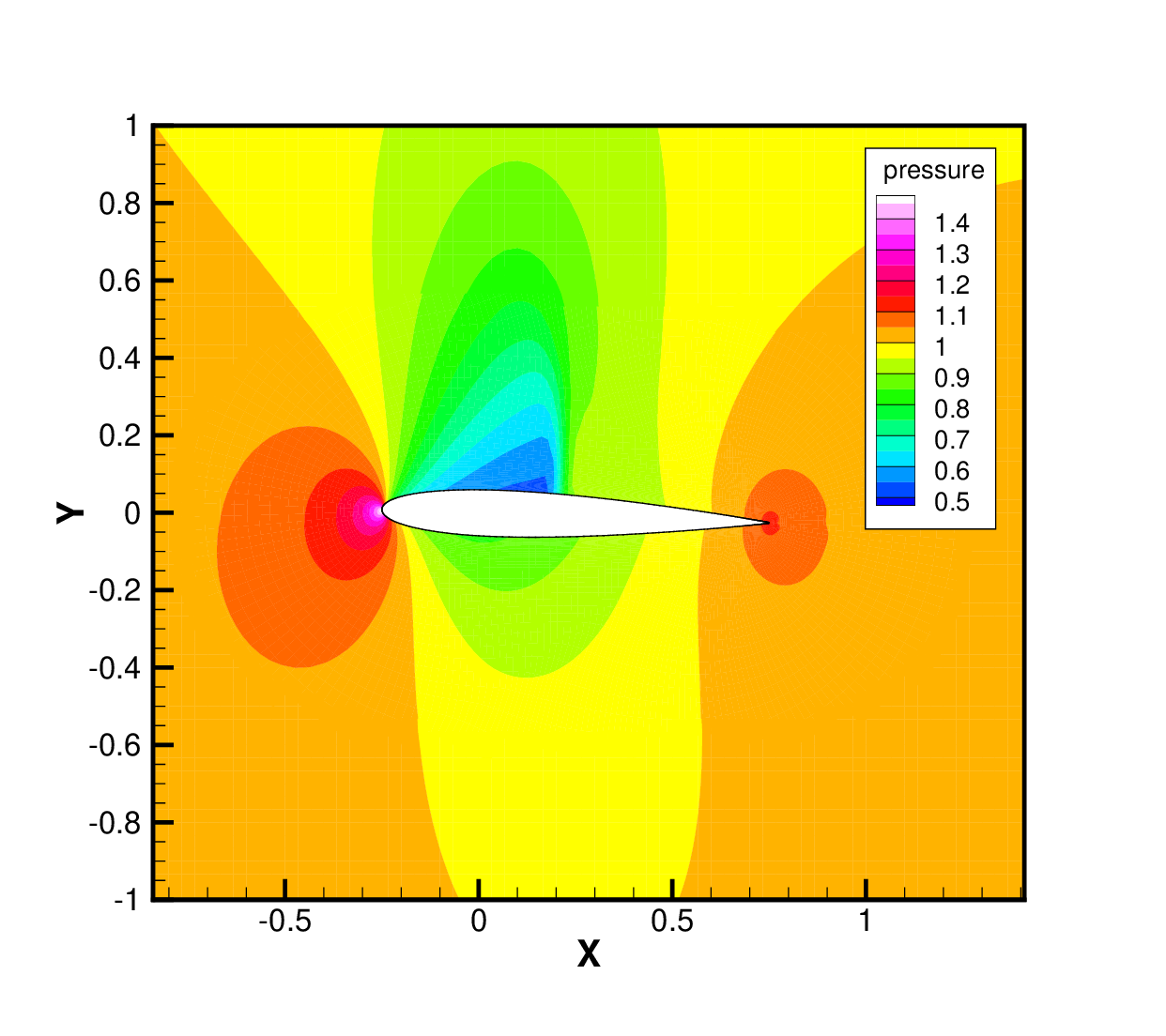}
    \includegraphics[trim={0 20 50 40},clip,scale=0.3]{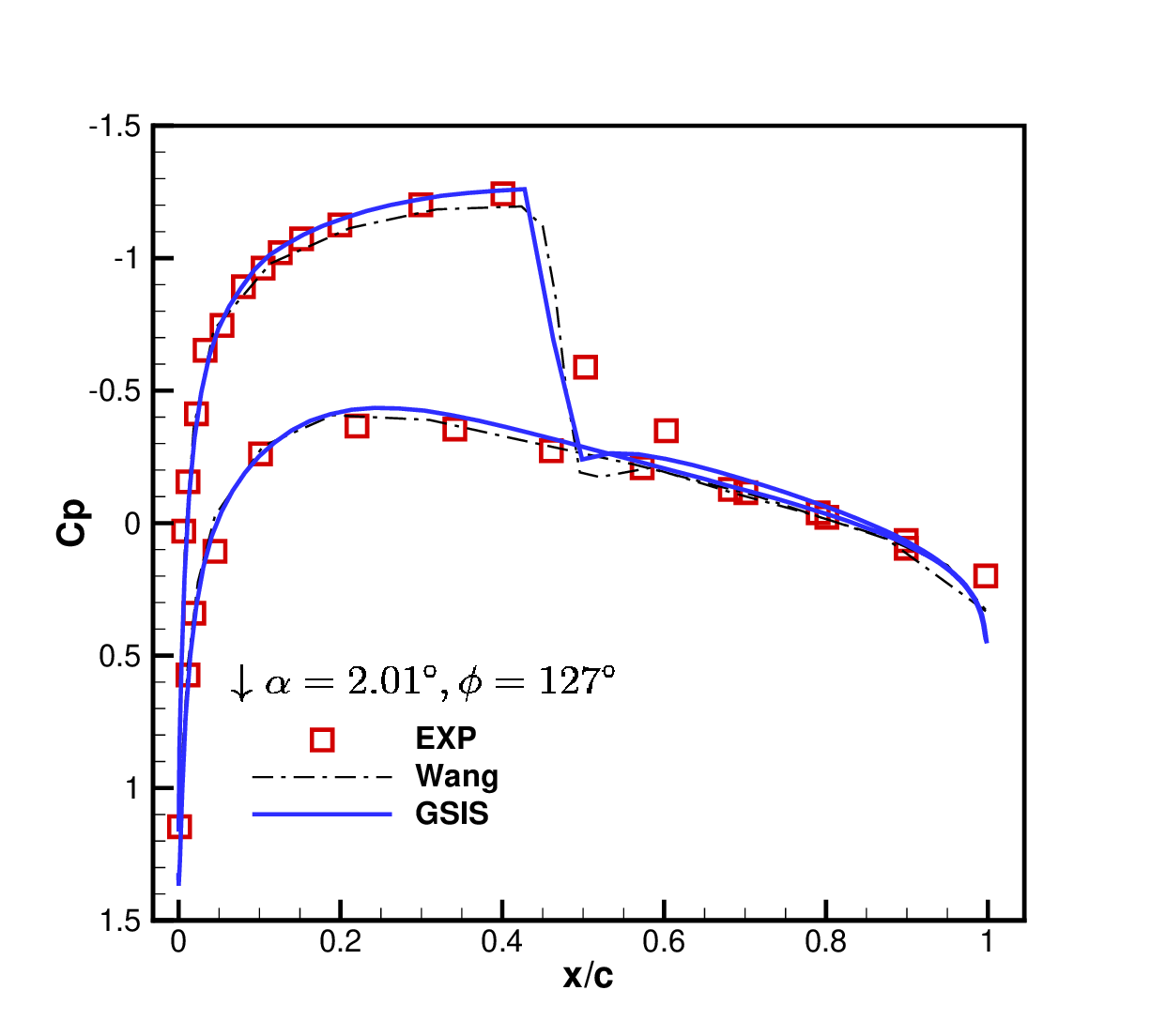}\\
    \includegraphics[trim={0 20 50 40},clip,scale=0.3]{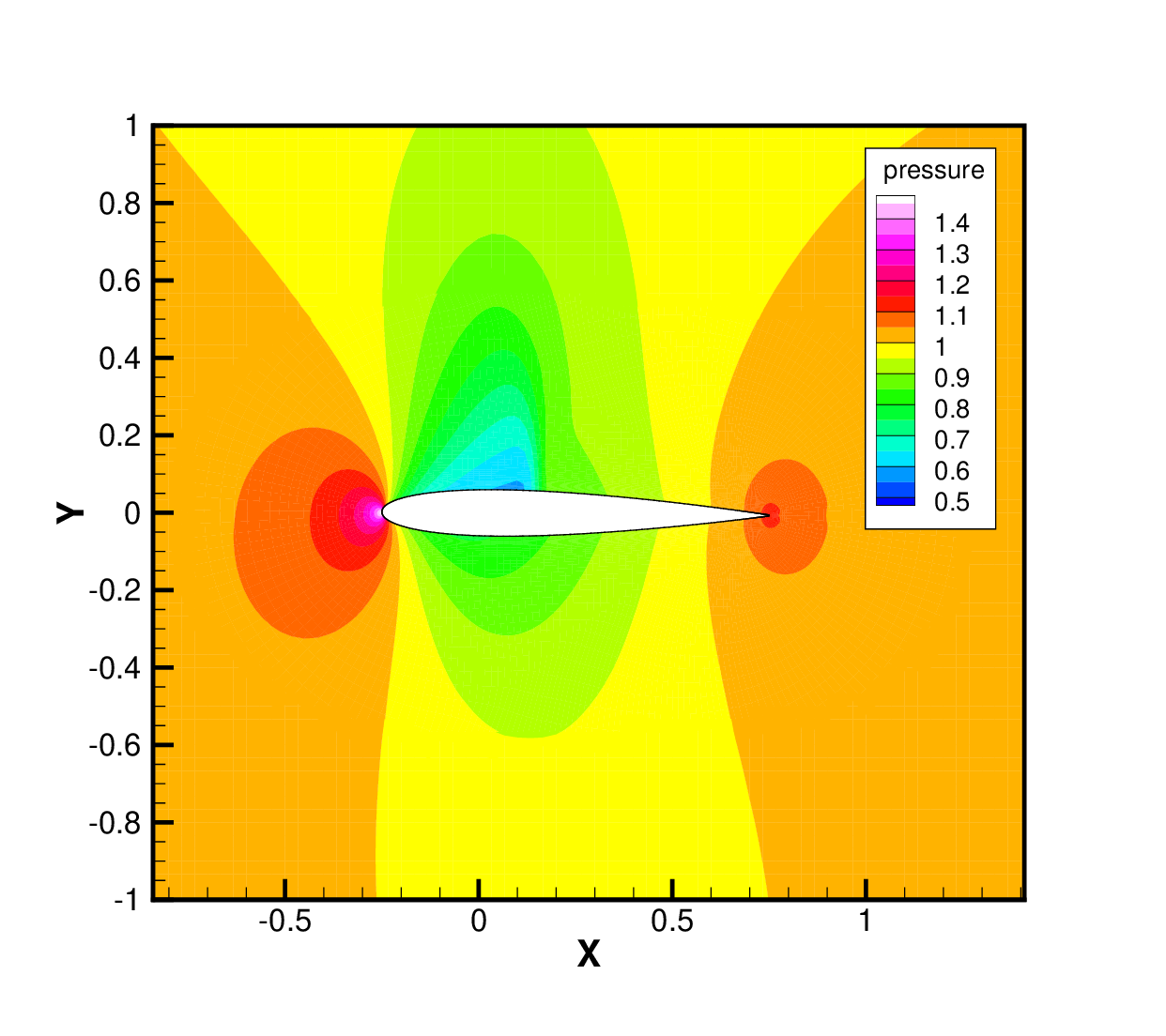}
    \includegraphics[trim={0 20 50 40},clip,scale=0.3]{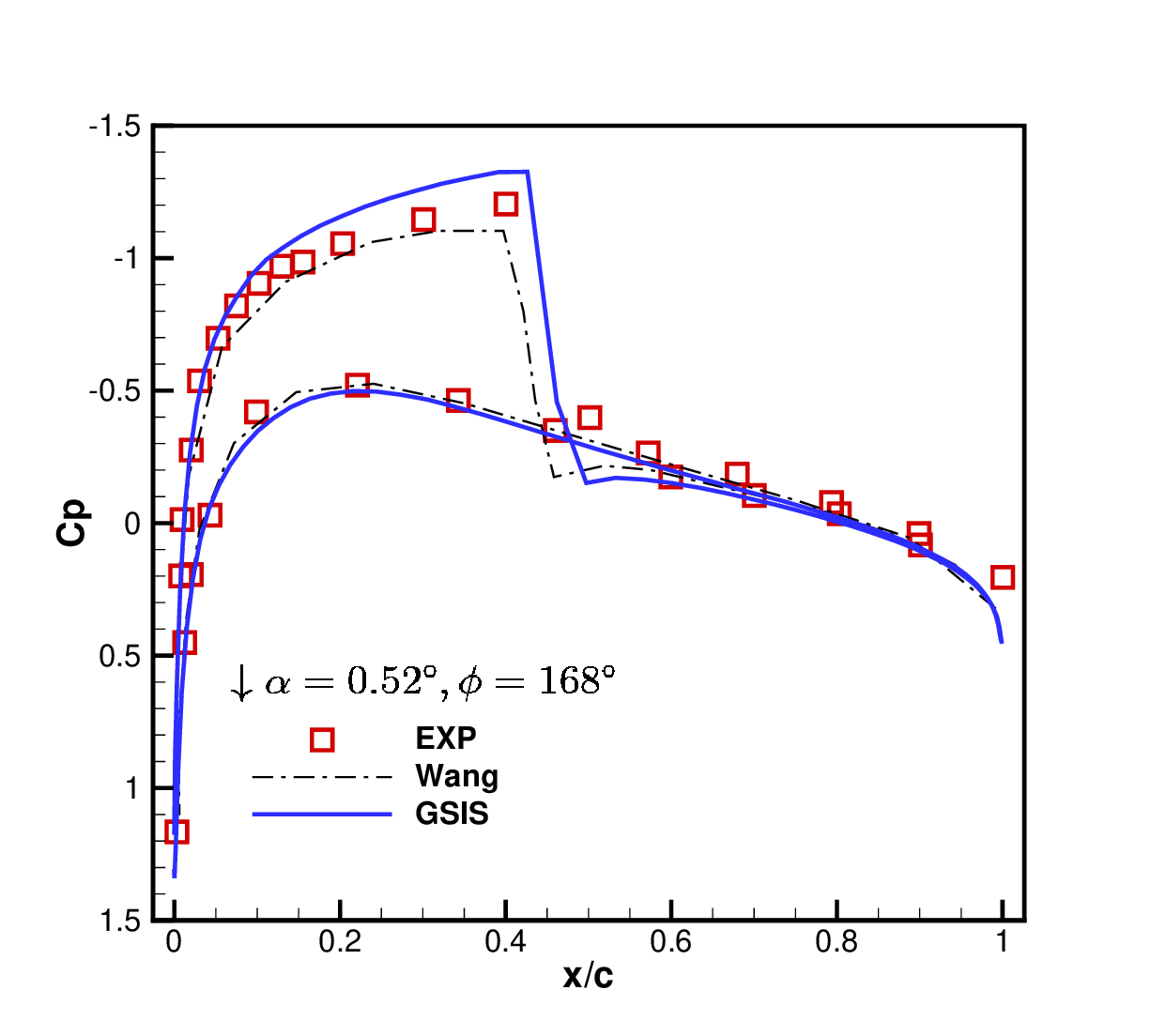}
    \caption{Pressure contours and pressure coefficients $C_p$ along the surface of airfoil at different phase angles for $\text{Ma}=0.755$.
    }
    \label{fig:diff_angles_1}
\end{figure}

\begin{figure}
    \centering
   \includegraphics[trim={0 20 50 40},clip,scale=0.3]{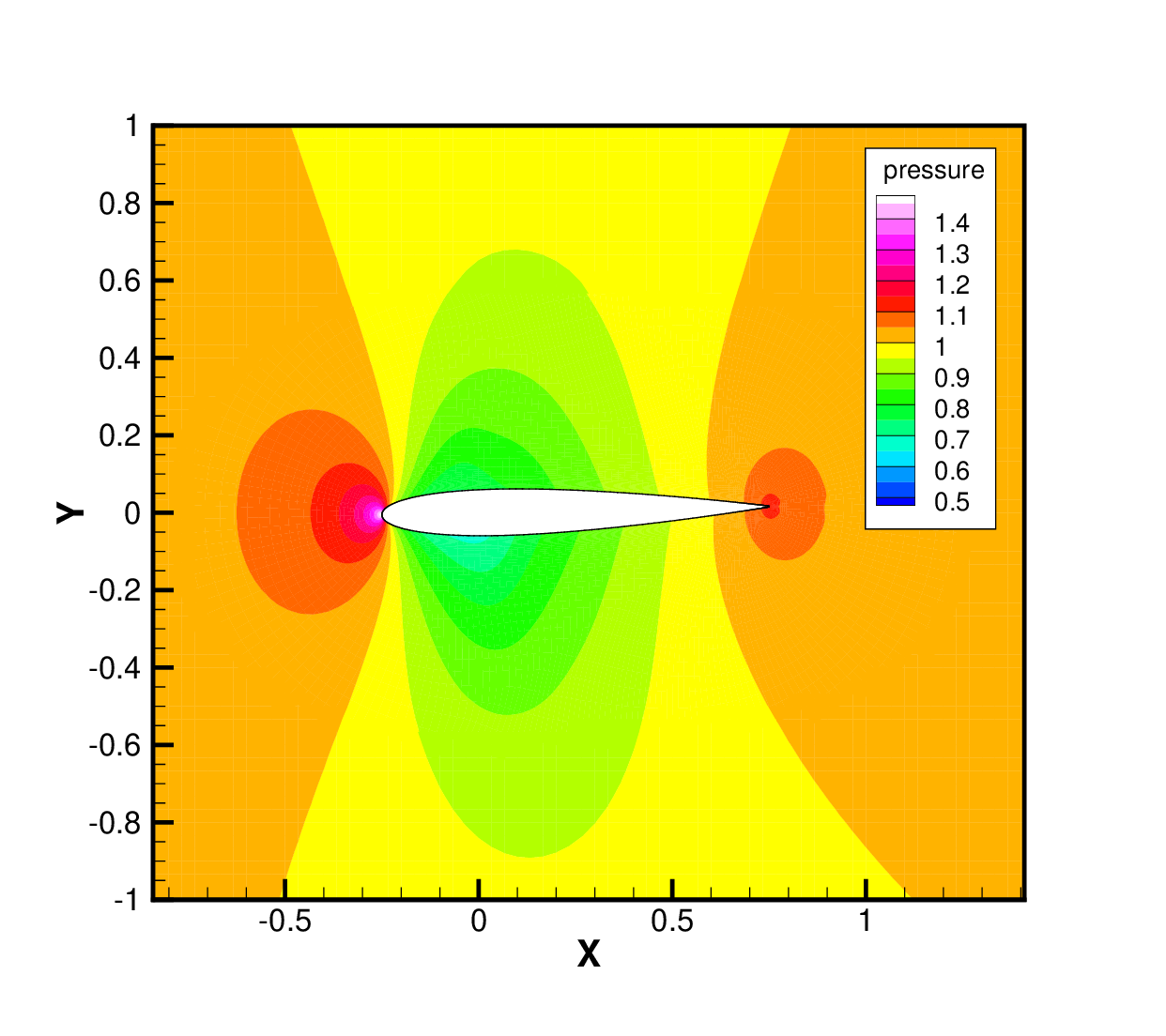}
    \includegraphics[trim={0 20 50 40},clip,scale=0.3]{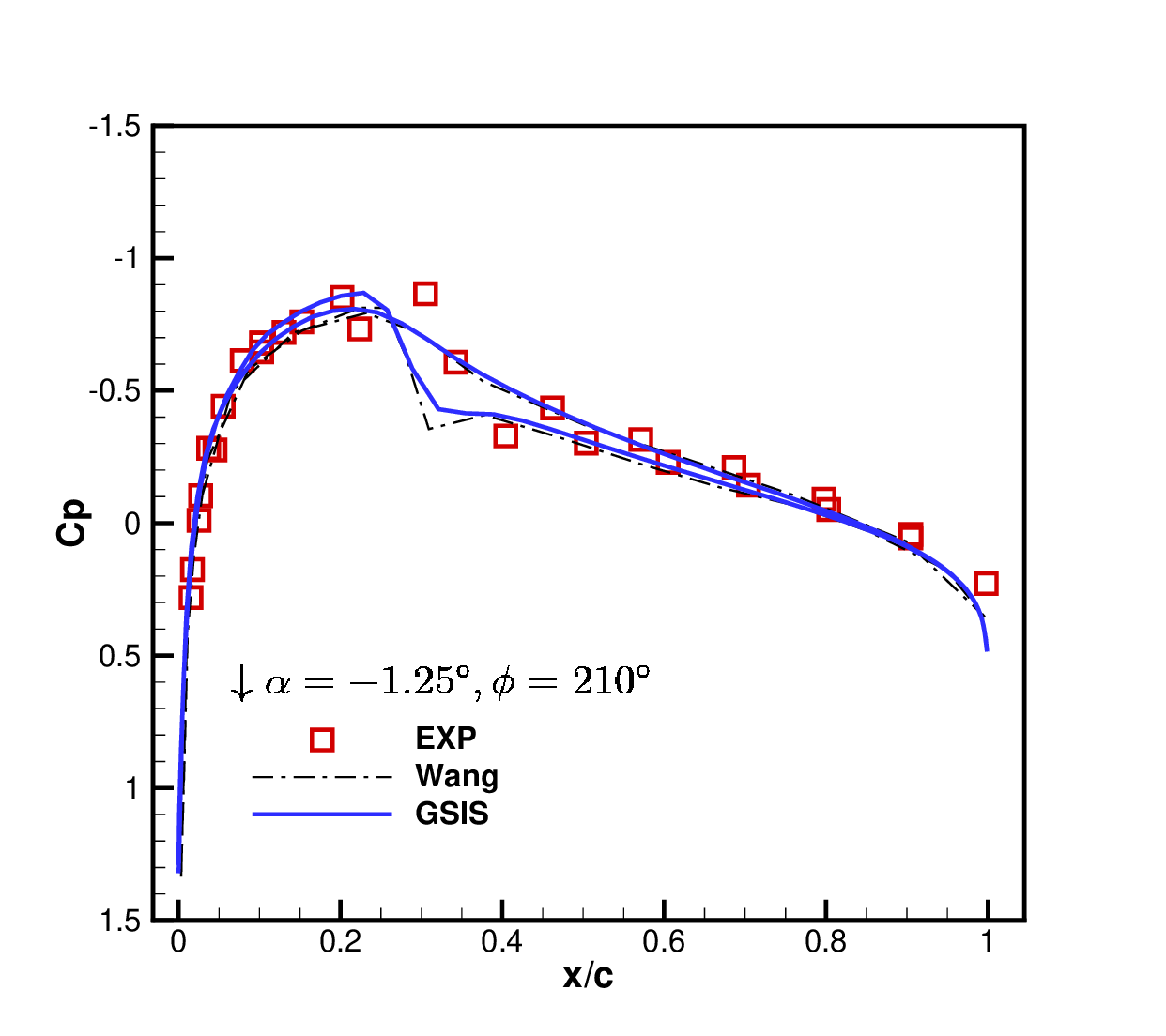}\\
   \includegraphics[trim={0 20 50 40},clip,scale=0.3]{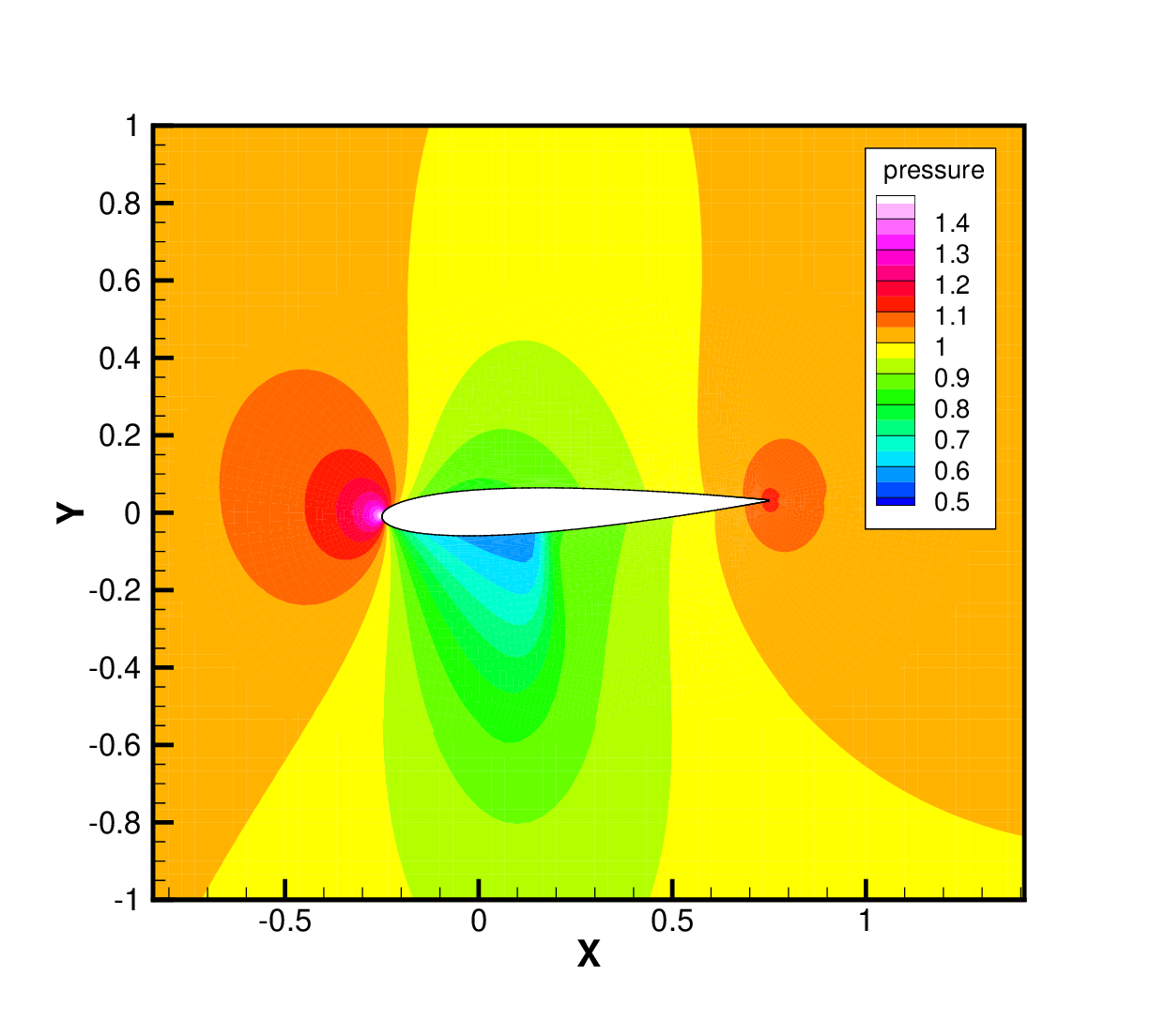}
\includegraphics[trim={0 20 50 40},clip,scale=0.3]{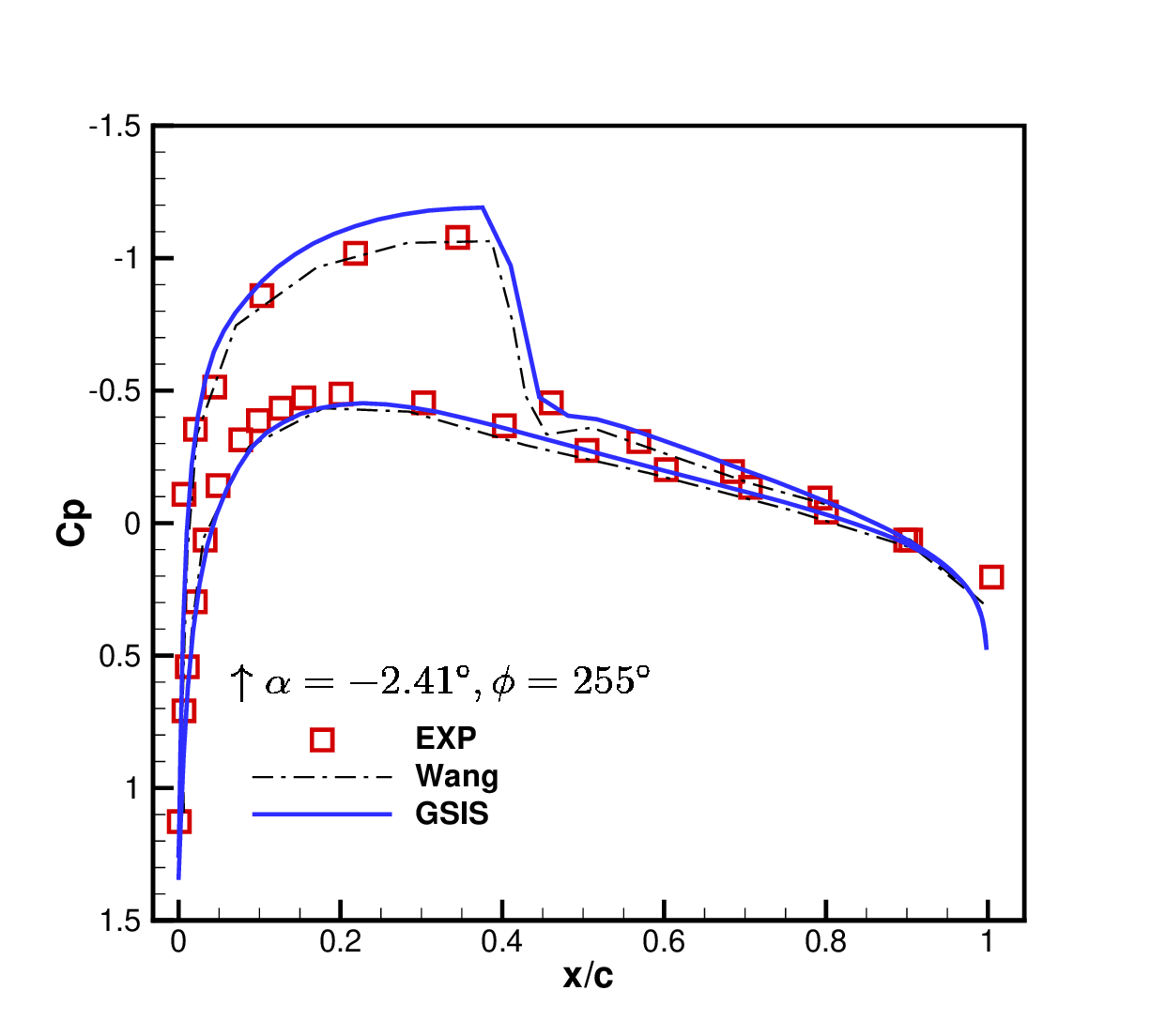}\\
\includegraphics[trim={0 20 50 40},clip,scale=0.3]{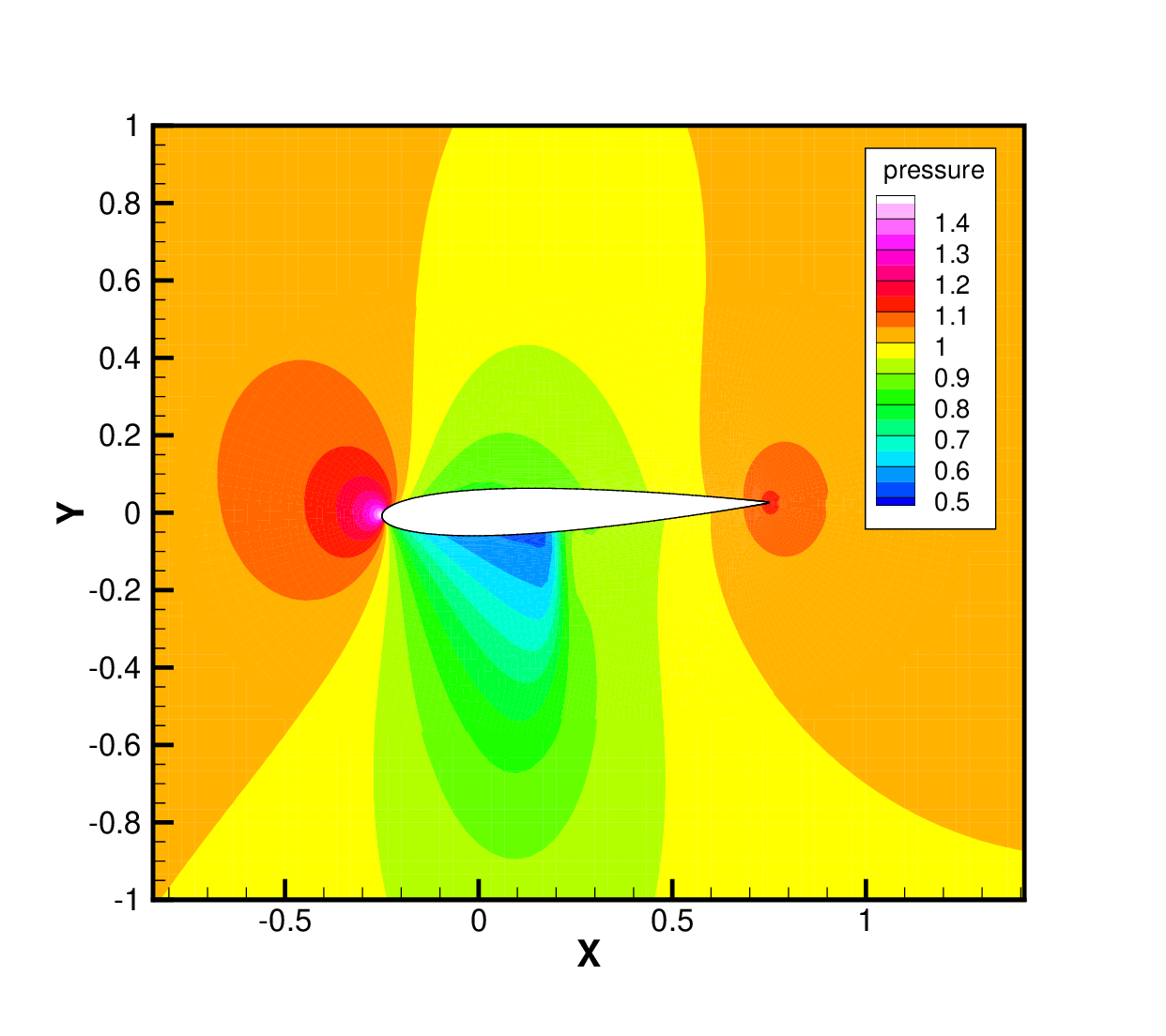}
    \includegraphics[trim={0 20 50 40},clip,scale=0.3]{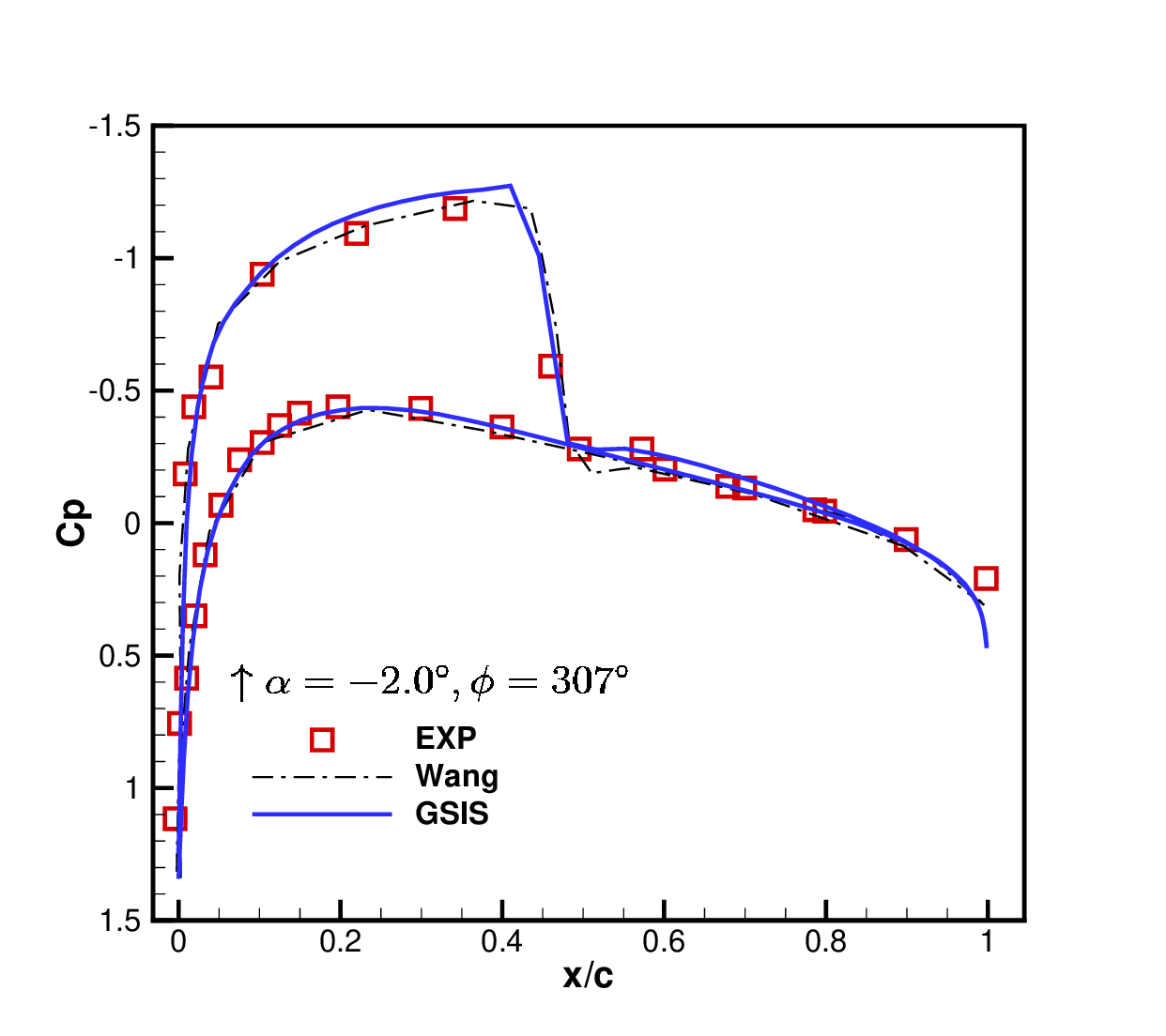}\\
    \includegraphics[trim={0 20 50 40},clip,scale=0.3]{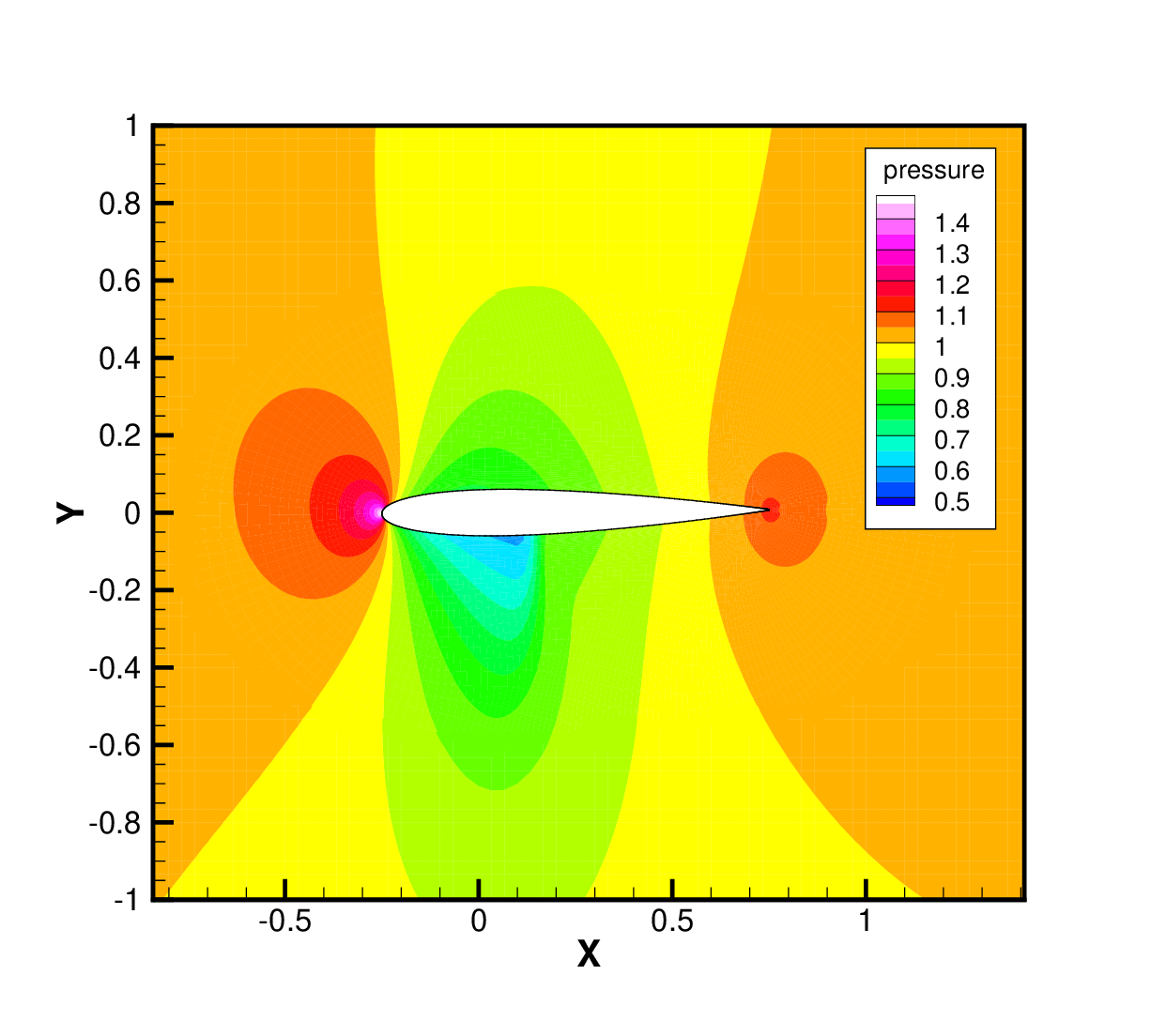}
        \includegraphics[trim={0 20 50 40},clip,scale=0.3]{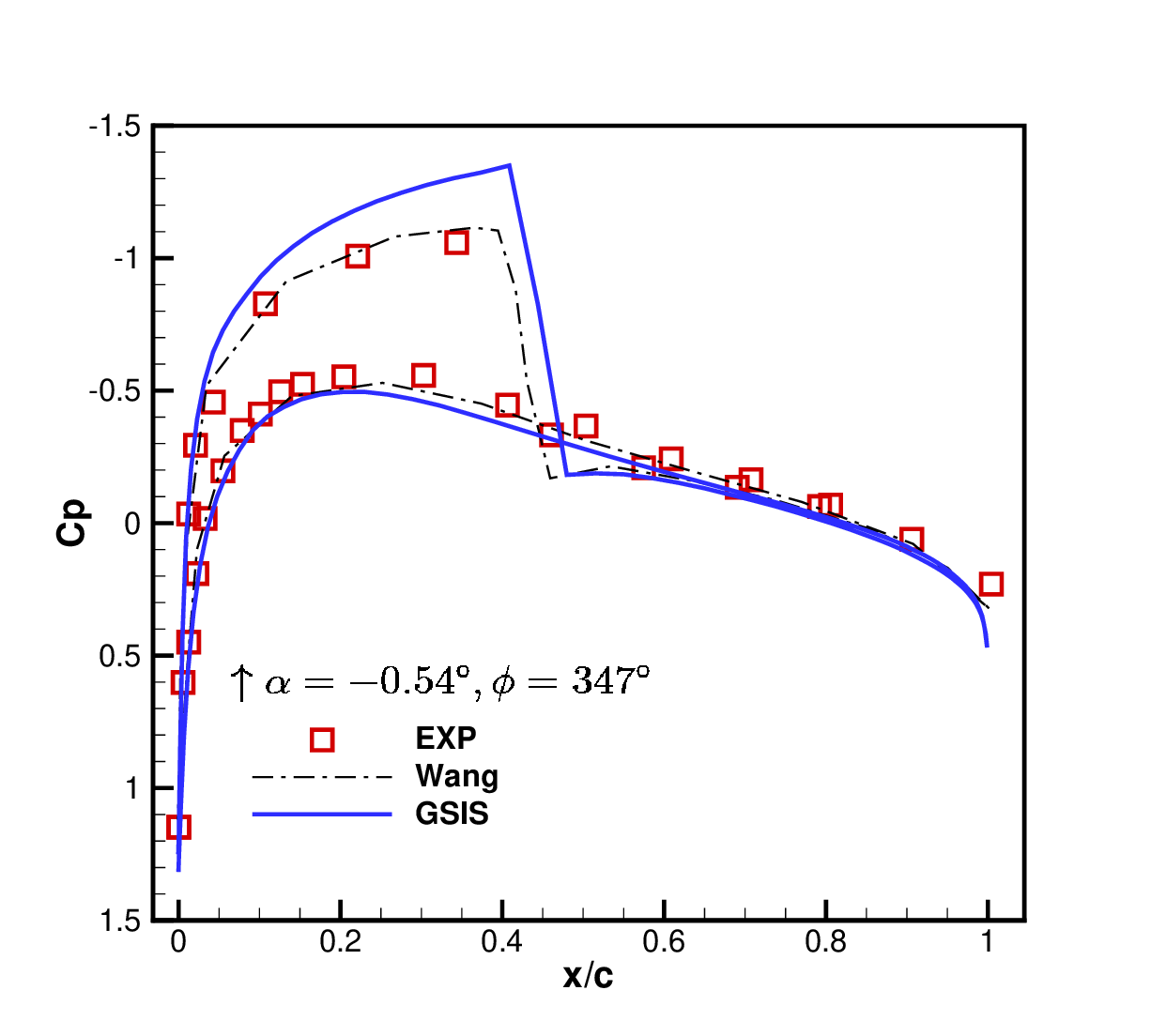}
    \caption{Same as Fig.~\ref{fig:diff_angles_1}, but at different AOAs.}
    \label{fig:diff_angles_2}
\end{figure}

Figures~\ref{fig:diff_angles_1} and~\ref{fig:diff_angles_2} compare the distribution of pressure coefficients on the airfoil surface at eight phase angles in one pitch cycle between GSIS and DUGKS. The pressure contours near the airfoil calculated by GSIS are also shown.  It can be seen that the shock wave occurs correspondingly on the top and bottom surfaces throughout the pitching cycle. In the first half cycle, pressure profiles on the lower surface are in good agreement with the experiment. However, there is a certain deviation in the location of the shock wave on the upper surface, especially at the position of $\phi=69^\circ,168^\circ$, which may explain why the lift coefficient in the first half cycle is slightly lower than the experimental value. In all four phases of the second half cycle, the pressure distribution obtained from GSIS is essentially the same as the experimental value. This example demonstrates that GSIS-ALE can provide more accurate results in the continuous regime for moving boundary problems.

\begin{figure}[t]
    \centering
    \subfloat[the evolution of $C_l$ under $\alpha$]{\includegraphics[scale=0.38,clip = true]{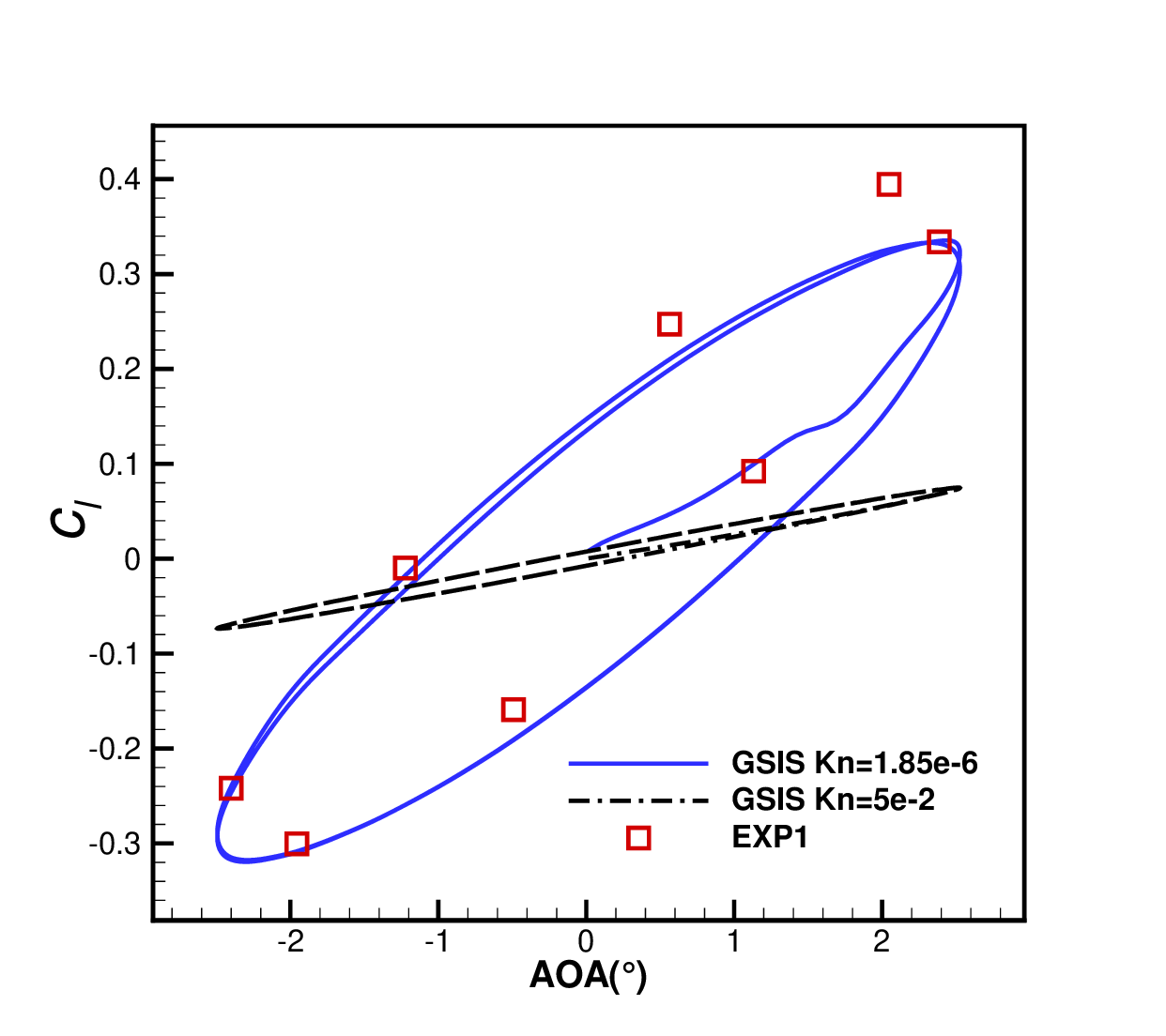}\label{fig:naca_diffkn_1}}
    \subfloat[the evolution of $C_l$ under $\phi$]{\includegraphics[scale=0.38,clip = true]{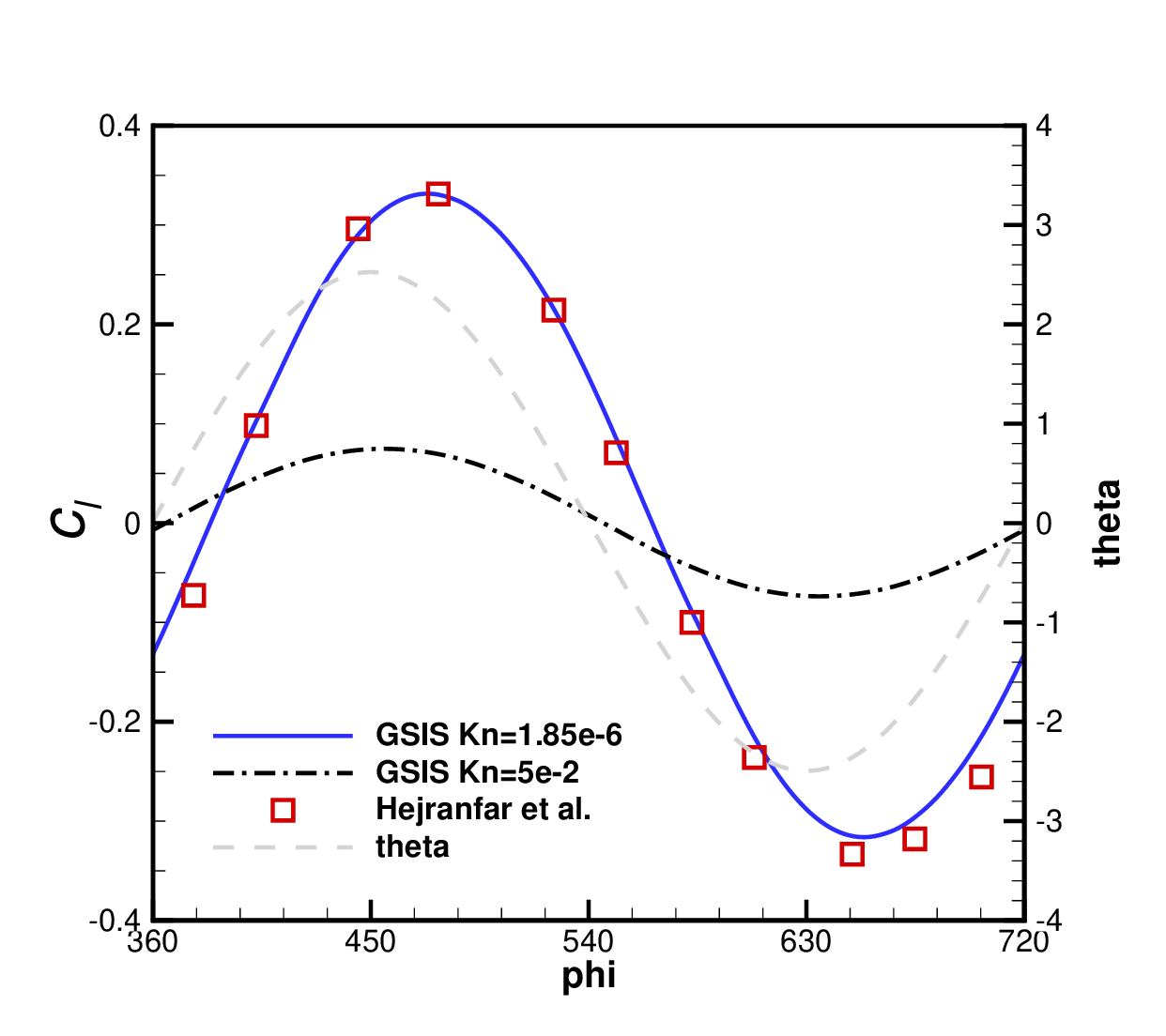}\label{fig:naca_diffkn_2}}
    \caption{The variation of the lift coefficient $C_l$ under the change of (a) angle of attack and (b) the phase, when the Knudsen numbers are Kn=1.85E-6 and Kn=0.05.  
    }
    \label{fig:naca0012_case1}
\end{figure}

Following the setup of Fig.~\ref{fig:lift_curve}, the airfoil pitching problem is further investigated for different Knudsen numbers. Figure~\ref{fig:naca_diffkn_1} shows the evolution of the lift coefficients as a function of the AOA, when Kn=1.85E-6 and 0.05. It is seen that the peak lift coefficient decreases from 0.35 to 0.074 with the increase of Kn. 
Figure~\ref{fig:naca_diffkn_2} shows the time evolution of the lift coefficient, and the gray dashed line represents the change of the AOA. 
In the continuum flow regime, the variation in lift coefficient consistently trails the variation in AOA, with an approximate phase lag of 24 degrees. Conversely, when Kn=0.05, the phase difference is minimal, at just 5 degrees. These observations suggest that the multi-scale unsteady aerodynamic characteristics merit further investigation in future studies, where the potential of the GSIS-ALE is highlighted for its applicability to scenarios like aeroelastic response analysis.

\subsection{Particle motion in lid-driven cavity flow}

\begin{figure}[t]
    \centering
    \subfloat[]{\includegraphics[scale=0.25,clip = true]{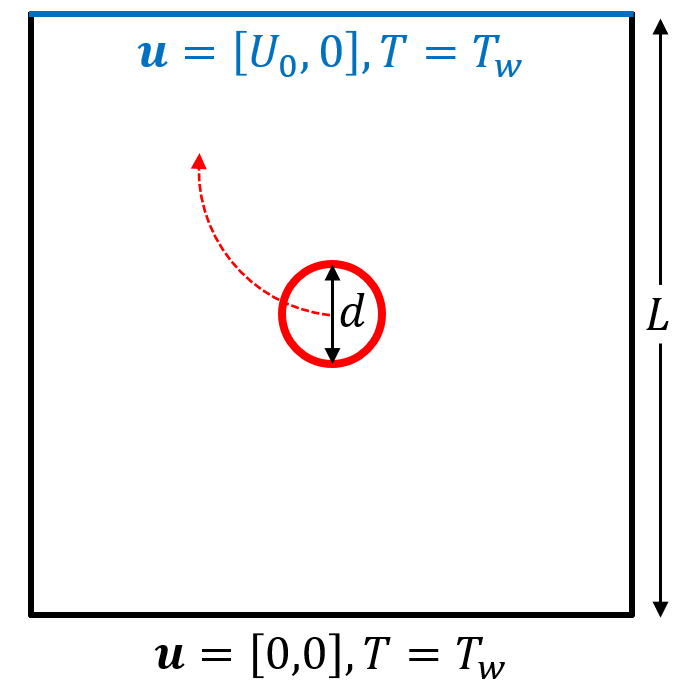}}
    \subfloat[]{\includegraphics[scale=0.15,clip = true]{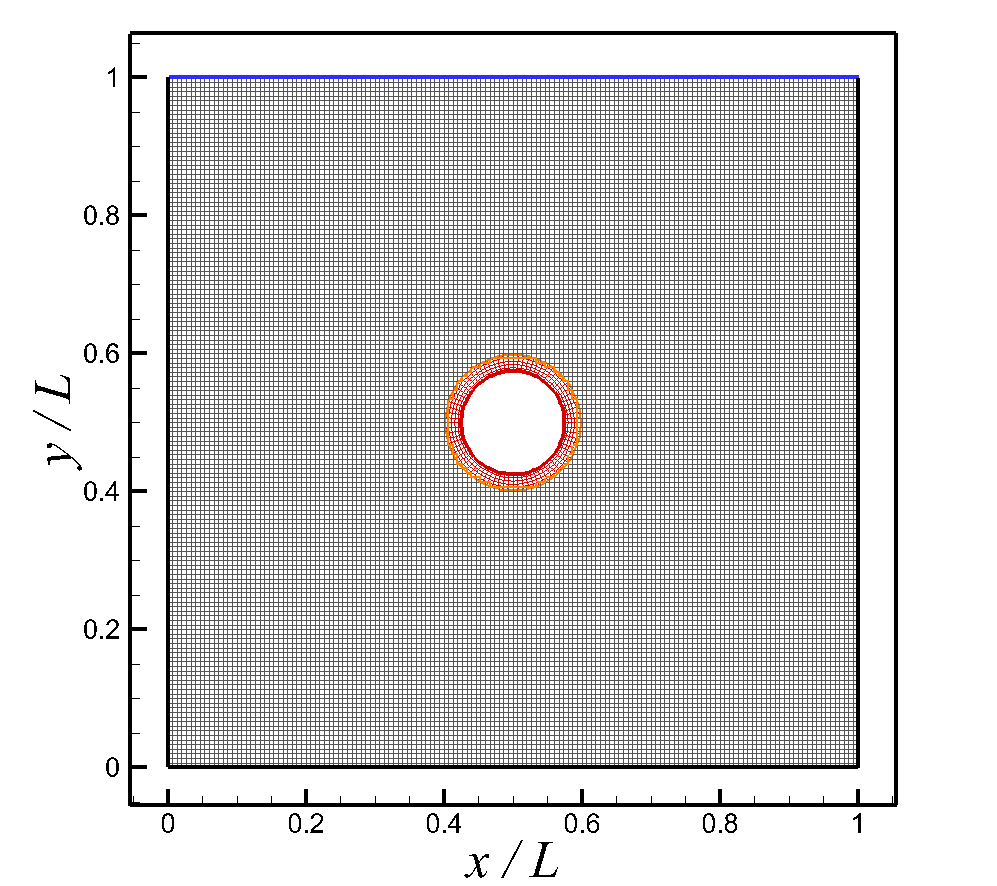}}
     \subfloat[]{\includegraphics[scale=0.2,clip = true]{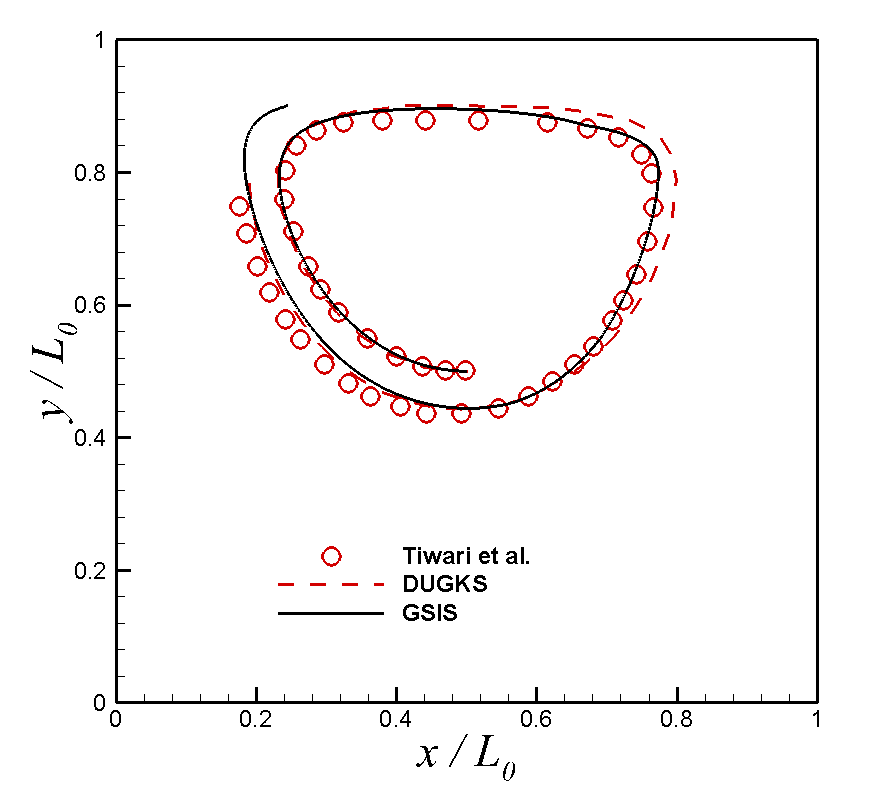}}
    \caption{(a) Schematic of the particle positioned at the center of the cavity, with the cavity length being five times the particle diameter. (b) Full domain of assembled overset mesh.  (c) The particle trajectory in the lid-driven cavity flow at Kn=0.1. 
    }
    \label{fig:paricle_mesh}
\end{figure}

The simulation of rarefied gas flows containing particles is a common test case, with a typical benchmark example being the particle-laden flow in a lid-driven cavity~\cite{tiwari2020interaction,he2024thermal}. As shown in Fig.~\ref{fig:paricle_mesh}(a), a spherical particle with a diameter $d$ and a density of $\rho_c=10\rho_f$ is placed at the center of a low-density $\rho_f$ square cavity with a side length of $L=5d$. Initially, a fixed horizontal velocity $\bm{u}_w$ is imposed on the top wall of the cavity. Upon commencement of the simulation, the particle moves freely within the cavity. The initial parameters are given by $T_0 = 300$ K, $p_0 = 100$ Pa, and $u_w = 30$ m/s. The Knudsen number is $\text{Kn}=0.1$. The walls of both the particle and the cavity are modeled using a diffuse boundary condition with a constant temperature $T_w = T_0$. Additionally, 108 mesh cells are employed to resolve the length of the cavity, see Fig.~\ref{fig:paricle_mesh}(b). In each direction of the discrete velocity space, 41 discretized grids are used.

The trajectory of the particle is plotted in Fig.~\ref{fig:paricle_mesh}(c). In the initial state, the particle moves towards the upper left corner of the cavity. Then, the particle's horizontal velocity reverses, and it migrates to the upper right corner of the cavity in an almost horizontal direction. Subsequently, the particle moves towards the lower left corner, reaching its lowest point around $ x = 0.5 $, thus completing one full cycle of motion. Although the particle's trajectory is similar to the streamlines of the lid-driven cavity flow, the particle does not return to its initial position after completing one cycle; instead, it shifts slightly downward. Our simulation results show good agreement with those obtained by Tiwari \textit{et al}~\cite{tiwari2020interaction}. The trajectory completed approximately $ \frac{4}{3} $ of a revolution, with a total of 2000 unsteady time steps computed. The computational cost for this simulation, performed in parallel (velocity space) across ten cores, is 43 minutes, with a total cost of 7 core-hours. In contrast, the explicit DUGKS took 23 days on a single core, with a total cost of approximately 552 core-hours. In this problem, GSIS reduces the core-hour expenditure by nearly two orders of magnitude.

\section{Applications}\label{application}

The proposed GSIS-ALE method is applied to two challenging problems, i.e., the two-body separation in two-dimensional supersonic flows where the interference of shock waves leads to counter-intuitive dynamics of block separation, and the three-dimensional lunar landing simulation. 

\subsection{Two-body separation in supersonic flow}

Two-body separation problems hold significant importance in various engineering applications, yet existing methods often fall short in simulating the entire separation process. Instead, steady-state results are computed at different separation heights and subsequently pieced together, which fails to capture the dynamic nature of the separation, particularly under high-speed flow conditions. 
The proposed GSIS-ALE is applied to study the drop of a square block under high-speed flow.

\begin{figure}[t]
    \centering
    \subfloat[velocity contour]{\includegraphics[trim={0 0 0 90},clip,scale=0.2]{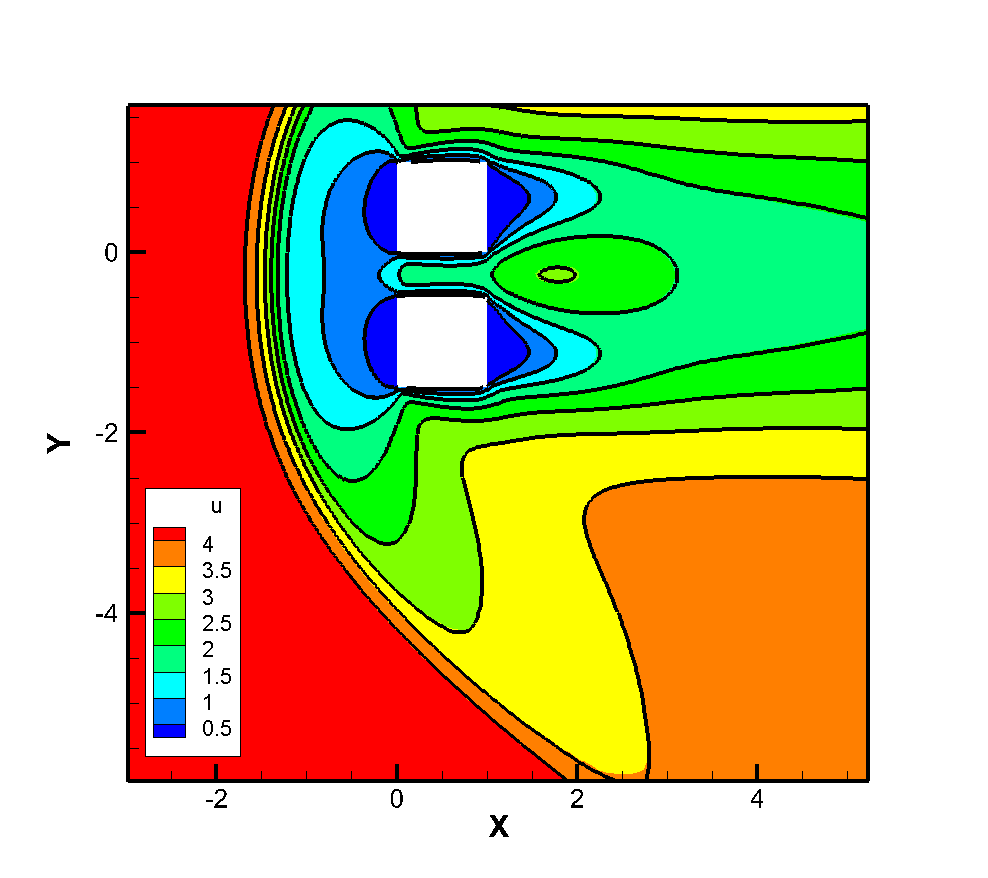}}
    \subfloat[temperature contour]{\includegraphics[trim={0 0 0 90},clip,scale=0.2]{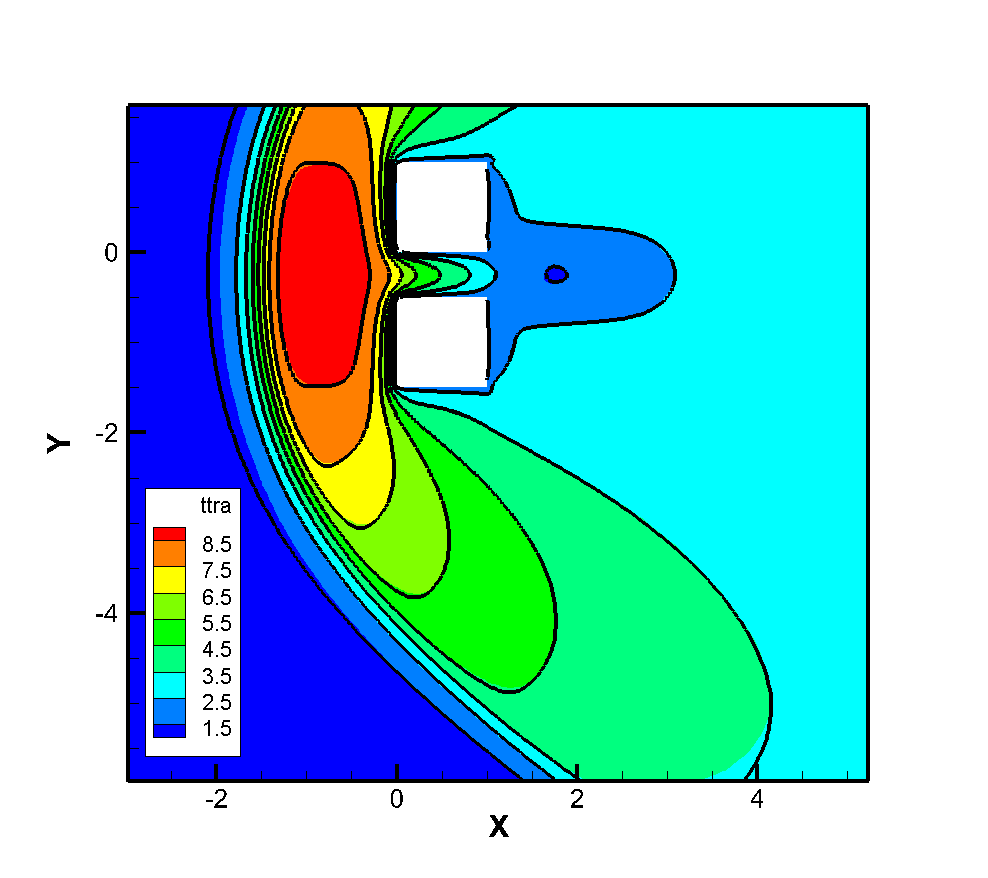}}\\
    \subfloat[velocity at $ y = -0.25 $]{\includegraphics[trim={0 0 0 80},scale=0.2,clip = true]{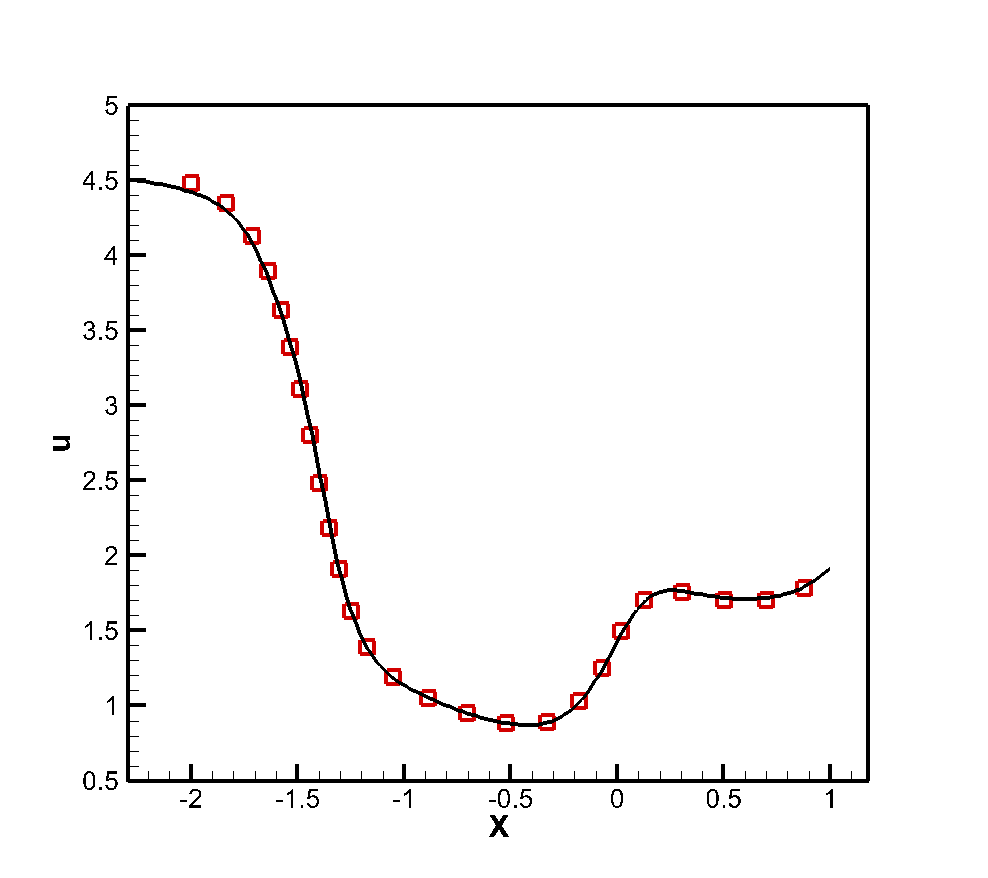}}
    \subfloat[temperature at $ y = -0.25 $]{\includegraphics[trim={0 0 0 80},scale=0.2,clip = true]{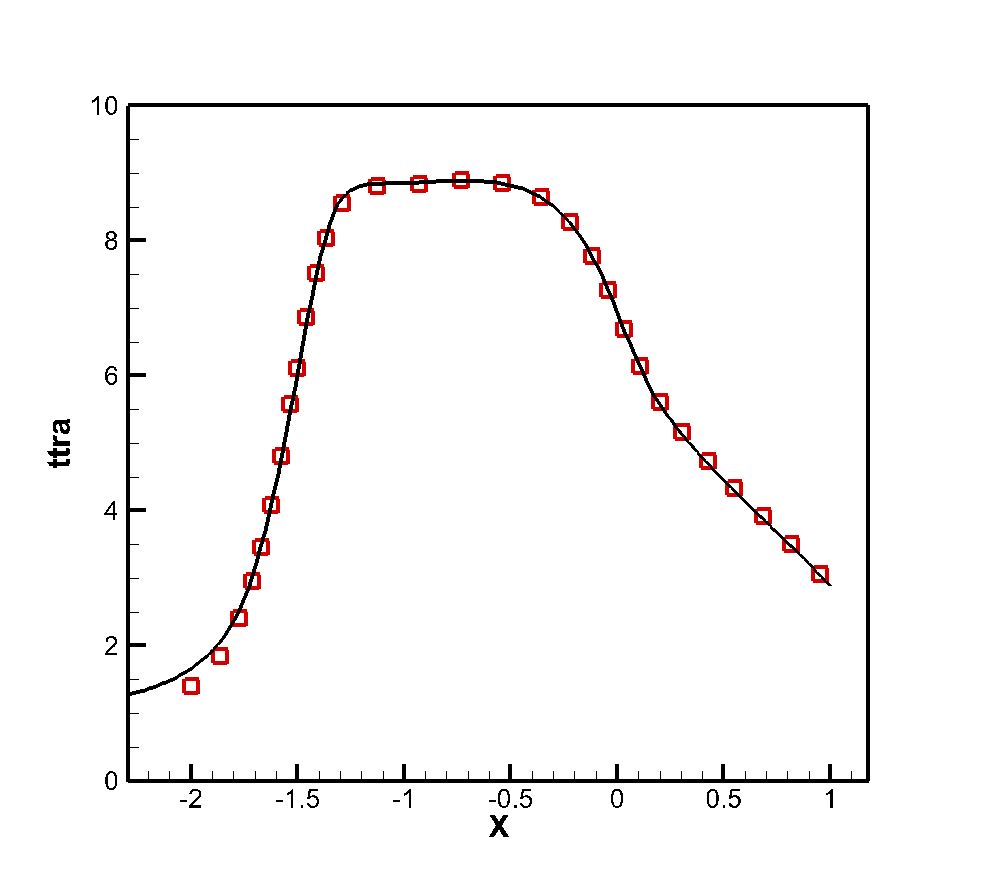}}
    \caption{(Top) The distribution of horizontal velocity and temperature at the initial stationary moment for the square drop problem with $\text{Ma}=5 $ and $ \text{Kn}=0.1$. 
    (Bottom) Horizontal velocity and temperature at fixed line. 
    Solid lines: GSIS; Contours/squares: DSMC. }
    \label{fig:tblock_contour}
\end{figure}


\begin{figure}[htbp]
    \centering
    \subfloat[$t=0.2$]{\includegraphics[width=0.4\textwidth,trim=30 30 100 100,clip]{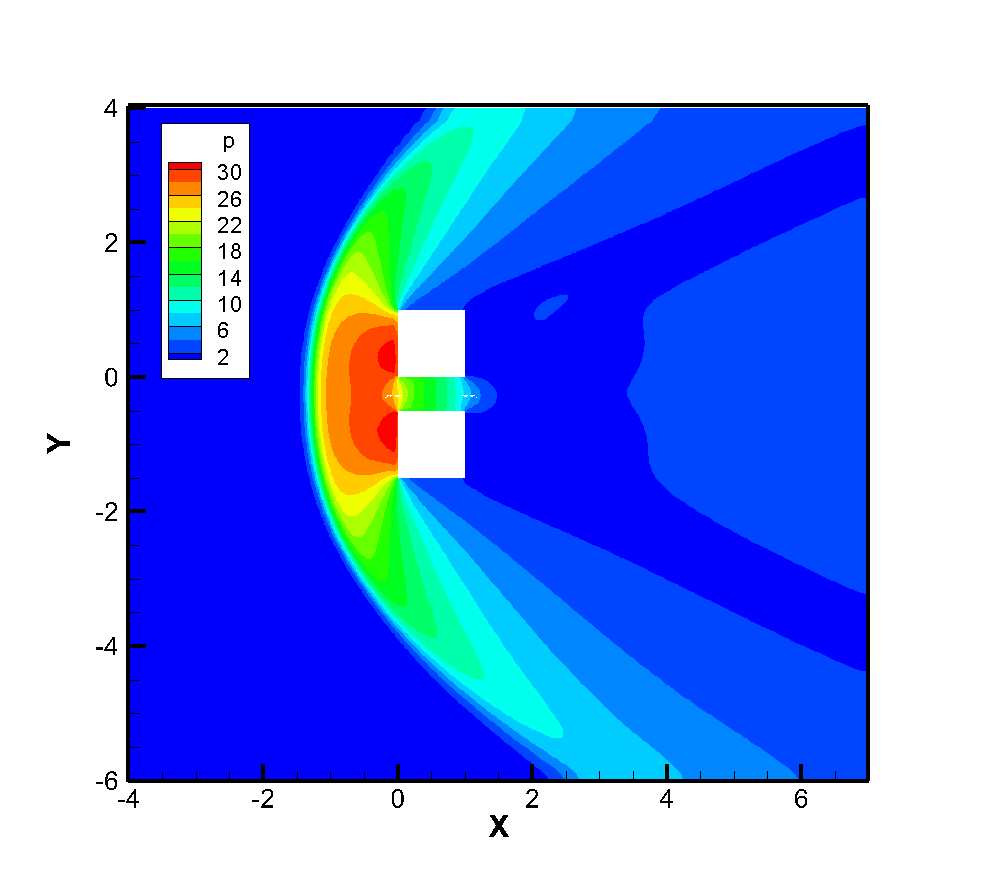}}
    \subfloat[$t=1.9$]{\includegraphics[width=0.4\textwidth,trim=30 30 100 100,clip]{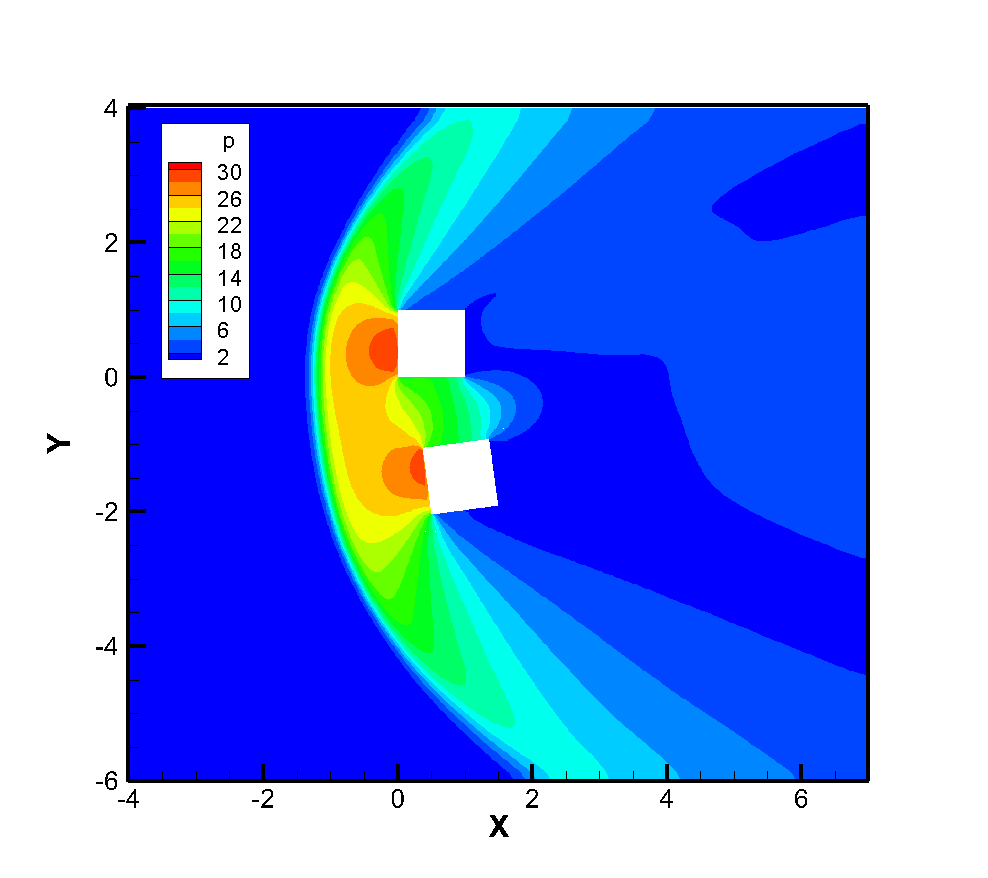}}\\
    \subfloat[$t=2.9$]{\includegraphics[width=0.4\textwidth,trim=30 30 100 100,clip]{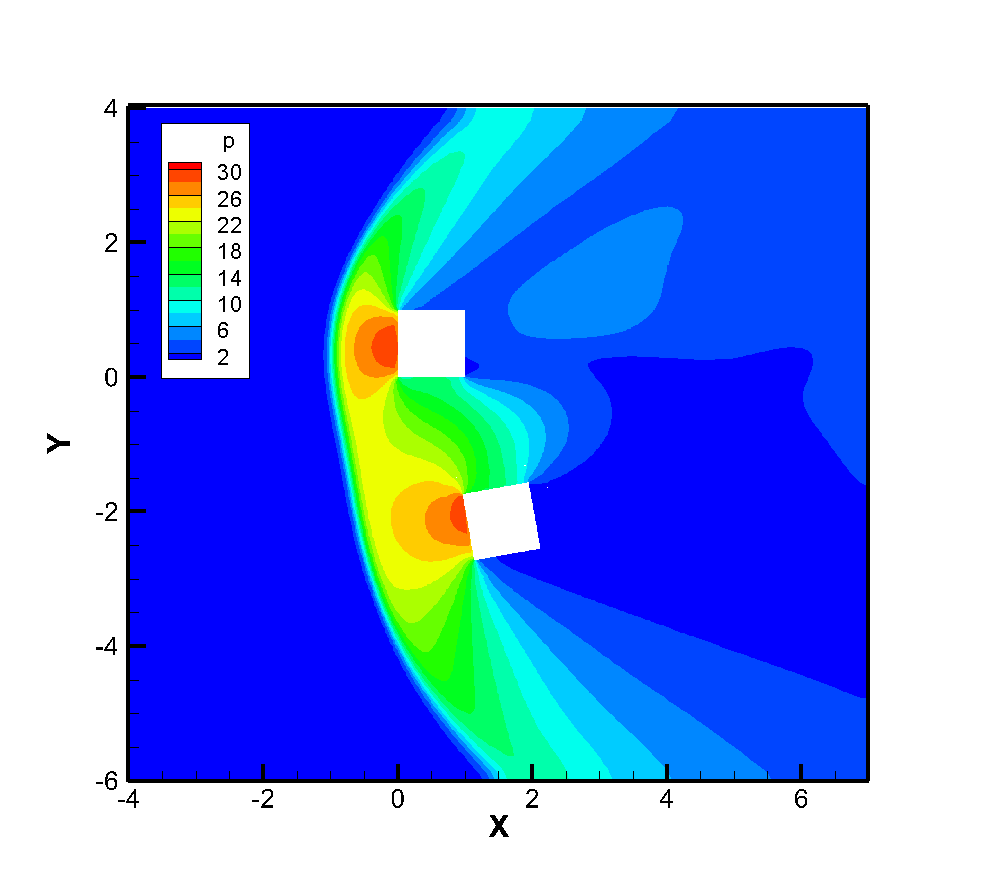}}
    \subfloat[$t=3.4$]{\includegraphics[width=0.4\textwidth,trim=30 30 100 100,clip]{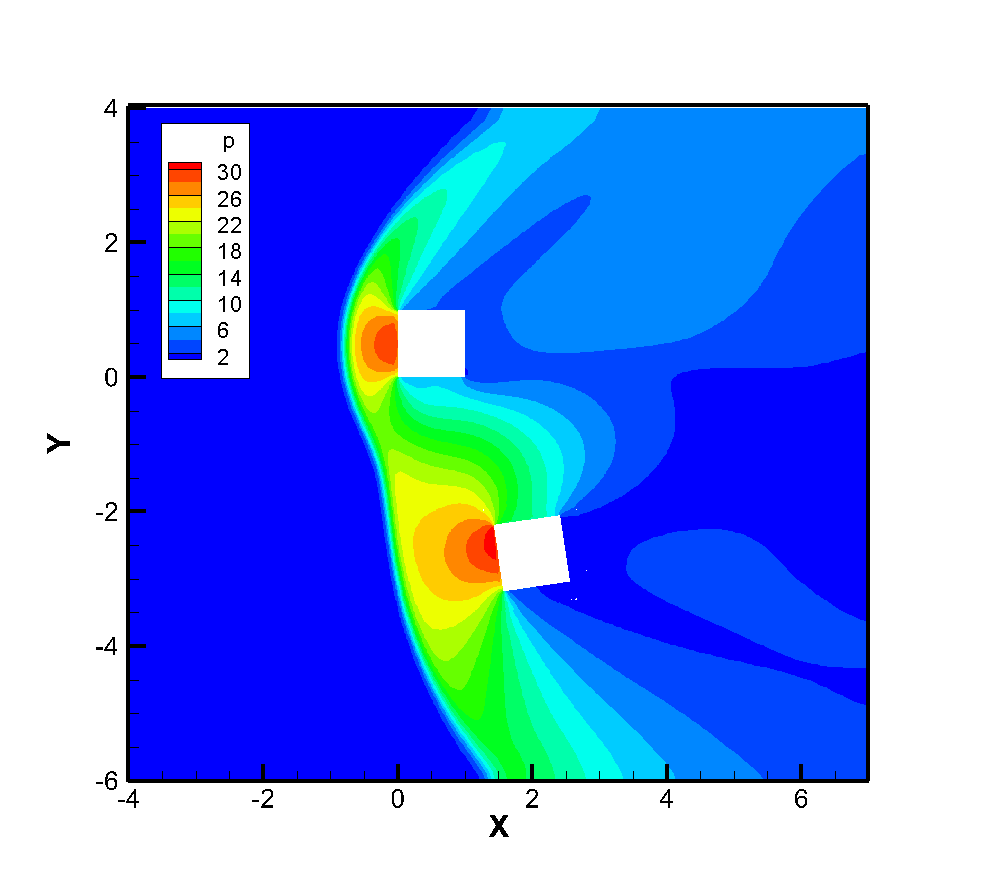}}\\
    \subfloat[$t=4.9$]{\includegraphics[width=0.4\textwidth,trim=30 30 100 100,clip]{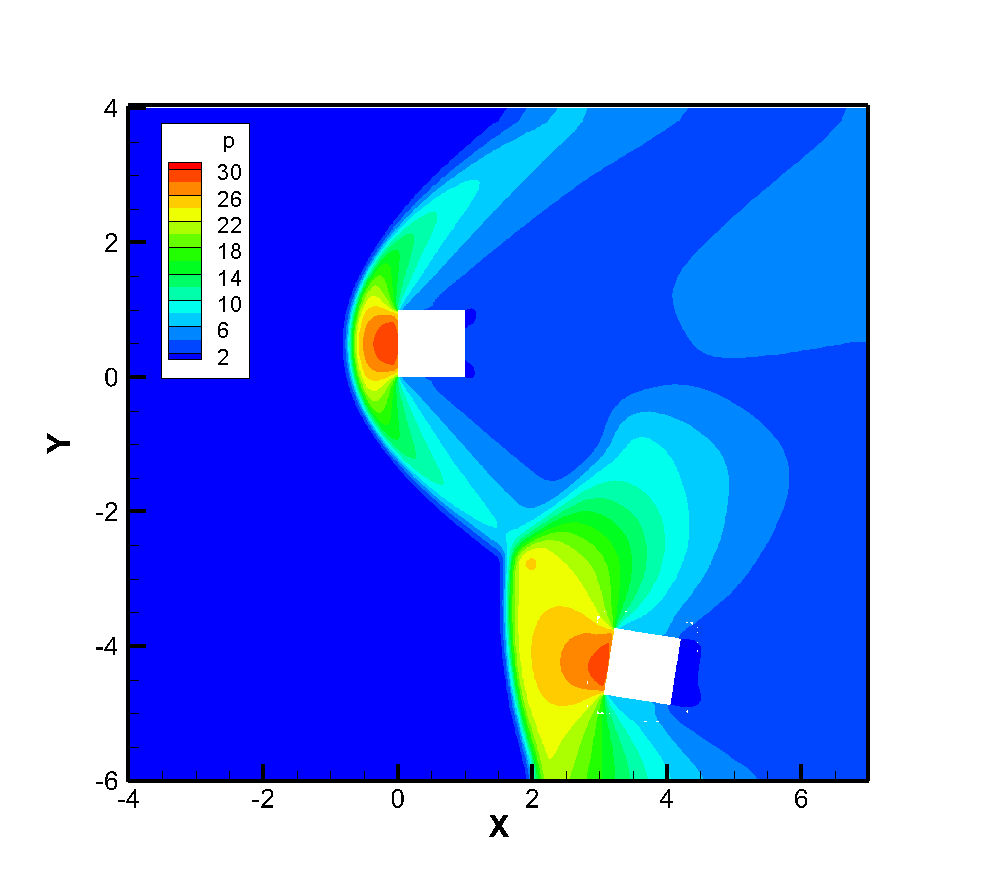}}
    \subfloat[positions of block.]{\includegraphics[width=0.4\textwidth,trim=20 20 90 80,clip]{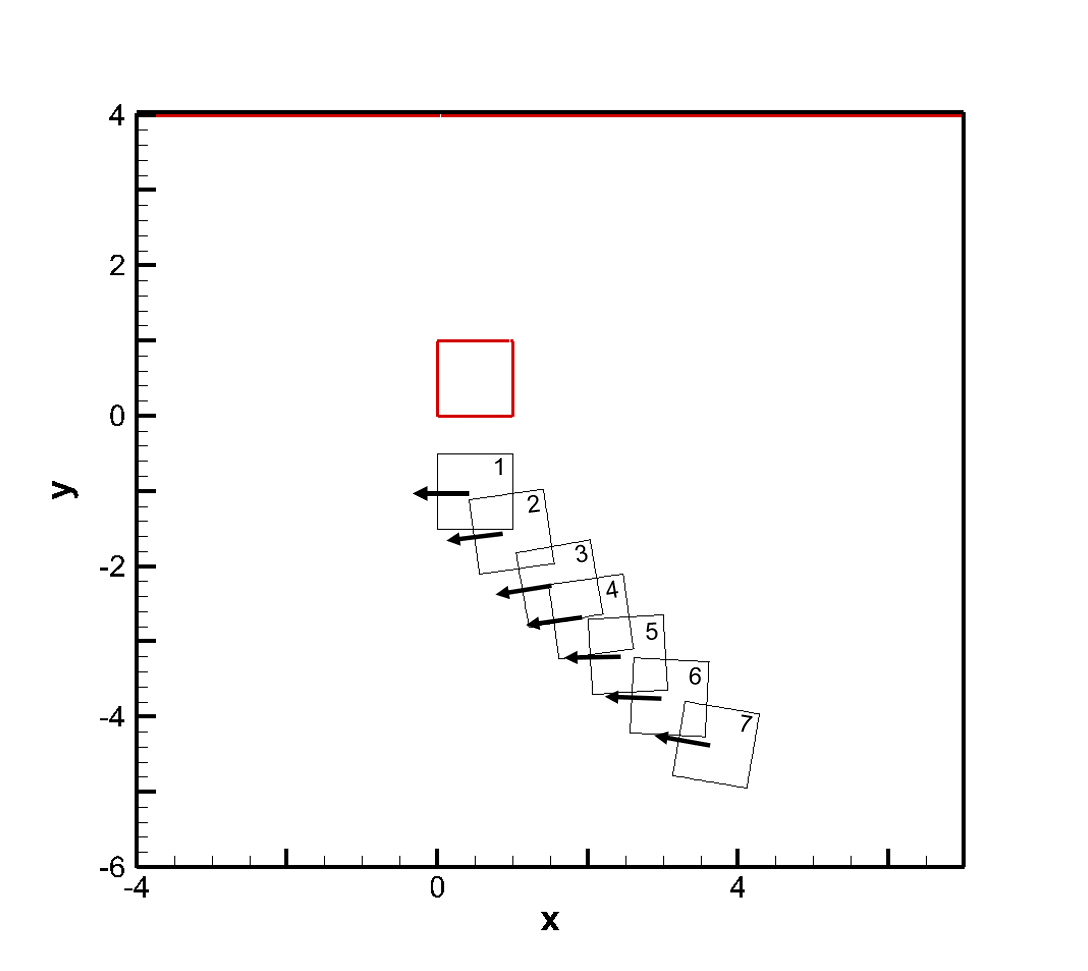}}
    \caption{The pressure distribution at different time steps and the trajectory for the two-body separation problem at $\text{Ma} = 5, \text{Kn}=0.1$. (f) From 1 to 7, the time is $t$ = (0.2, 1.9, 2.9, 3.4, 3.9, 4.4, 4.9) $t_{ref}$, respectively. }
    \label{fig:tblock_track}
\end{figure}

To ensure the accuracy of the overset assembly, a stationary block case is first simulated with inflow conditions of Ma=5 and Kn=0.1. The side length of the square block is $L$, and the centers of the two horizontally placed blocks are located at $(0.5L,0.5L)$ and $(0.5L,-L)$, respectively. The block's sides are resolved by 30 grid points, with a background grid containing 33,100 cells and a body-fitted grid with 2,520 cells. The velocity space is discretized in two directions over the ranges $[16,14]$ with $72\times64$ uniform discrete points.
Fig.~\ref{fig:tblock_contour} shows that the normalized velocity and temperature obtained from the DSMC and GSIS. The distributions of macroscopic quantities along the central line is also compared. The velocity profile exhibits an initial rise followed by a subsequent decline, with the turning point located near $x=-0.5$. As the gas enters the channel between the two blocks, the velocity gradually increases, and the temperature decreases. When $x>1$, the velocity continues to rise due to the rapid expansion outside the channel. Again, GSIS results with the overset approach match well with the DSMC results. For the initial case, the GSIS-ALE requires only 90 iteration steps to reach a steady state.

Subsequently, the steady-state results from the stationary case are used as the initial solution for an unsteady simulation of the free movement of the lower block, employing an implicit second-order time discretization scheme. Due to the limitations of the background grid computational domain, the drop process is simulated for only 5 $t_{ref}$. Under the ten-core parallel computation of the velocity space, the total computation time for this process is about one hour. Fig.~\ref{fig:tblock_track} shows the pressure contour and trajectory at different times. During the initial stage of separation, the block rotates counterclockwise, but as the separation height increases, the block's rotation direction changes to clockwise. This abnormal change in angle may be related to the shock wave interference between the two blocks. In the later stage of free motion $(t > 2)$, the pressure on the upper side of the windward face of the block is slightly higher than on the lower side, resulting in a clockwise angular acceleration.
At $t=4.9$, the flow interference between the two blocks diminishes, and the pressure distribution on the upper block nearly returns to an up-down symmetrical state. 

\subsection{Three-dimensional lunar landing}

\begin{figure}[t]
    \centering
    \subfloat[geometry and boundary condition]{\includegraphics[width=0.4\textwidth,clip = true]{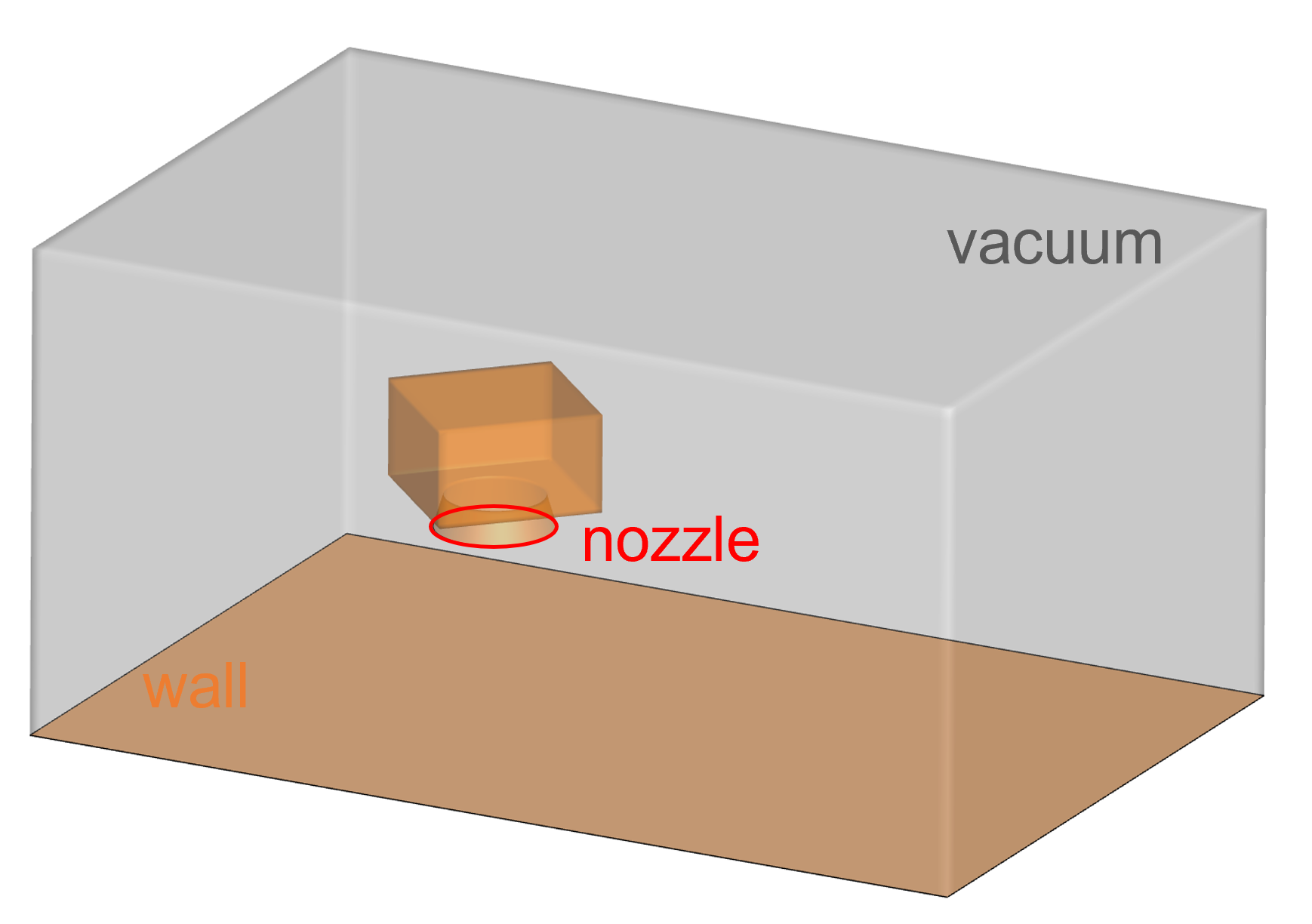}}\\
    \subfloat[Assembly overset mesh]{\includegraphics[width=0.7\textwidth,clip = true]{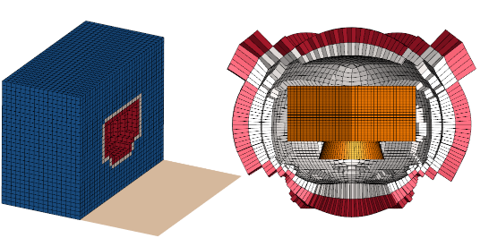}}
    \caption{Assembled overset mesh for the unsteady simulation of a 3D lander. The background mesh (173,304 cells) includes vacuum outlets and wall boundaries, while the body-fitted mesh (52,880 cells) encompasses pressure inlet boundaries and wall boundaries. In the mesh assembly result, white and red cells represent interpolation cells with $celltype=2$ and 4, respectively.}
    \label{fig:lander_mesh}
\end{figure}

In the conceptual design of the Chang'e 7  lunar exploration mission, the leaper is equipped to perform surface hops on the Moon using its jet engines. During hopping process, both the attitude of the leaper and the thrust of its engines change frequently. To accurately assess the impact of engine plume contamination on lunar soil sampling, it is crucial to precisely predict the engine plume flow-field and its impingement effects during the unsteady hopping process.

\begin{figure}[p]
    \centering
    \includegraphics[scale=1.0,clip = true]{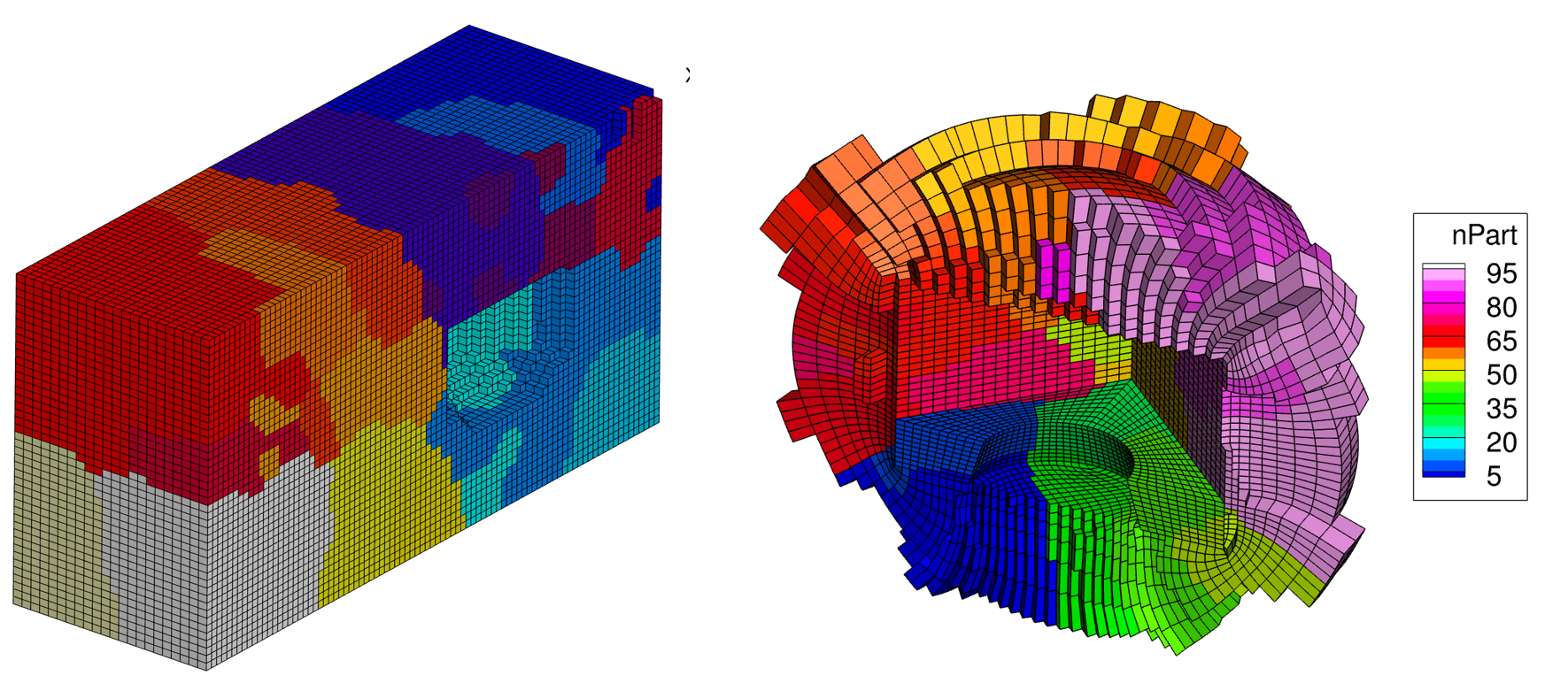}
    \caption{Schematic diagram of physical mesh partitioning, with the left and right images representing the background mesh and body-fitted mesh, respectively. The variable $ n_{\text{Part}} $ denotes the physical partition number.}
    \label{fig:lander_metis}
\end{figure}

\begin{figure}[p]
    \centering
    \subfloat[$\phi = 3.6 ^\circ$]{\includegraphics[trim={0 70 0 20},scale=0.18,clip = true]{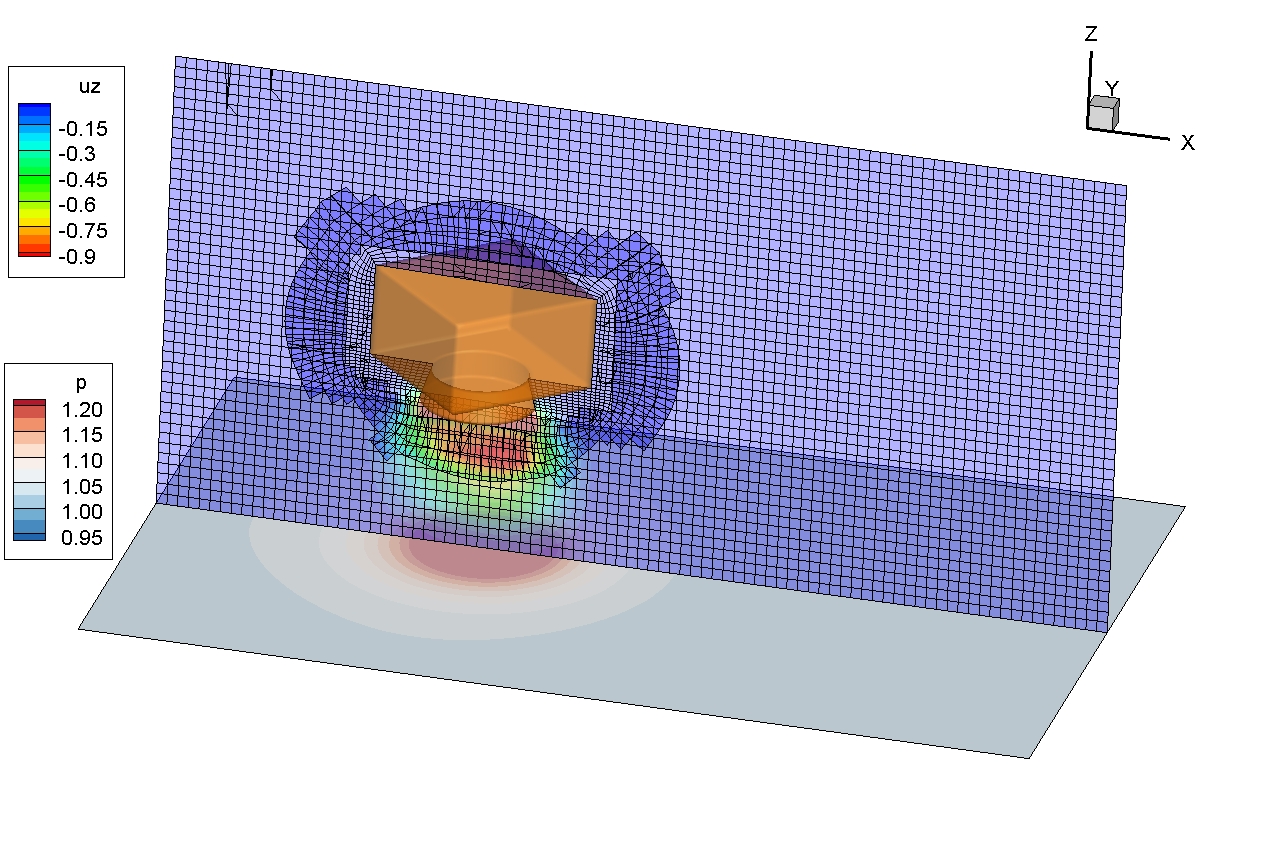}}
    \subfloat[$\phi = 90 ^\circ$]{\includegraphics[trim={0 70 0 20},scale=0.18,clip = true]{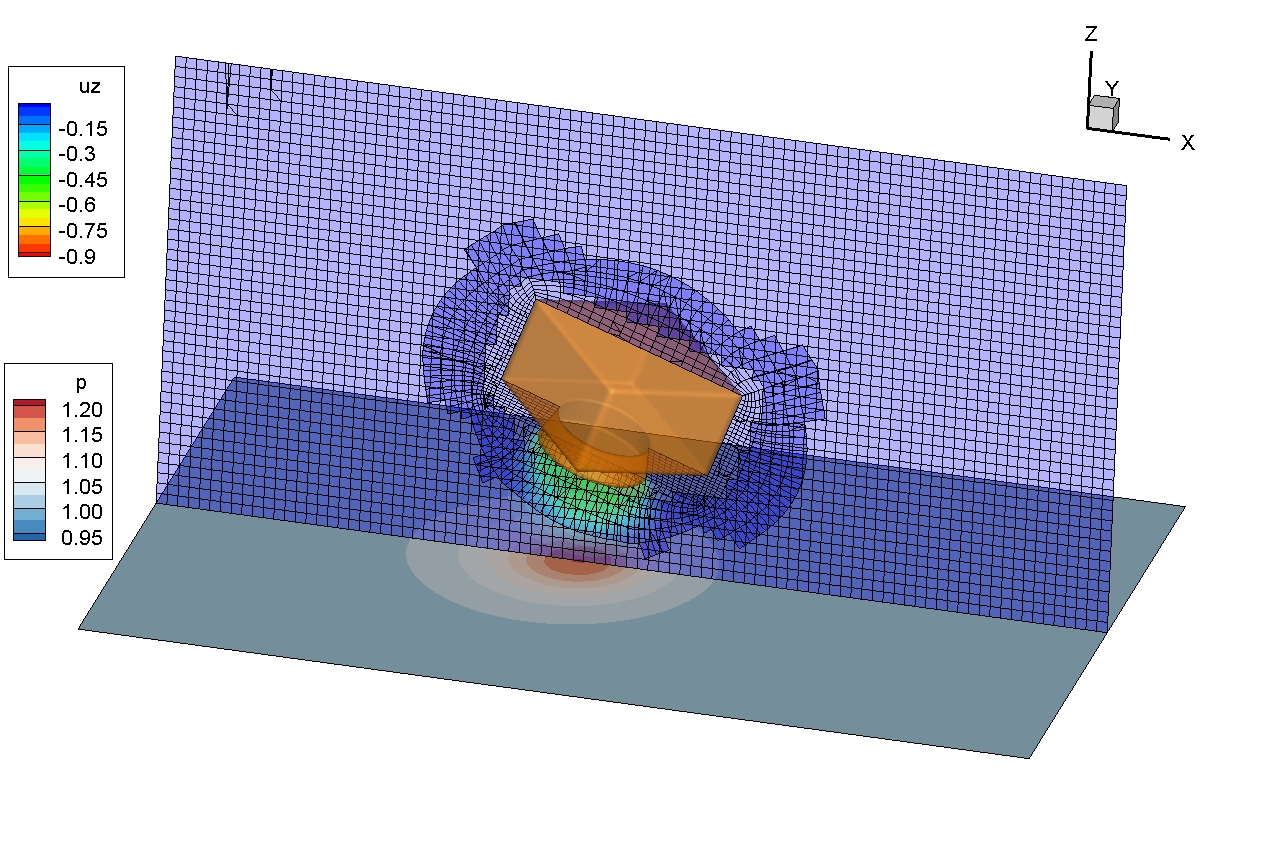}}\\
    \subfloat[$\phi = 180 ^\circ$]{\includegraphics[trim={0 70 0 20},scale=0.18,clip = true]{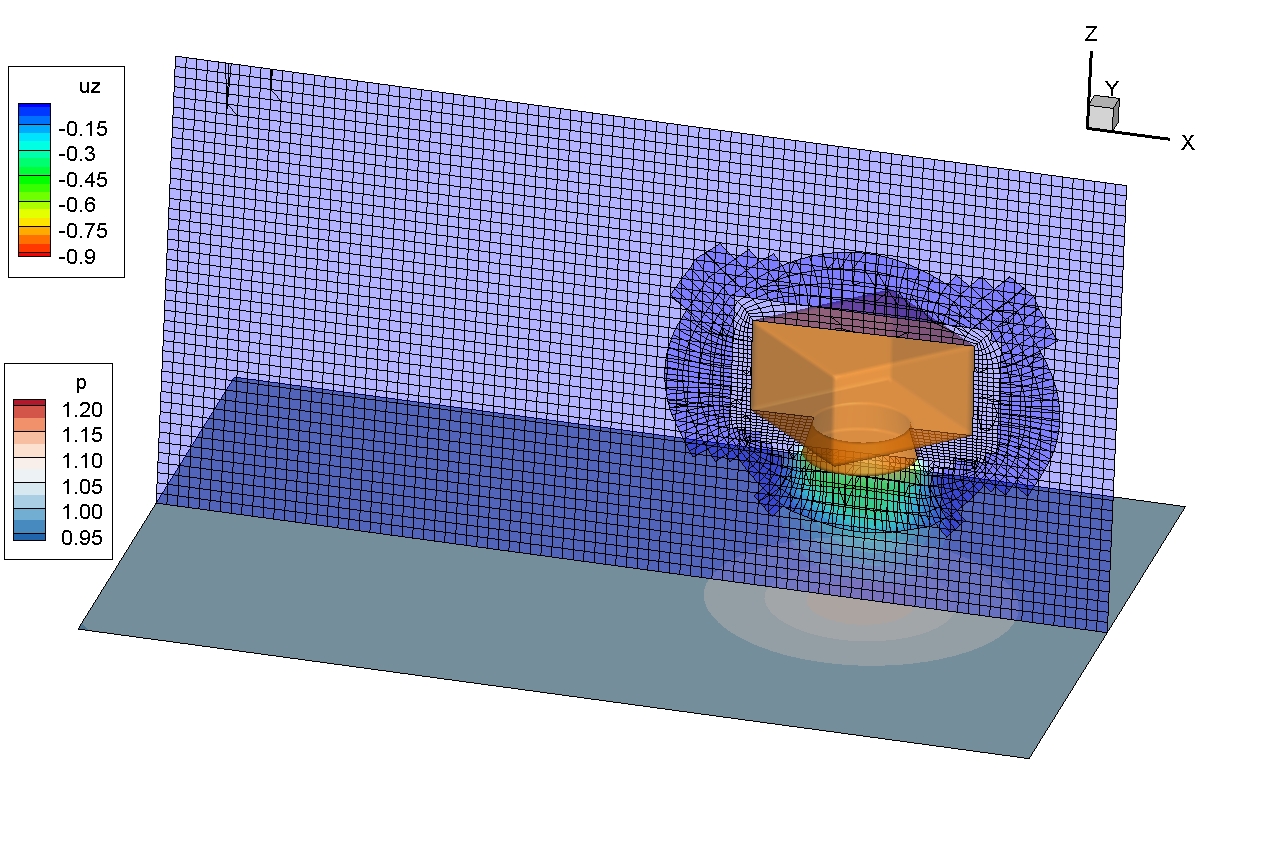}}
    \subfloat[$\phi = 270 ^\circ$]{\includegraphics[trim={0 70 0 20},scale=0.18,clip = true]{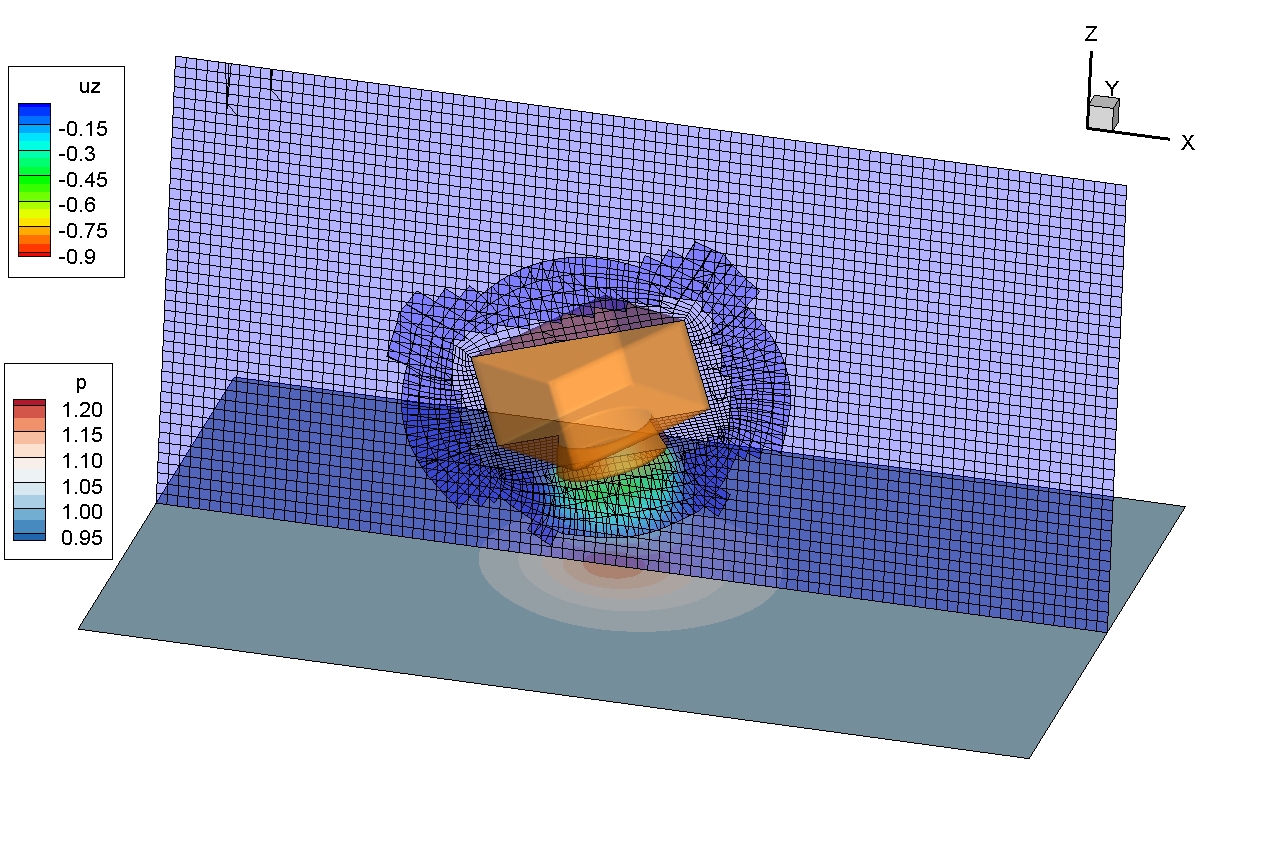}}
    \caption{The attitude and ground pressure distribution of the lander at different time steps.}
    \label{fig:lander_track}
\end{figure}

In the preliminary demonstration, the solar sail structure was simplified, retaining only the leaper's body and nozzle. The geometry and meshes for the leaper's movement are shown in Fig.~\ref{fig:lander_mesh}, where the yellow part represents the 300K wall surface, gray part indicates the vacuum outlet, and the red circle denotes the thrust of the engine with a pressure inlet boundary condition. The exit diameter of the engine is 36 mm, with pressure and temperature set to 300 Pa and 300 K, respectively. The velocity domain is truncated to a cube with length $10v_0$, corresponding to $24\times24\times24$ uniformly distributed velocity discretization points. The motion trajectory of the center of mass in the $xoz$-plane($y(t)=0$) is defined as follows:
\begin{equation}
\begin{aligned}
    x(t) &= - 60 \cos\phi + 60,\\
    z(t) &=\max(55, - 20 |\sin(\omega t)| + 70), \\
    \alpha(t) &= -20^\circ \sin(\omega t),\quad\text{with} \quad \phi = \pi + kt,
    \end{aligned}
\end{equation}
where $k = 0.64$ is defined as frequency of the periodic motion.

\begin{figure}[t]
    \centering
\includegraphics[width=0.5\textwidth,clip = true]{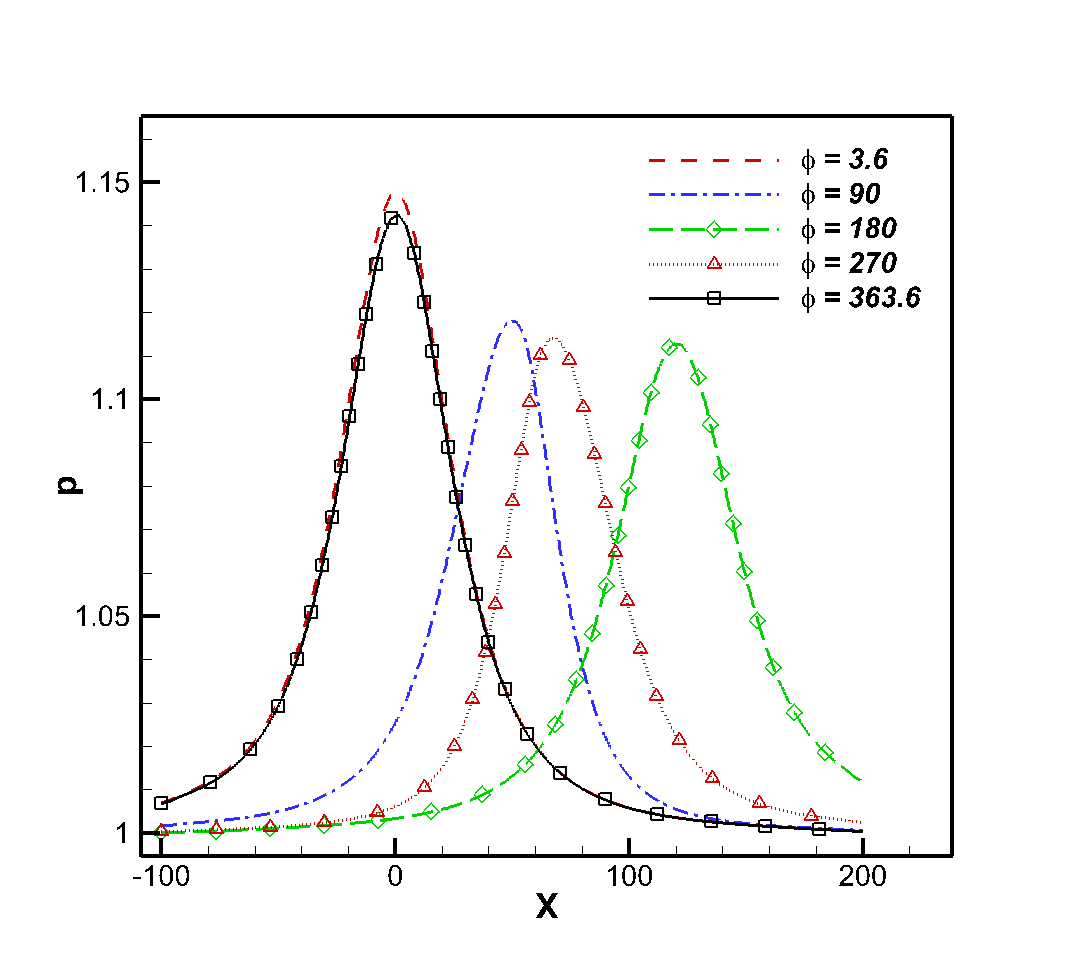}
\caption{The pressure distribution along the ground symmetry axis (y=0) at different time steps.
    }
    \label{fig:lander_pressure}
\end{figure}

In addition to the velocity space parallelization employed in the two-dimensional problem, METIS technology is utilized to decompose the physical grid in the three-dimensional problems, see detailed implementation in Ref.~\cite{zhang2023efficient}. Fig.~\ref{fig:lander_metis} illustrates the results of physical partitioning during the leaper's movement, where $nPart$ denotes the identifiers of different processes. It can be observed that each process contains the physical data of partitioned grid blocks from all components. In the current study, grid assembly is performed serially on the root process. Prior to updating the interpolation boundary conditions, physical quantities from different processes are gathered to the root process, then distributed back to all processes. Finally, the interpolation cells within each process are updated. In order to limit the range of motion, one cycle loop is discretized into 100 steps, with an angular increment of $3.6^\circ$ per step. Under a 100-core physical space parallel setup, GSIS-ALE requires 21.5 hours to complete the computation for one cycle. Note that the assembly method used in this paper is a serial assembly method based on wall distance calculation. As the number of grids increases, its time complexity also rises accordingly. In this section with 220,000 physical grids, the assembly overhead is roughly equal to the flow field calculation overhead. To improve the computational efficiency of assembly from different perspectives, various strategies can be referenced, such as parallelization~\cite{hu2021robust}, cell classification algorithms~\cite{xia2021highly}, and fast search for interpolation cells~\cite{roget2014robust}. According to estimates from these studies, assembly time can be reduced by more than an order of magnitude.

Figure~\ref{fig:lander_track} shows the ground pressure distribution at different moments, the velocity distribution at the $y=0$ cross-section, and the assembled grid. It can be observed that, within one motion cycle, the assembly method correctly constructs the hole boundary under different pitch angles, and there is a buffer zone of about four layers between the two sets of grids, which is sufficient to meet the calculation requirements for second-order interpolation. Additionally, the pressure distribution along the axis of symmetry at different times is provided in Fig.~\ref{fig:lander_pressure}. It can be seen that the pressure on the windward side decreases more rapidly, and at $x=0$  (close to the left endpoint of the trajectory), the pressure peak is not entirely the same, possibly due to the flow field not being fully stabilized. 
\section{Conclusion}\label{sec:5}


In this work, the original GSIS method on a static mesh has been further extended to the GSIS-ALE method, designed for simulating moving boundary problems across a wide range of flow speeds and Knudsen numbers. This method employs overset mesh techniques to handle mesh motion and introduces mesh motion velocity to implement the ALE framework. The rigid body motion is solved using the six degrees of freedom equations. The study considered the pitching of an airfoil and the particle flow problem within a lid-driven cavity. The results from the GSIS-ALE method show good agreement with experimental data and other numerical results, validating the accuracy and effectiveness of the current GSIS-ALE method, even for a large physic time step. Additionally, numerical studies on a two-body separation problem and a three-dimensional lander motion problem were conducted, demonstrating that the GSIS-ALE method is a promising new choice for simulating moving boundary problems.


This work represents the first step in developing the GSIS solver for moving boundaries involving relative motion of bodies. Looking ahead, surface response solvers could be integrated into body-fitted subgrids to facilitate the simulation of fluid-structure interaction such as aerodynamic elasticity and ablative effects. Furthermore, while this work utilizes the Shakhov model to simulate monatomic gas flows, the current methodology can be straightforwardly extended to simulate molecular gases or multi-species gas flows. This extension may offer potential applications in accurately simulating non-equilibrium moving boundary flows with high-temperature gas effects.

\section*{Acknowledgments}

This work is supported by the National Natural Science Foundation of China (12172162) and the Stable Support Plan (80000900019910072348). Special thanks are given to the Center for Computational Science and Engineering at the Southern University of Science and Technology.
\appendix
\section{Dimensional reduction}\label{sec:reduction}

In two-dimensional problems, the velocity space can be reduced from three-dimensional to two-dimensional to save the computational cost. To achieve this, reduced velocity distribution functions, denoted as $\hat{f}_{xy}$ and $\hat{f}_{z}$, are introduced:
\begin{equation}
        \left(\hat{f}_{xy}, \hat{f}_{z}\right) = \int \left(f, \xi_z^2 f\right) \myd\xi_z,
\end{equation}
and the dimensionless macroscopic quantities can be calculated as:
\begin{equation}
    \begin{aligned}
&\left(\rho,\rho\bm{u}\right)=\int\left(1,\bm{\xi}\right) \hat f_{xy} \myd \bxi,\quad
\frac{3}{2}\rho T=\int\frac{1}{2}\left(|\bxi - \bu|^2\hat f_{xy}+\hat f_{s,z}\right)\myd \bxi,\\
&\bm{P}=\int (\bxi - \bu)\otimes(\bxi - \bu)\hat f_{xy} \myd \bxi,\quad 
\bq=\int \frac{1}{2}\left(|\bxi - \bu|^2\hat f_{xy}+\hat f_{s,z}\right)(\bxi - \bu)\myd \bxi,
\end{aligned}
\end{equation}
where all vectors and tensors $\bm{\xi},~\bm{u},~\bm{q},~\bm{P}$ are in two-dimensional space. The dimensionless model equations of the reduced velocity distribution functions can be written as:
\begin{equation}
    \begin{aligned}
    &\frac{\partial \hat f_{xy}}{\partial t}+\bm{\xi}\cdot \frac{\partial \hat f_{xy}}{\partial \bm{x}} = \frac{g_{xy}-\hat f_{xy}}{\tau}, \\
    &\frac{\partial \hat f_{z}}{\partial t}+\bm{\xi}\cdot \frac{\partial \hat f_{z}}{\partial \bm{x}} = \frac{g_{z}-\hat f_{z}}{\tau}, \\
    \end{aligned}
\end{equation}
where the reduced reference velocity distribution functions are given by:
\begin{equation}\label{eq:reference_velocity distribution function_reduced}
    \begin{aligned}
&g_{xy}=\rho\left(\frac{1}{2\pi T}\right)\exp\left(-\frac{|\bxi-\bu|^2}{2T}\right) \left[1+\frac{2\bq \cdot \left(\bxi-\bu\right)}{5 \rho T^2}\left(\frac{ |\bxi-\bu|^2}{2 T}-2\right)\right],\\
&g_{z}=\rho\left(\frac{1}{2\pi}\right)\exp\left(-\frac{|\bxi-\bu|^2}{2T}\right) \left[1+\frac{2\bq \cdot \left(\bxi-\bu\right)}{5 \rho T^2}\left(\frac{ |\bxi-\bu|^2}{2 T}-1\right)\right].
    \end{aligned}
\end{equation}

\bibliographystyle{elsarticle-num}
\bibliography{ref}
\end{document}